\documentclass[reprint,
superscriptaddress,
amsmath,amssymb,
aps,
prx,
longbibliography
]{revtex4-1}

\usepackage{graphicx}
\usepackage{comment}
\usepackage{dcolumn}
\usepackage{color}
\usepackage{bm}
\usepackage{url}
\usepackage{braket}
\usepackage{amsfonts,amsmath,mathtools}
\usepackage{dsfont}
\usepackage{hyperref}
\usepackage[all]{hypcap}
\setcitestyle{numbers,square}

\newcommand{\ee}{\mathrm{e}}
\newcommand{\ii}{\mathrm{i}}
\newcommand{\dd}{\mathrm{d}}
\newcommand{\Tr}{\mathrm{Tr}}
\renewcommand{\Im}{\,\mathrm{Im}}

\newcommand*{\defeq}{\mathrel{\vcenter{\baselineskip0.5ex\lineskiplimit0pt\hbox{\scriptsize.}\hbox{\scriptsize.}}}=}

\begin{document}

\title{Tensor-network method to simulate strongly interacting quantum thermal machines}

\author{Marlon Brenes}
\email{Corresponding author: brenesnm@tcd.ie}
\affiliation{School of Physics, Trinity College Dublin, College Green, Dublin 2, Ireland}
\author{Juan Jos\'e Mendoza-Arenas}
\affiliation{Departamento de F\'isica, Universidad de los Andes, A.A. 4976, Bogot\'a D. C., Colombia}
\author{Archak Purkayastha}
\affiliation{School of Physics, Trinity College Dublin, College Green, Dublin 2, Ireland}
\author{Mark T. Mitchison}
\affiliation{School of Physics, Trinity College Dublin, College Green, Dublin 2, Ireland}
\author{Stephen R. Clark}
\affiliation{H. H. Wills Physics Laboratory, University of Bristol, Bristol BS8 1TL, United Kingdom}
\author{John Goold}
\affiliation{School of Physics, Trinity College Dublin, College Green, Dublin 2, Ireland}

\date{\today}

\begin{abstract}
We present a methodology to simulate the quantum thermodynamics of thermal machines which are built from an interacting working medium in contact with fermionic reservoirs at fixed temperature and chemical potential. Our method works at finite temperature, beyond linear response and weak system-reservoir coupling, and allows for non-quadratic interactions in the working medium. The method uses mesoscopic reservoirs, continuously damped towards thermal equilibrium, in order to represent continuum baths and a novel tensor network algorithm to simulate the steady-state thermodynamics. Using the example of a quantum-dot heat engine, we demonstrate that our technique replicates the well known Landauer-B\"{u}ttiker theory for efficiency and power. We then go beyond the quadratic limit to demonstrate the capability of our method by simulating a three-site machine with non-quadratic interactions. Remarkably, we find that such interactions lead to power enhancement, without being detrimental to the efficiency. Furthermore, we demonstrate the capability of our method to tackle complex many-body systems by extracting the super-diffusive exponent for high-temperature transport in the isotropic Heisenberg model. Finally, we discuss transport in the gapless phase of the anisotropic Heisenberg model at finite temperature and its connection to charge conjugation-parity, going beyond the predictions of  single-site boundary driving configurations.
\end{abstract}

\maketitle

\section{Introduction}
\label{sec:intro}

The miniaturisation of technologies in combination with the exquisite control now available over nanoscale systems has motivated increasing interest in thermal machines that operate in the quantum regime~\cite{Kosloff2014, Goold2016, Benenti2017, Binder2018, Mitchison2019}. While recent demonstrations with trapped ions~\cite{Rossnagel2016,Maslennikov2019,Horne2018, Lindenfels2019}, nanomechanical oscillators~\cite{Klaers2017} and diamond colour centres~\cite{Klatzkow2019} serve as impressive proofs of principle, practical applications such as thermoelectric power generation call for electronic devices. To that end, the focus of experiments in mesoscopic physics has expanded beyond traditional questions of charge transport to include the manipulation of heat currents in platforms such as semiconductor quantum dots~\cite{Linke2018}, superconducting circuits~\cite{Ronzani2018} and molecular junctions~\cite{Mosso2019}. Understanding the non-equilibrium thermodynamics of these systems is a formidable theoretical challenge, due to the simultaneous presence of strong system-reservoir coupling, interparticle interactions and finite temperatures. 

Existing approaches to modelling energy transport in complex quantum systems typically depend on perturbative arguments, which require a clear separation of energy or time scales. For example, a quantum master equation can be derived under the assumption of weak system-reservoir coupling~\cite{BreuerPetruccione}. However, the approximations needed to ensure positivity of the density matrix may fail to capture quantum coherences far from equilibrium~\cite{Wichterich2007,Purkayastha2016,Kirsanskas2018,Mitchison2018}, while a first-principles derivation requires full diagonalisation of the system Hamiltonian and thus becomes infeasible for large open systems. A more tractable approach for many-body problems is a local master equation, where incoherent sinks and sources create and remove excitations at the system's boundaries. This method has been successfully applied to study infinite-temperature transport in strongly interacting systems~\cite{Prosen_2015}, but its finite-temperature predictions may violate basic thermodynamic laws~\cite{Levy2014,Stockburger_2016,Gonzalez2017,Hofer2017} unless a specific kind of periodically modulated system-bath interaction is assumed~\cite{karevski2009quantum,Clark_2010,barra2015thermodynamic,Strasberg2017,De_Chiara_2018}. Alternatively, non-equilibrium Green functions~\cite{Stefanucci_2009} can be used to model energy transport under strong system-reservoir coupling, but at the cost of treating many-body interactions within the system perturbatively~\cite{Wang_2013,Talarico2019}. Another possibility is the numerical renormalisation group, which can handle strong interactions but is typically limited to near-equilibrium transport properties~\cite{Bulla2008}. The related chain representation of unitary system-bath dynamics~\cite{Prior2010} is also capable of non-perturbative transport calculations~\cite{Nuesseler2019} at finite temperatures~\cite{Tamascelli2019} but its scalability to large system size remains unclear.

In this work, we put forward a general and efficiently scalable numerical approach to quantum thermodynamics that can deal with simultaneously strong intra-system and system-bath interactions and which works arbitrarily far from equilibrium. We focus on autonomous thermal machines, where macroscopic fermion reservoirs held at different temperatures and chemical potentials drive currents through a complex quantum working medium
. 
We model the macroscopic reservoirs by a finite collection of fermionic modes that are continuously damped towards thermal equilibrium by an appropriate Lindblad master equation. We use a purification scheme based on auxiliary ``superfermion'' modes~\cite{Dzhioev2011} to compute the non-equilibrium steady states of both non-interacting and interacting working media. For interacting systems, we develop a tensor-network algorithm to efficiently simulate the real-time dynamics of the entire configuration, working directly in the energy eigenbasis of the reservoirs. Our approach is well suited to far-from-equilibrium problems in which all energy scales are comparable, such that perturbative or linear-response theories fail. To exemplify this, we demonstrate that the efficiency of a three-site quantum heat engine is enhanced by repulsive interactions and is further improved when the system-reservoir coupling is increased.

The concept of modelling infinite baths by a finite set of damped modes has been widely adopted and adapted since the seminal work of Imamoglu~\cite{Imamoglu1994} and Garraway~\cite{Garraway1997a,Garraway1997b}. In the context of open quantum systems coupled to bosonic reservoirs, this representation has been placed on a mathematically rigorous footing~\cite{Tamascelli2018,Mascherpa2019}, while its amenability to tensor-network simulations has been demonstrated~\cite{Somoza2019}. Related approaches have been used to study quantum heat engines~\cite{Strasberg_2016,Newman2017} and thermalisation in few-level~\cite{IlesSmith2014} and many-particle systems~\cite{Uzdin2018,Reichental2018}. In the fermionic setting, conditions under which continuum baths can be modelled by mesoscopic reservoirs have been recently discussed in Refs.~\cite{Gruss2016,Elenewski2017,Chen2019}. Such mesoscopic reservoirs have been used quite extensively over the last several years for studying transport in non-interacting systems~\cite{Dzhioev2011, Ajisaka2012, Ajisaka2013, Zelovich2014, Guimares2016,Gruss2016, Elenewski2017}, including under time-dependent driving fields~\cite{Oz2019}. For interacting systems, a mesoscopic-reservoir description was recently applied to study particle transport and Kondo phenomena in impurity models~\cite{Schwarz2016,Schwarz2018}, while a related approach to simulating non-equilibrium many-body problems via an auxiliary master equation has been reported~\cite{Dorda2015, Titvinidze2015}. 

A key feature of our work that differs from previous approaches is a novel tensor-network algorithm that exploits the superfermion representation to simulate Lindblad dynamics directly in the energy eigenbasis of the baths (the so-called star geometry). This configuration is particularly favourable in fermionic systems, where only a limited energy window participates in the dynamics at finite temperature due to Pauli exclusion effects at low energies. Although we focus here on steady states of autonomous machines, our methods can be adapted to study transient dynamics or time-dependent Hamiltonians. Moreover, our tensor-network algorithm is inherently scalable to many-body problems, as we demonstrate by first extracting the super-diffusive transport exponents of the isotropic Heisenberg model at high temperature, and then by studying finite-temperature regimes in the gapless phase of the anisotropic Heisenberg model beyond the predictions of single-site boundary driving configurations. Our work thus paves the way for simulations of heat transport in strongly correlated systems that probe heretofore inaccessible regimes of temperature and system size.

In the remainder of the article, we build our methodology step by step. We begin with an introduction to autonomous thermal machines in Sec.~\ref{sec:ATMs}, where the problem to be solved is precisely defined. We then outline the mesoscopic-reservoir approach and demonstrate its connection to the infinite-bath scenario in Sec.~\ref{sec:mesoscopic_leads}. Subsequently, in Sec.~\ref{sec:superfermion} we detail the superfermion representation and use it to find an analytical expression for the non-equilibrium steady state of a non-interacting (quadratic) system. In Sec.~\ref{sec:thermo_meso} we explain how to compute particle and energy currents within our framework.  Equipped with the exact solution for quadratic systems, in Sec.~\ref{sec:non_interacting_example} we study a non-interacting quantum-dot heat engine and compare the results with Landauer-B\"uttiker theory in order to identify the number and distribution of modes in the mesoscopic reservoirs needed to accurately reproduce the continuum limit. Next, in Sec.~\ref{sec:tensor} we detail our tensor-network algorithm for studying interacting problems. We then apply this algorithm in Sec.~\ref{sec:interacting} to study a three-site interacting heat engine and a many-body Heisenberg spin model at infinite and finite temperatures. Finally, we summarise and conclude in Sec.~\ref{sec:conclusions}.

\section{Autonomous quantum thermal machines}
\label{sec:ATMs}

This work is concerned with autonomous thermal machines whose working medium is a quantum system $\tt S$, which may be a complex entity comprising many interacting subsystems. The working medium is connected to multiple fermionic reservoirs labelled by the index $\alpha$. These reservoirs are macroscopic systems described by equilibrium temperatures $T_\alpha = 1/\beta_\alpha$ and chemical potentials $\mu_\alpha$ (we set $k_{\rm B} = 1 = \hbar$). The total Hamiltonian of such a setup takes the form 
\begin{equation}
    \label{eq:global_Hamitonian}
    \hat{H}_{\rm tot} = \hat{H}_{\tt S} + \sum_{\alpha}\left( \hat{H}_\alpha + \hat{H}_{\tt S \alpha}\right),
\end{equation}
where $\hat{H}_{\tt S}$ is the system Hamiltonian, $\hat{H}_\alpha$ is the Hamiltonian of bath $\alpha$ and $\hat{H}_{\tt S \alpha}$ describes its coupling to the system. We will consider exclusively Hamiltonians $\hat{H}_{\rm tot}$ that conserve fermion number $\hat{N} = \hat{N}_{\tt S} + \sum_\alpha \hat{N}_\alpha$, where $\hat{N}_{\tt S}$ and $\hat{N}_\alpha$ are the total particle number operators for the system and each bath $\alpha$, respectively.

Crucially, the baths are taken to have an infinite volume and heat capacity, implying a diverging number of degrees of freedom, $N\to\infty$. Moreover, it is typical to assume a factorised initial state of the form 
\begin{equation}
\label{eq:product_state}
    \hat{\rho}_{\rm tot}(0) = \hat{\rho}(0) \hat{\rho}_{\tt B},
\end{equation}
where $\hat{\rho}(0)$ is the initial system state and $\hat{\rho}_{\tt B}  = \prod_\alpha \hat{\rho}_\alpha$, with $\hat{\rho}_\alpha  = \ee^{-\beta_\alpha(\hat{H}_\alpha - \mu_\alpha \hat{N}_\alpha)}/Z_\alpha$ a thermal state and $Z_\alpha$ the partition function of each reservoir. Evolving into the long-time limit the system ${\tt S}$ will generically relax to a steady state given by 
\begin{equation}
    \label{eq:NESS_unitary}
    \hat{\rho}(\infty) = \lim_{t\to \infty} \lim_{N\to\infty} \Tr_{\tt B}\left[ \ee^{-\ii \hat{H}_{\rm tot}t} \hat{\rho}_{\rm tot}(0) \ee^{\ii \hat{H}_{\rm tot}t} \right],
\end{equation}
where $\Tr_{\tt B}$ denotes the trace over all bath degrees of freedom. If the temperatures or chemical potentials of the reservoirs differ, $\hat{\rho}(\infty)$ will be a non-equilibrium steady state (NESS) possessing currents of particles and energy.

\begin{figure}[t]
\fontsize{13}{10}\selectfont 
\centering
\includegraphics[width=0.8\columnwidth]{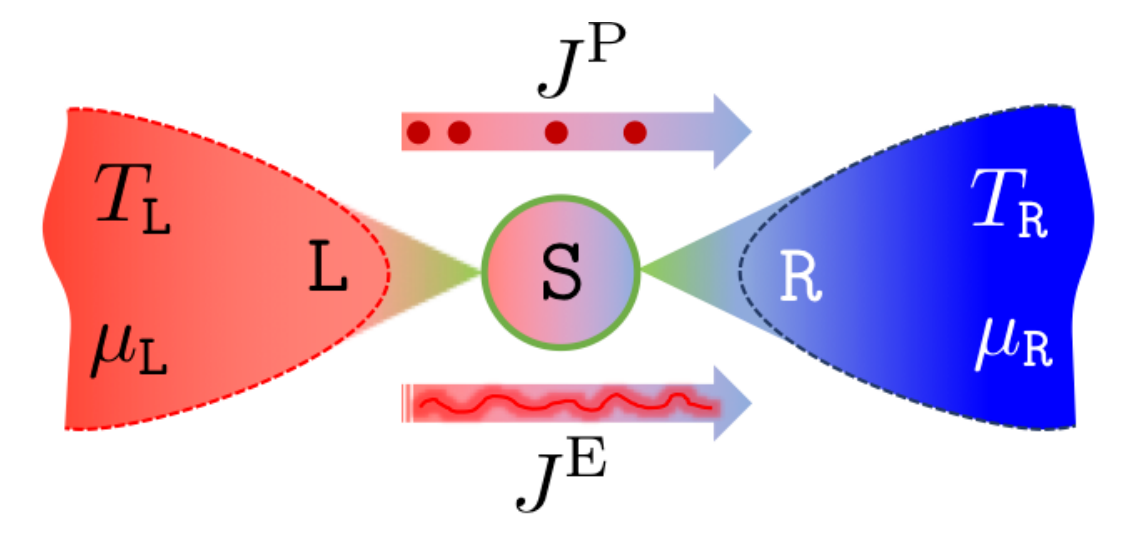}
\caption{A simple thermal machine scenario in which the system $\tt S$ is coupled to two reservoirs $\tt L$ and $\tt R$ at temperatures $T_{\tt L}> T_{\tt R}$ and possessing a chemical-potential difference $\mu_{\tt R} - \mu_{\tt L}> 0$. A particle $J^{\rm P}$ and energy $J^{\rm E}$ current is thus sustained through $\tt S$.}
\label{fig:thermal_machine}
\end{figure}

We focus especially on the simplest scenario depicted in Fig.~\ref{fig:thermal_machine}, with two reservoirs labelled by $\alpha = {\tt L},{\tt R}$. The sustained fluxes of particles and energy in this setup can be exploited, for example by operating the device as an autonomous heat engine. In this case a temperature gradient, $T_{\tt L}> T_{\tt R}$, drives a current that performs work by moving fermions against a chemical-potential difference $V = \mu_{\tt R} - \mu_{\tt L} > 0$. The power developed per unit time is given by
\begin{equation}
    \label{eq:power_def}
    P = V J^{\rm P},
\end{equation}
where $J^{\rm P}$ is the particle current, defined to be positive when flowing from left to right. The concomitant energy current $J^{\rm E}$ (also from left to right) transfers heat out of the left lead and into the right lead at a rate~\cite{Benenti2017}
\begin{equation}
    \label{eq:heat_current_def}
    \dot{Q}_\alpha = J^{\rm E} - \mu_\alpha J^{\rm P},
\end{equation}
so that the first law of thermodynamics can be written as $P = \dot{Q}_{\tt L} - \dot{Q}_{\tt R}$. The second law of thermodynamics imposes the relation $\beta_{\tt R} \dot{Q}_{\tt R} \geq \beta_{\tt L} \dot{Q}_{\tt L}$. The efficiency of heat-to-work conversion is thus given by 
\begin{equation}
    \label{eq:efficiency_def}
    \eta = \frac{P}{\dot{Q}_{\tt L}}  = 1 - \frac{\dot{Q}_{\tt R}}{\dot{Q}_{\tt L}}\leq \eta_{\rm C},
\end{equation}
where $\eta_{\rm C} = 1 - T_{\tt R}/T_{\tt L}$ is the Carnot efficiency. Thus, the performance of an autonomous thermal machine depends on the currents and their relationship to the thermodynamic properties of the reservoirs. 

Evaluating the currents requires finding the NESS of the quantum system. In general, however, the computation of Eq.~\eqref{eq:NESS_unitary} is a difficult task. Analytical solutions are available only if the global Hamiltonian is non-interacting, while a direct numerical approximation with finite baths may require prohibitively large values of $N$ in order to avoid Poincar\'e recurrences within the timescale of relaxation. On the other hand, perturbative schemes are limited to cases where either the internal interactions within $\tt S$ or its couplings to the reservoirs are weak. We thus take an alternative approach, in which the macroscopic reservoirs are replaced with mesoscopic leads comprising $L$ sites, which are continuously damped towards thermal equilibrium by dissipative processes. As a consequence, convergence can be obtained with only moderate values of $L$, bringing the non-equilibrium thermodynamics of complex many-body quantum systems within reach.

\section{From macroscopic reservoirs to mesoscopic leads}
\label{sec:mesoscopic_leads}

In this section, we detail our approach to studying the problem described in Sec.~\ref{sec:ATMs}, where an infinite bath is replaced by a finite collection of damped modes. Here we outline the idea, leaving the mathematical details in Appendix~\ref{app:meso_equivalence}.

\begin{figure}[b]
\fontsize{13}{10}\selectfont 
\centering
\includegraphics[width=0.8\columnwidth]{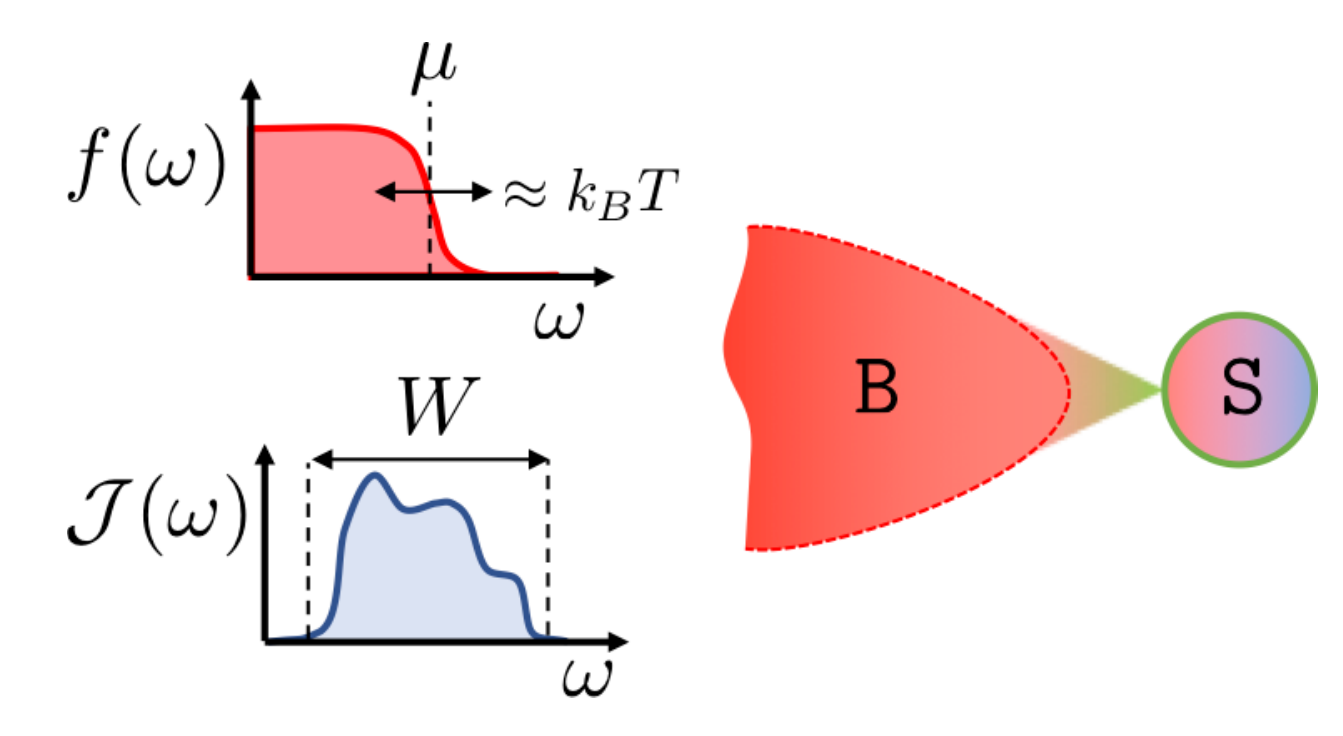}
\caption{The dynamics of a system coupled to a single thermal bath is determined by the bath's spectral density $\mathcal{J}(\omega)$, with a bandwidth $W$, and the Fermi-Dirac distribution $f(\omega)$ corresponding to its chemical potential $\mu$ and temperature $T$.}
\label{fig:single_bath}
\end{figure}

The system $\tt S$ is assumed to be a lattice of $D$ sites, with arbitrary geometry and interactions, while the baths are modelled by infinite collections of non-interacting spinless fermionic modes. To illustrate the approach, we consider first the case of a single bath $\tt B$, as shown in Fig.~\ref{fig:single_bath}, described by the Hamiltonian
\begin{align}
\label{eq:H_B_infinite}
\hat{H}_{\tt B} = \sum_{m=1}^\infty \omega_m \hat{b}^{\dagger}_m \hat{b}_m,
\end{align}
where $\hat{b}^\dagger_m$ creates a fermion with energy $\omega_m$. Each site $j$ of the system is described by a fermionic operator $\hat{c}_j$. A particular site $p$ of the system exchanges particles and energy with the bath via a tunnelling interaction
\begin{align}
\label{eq:H_SB_infinite}
\hat{H}_{\tt SB} = \sum_{m=1}^{\infty} \left( \lambda_{m} \hat{c}^{\dagger}_p \hat{b}_m + \lambda^*_{m} \hat{b}^{\dagger}_m \hat{c}_p \right),
\end{align}
where $\lambda_m$ is its coupling to bath mode $m$.

The Heisenberg equation for the system operators reads as
\begin{equation}
    \label{eq:Heisenberg_cj}
    \frac{{\rm d}}{{\rm d} t}\hat{c}_j(t) = \ii [\hat{H}_{\tt S}, \hat{c}_j(t)] + \delta_{jp} \left[\hat{\xi}(t) - \int_0^t {\rm d}t'\, \chi(t-t') \hat{c}_p(t')\right].
\end{equation}
Here, we have defined the noise operator
\begin{equation}
\label{eq:noise_operator}
\hat{\xi}(t) = -{\rm i}\sum_m \lambda_{m}\ee^{-\ii\omega_m t}\hat{b}_m,
\end{equation}
and the memory kernel $\chi(t-t') = \langle \{\hat{\xi}(t),\hat{\xi}^\dagger(t')\}\rangle $. The Gaussian statistics of the noise operator with respect to the initial product state Eq.~\eqref{eq:product_state} are defined by $\langle \hat{\xi}(t)\rangle  = 0$ and
\begin{align}
    \label{eq:memory_kernel}
    \langle \{\hat{\xi}(t),\hat{\xi}^\dagger(t')\}\rangle & = \int \frac{{\rm d}\omega}{2\pi} \mathcal{J}(\omega) \ee^{-\ii \omega(t-t')},\\
    \label{eq:noise_correlations}
    \langle \hat{\xi}^\dagger(t)\hat{\xi}(t')\rangle & = \int \frac{{\rm d}\omega}{2\pi} \mathcal{J}(\omega) f(\omega) \ee^{\ii \omega(t-t')},
\end{align}
where we have defined the spectral density as
\begin{equation}
    \label{eq:spectral_density_def}
    \mathcal{J}(\omega) = 2\pi\sum_{m=1}^\infty |\lambda_{m}|^2 \delta(\omega - \omega_m),
\end{equation}
and introduced the Fermi-Dirac distribution
$f(\omega) = (\ee^{\beta(\omega-\mu)} + 1)^{-1}$. The average system-bath coupling strength is typically quantified as
\begin{equation}
    \label{eq:coup_spectral_density}
    \Gamma = \frac{1}{2W}\int_{-\infty}^\infty {\rm d}\omega\, \mathcal{J}(\omega),
\end{equation}
where $2W$ denotes the reservoir bandwidth, namely the size of the energy range over which $\mathcal{J}(\omega)$ has support [see Eq.~\eqref{eq:wideband}, for example]. The state of $\tt S$ is completely determined by $f(\omega)$ and $\mathcal{J}(\omega)$ via the noise statistics, since for an overall closed system the solution of Eq.~\eqref{eq:Heisenberg_cj} is sufficient to reconstruct all $n$-point correlation functions.

Our approach is based on a key insight. Namely, that the open-system dynamics in Eq.~\eqref{eq:Heisenberg_cj}, induced by an infinite bath with spectral function $\mathcal{J}(\omega)$, can be accurately approximated by instead coupling the system to a finite collection of damped modes. Indeed, let us consider a {\em lead} of size $L$ coupled to site $p$ of the system, described by the Hamiltonian
\begin{align}
\label{eq:H_lead}
    \hat{H}_{\tt L} & = \sum_{k=1}^L \varepsilon_k \hat{a}^\dagger_k \hat{a}_k,\\
    \label{eq:H_lead_sys}
    \hat{H}_{\tt SL} & = \sum_{k=1}^{L} \left( \kappa_{kp} \hat{c}^{\dagger}_p \hat{a}_k + \kappa^*_{kp} \hat{a}^{\dagger}_k\hat{c}_p \right),
\end{align}
where $\hat{a}^\dagger_k$ creates a fermion in the lead with energy $\varepsilon_k$, and $\kappa_{kp}$ is the coupling strength. Each energy eigenmode $k$ of the lead is coupled to an independent thermal bath modelled by an infinite non-interacting fermion reservoir ${\tt B}_k$, as illustrated in Fig.~\ref{fig:meso_lead} (see Appendix~\ref{app:meso_equivalence} for details). These baths have identical temperatures and chemical potentials, but crucially they are characterised by a structureless frequency-independent spectral density $\mathcal{J}_k(\omega) = \gamma_k$, where $\gamma_k$ is a characteristic damping rate whose value may be different for each bath.

\begin{figure}
\fontsize{13}{10}\selectfont 
\centering
\includegraphics[width=1.0\columnwidth]{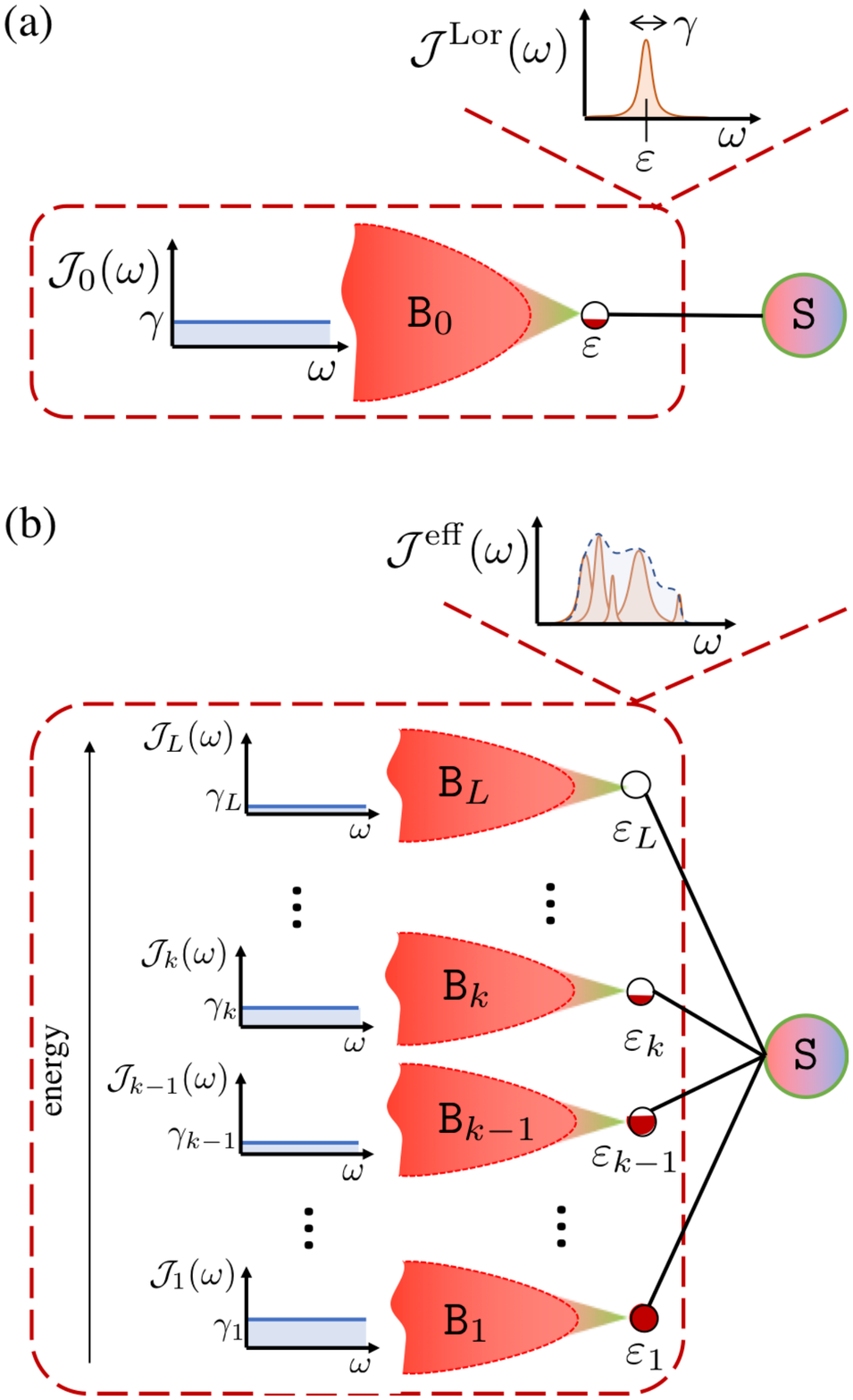}
\caption{(a)~A Lorentzian spectral density $\mathcal{J}^{\rm Lor}(\omega)$ is equivalent to coupling the system to a single auxiliary mode damped by a structureless reservoir. (b)~A mesoscopic reservoir comprising many damped modes gives rise to an effective spectral density $\mathcal{J}^{\rm eff}(\omega)$ that is a sum of Lorentzians. By tuning the damping of each mode and its coupling to the system $\mathcal{J}^{\rm eff}(\omega)$ can approximate $\mathcal{J}(\omega)$ of the infinite bath depicted in Fig.~\ref{fig:single_bath}.}
\label{fig:meso_lead}
\end{figure}

\begin{figure}[t]
\fontsize{13}{10}\selectfont 
\centering
\includegraphics[width=0.75\columnwidth]{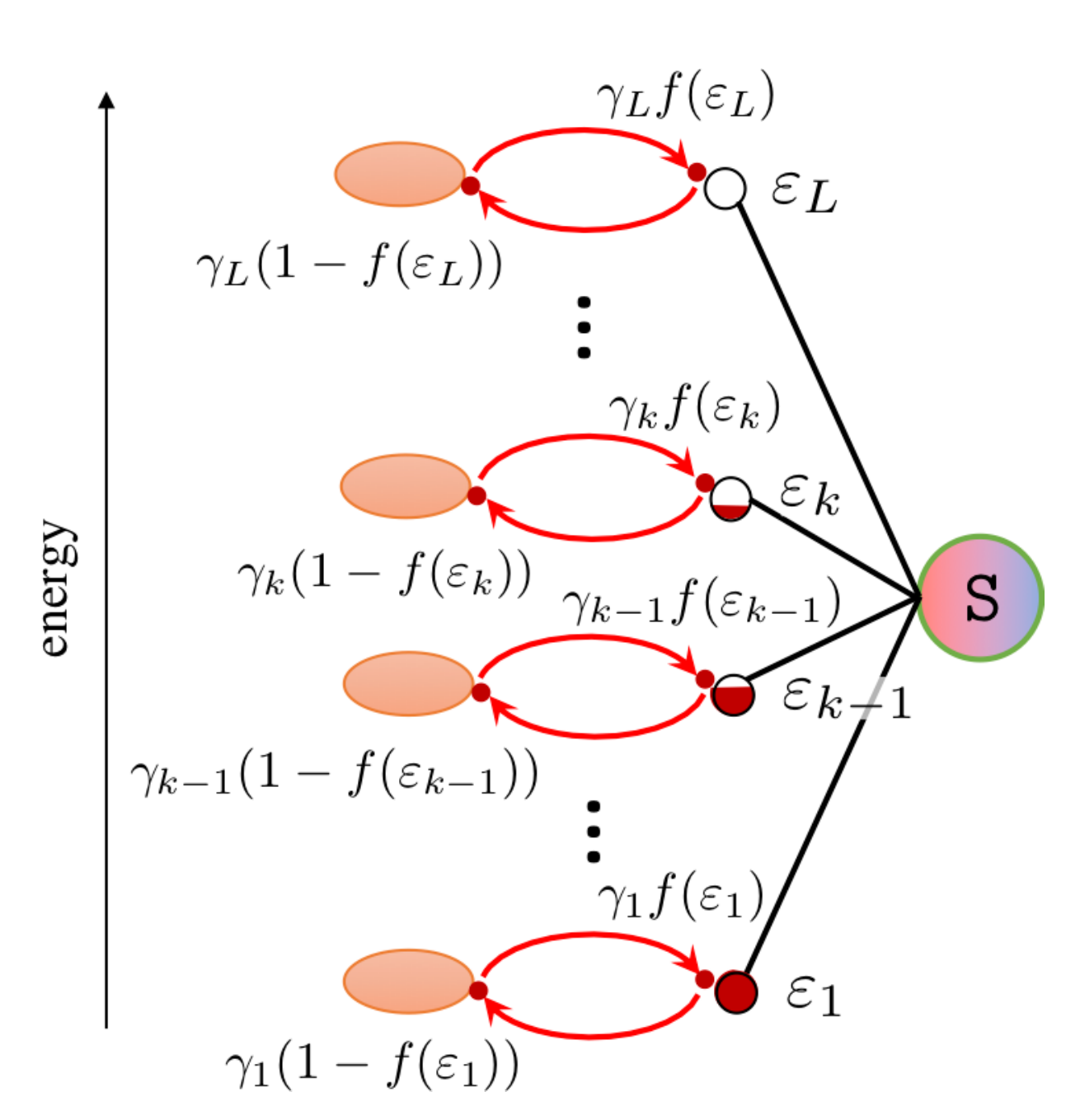}
\caption{In the limit $L\gg 1$ modes in the lead each bath ${\tt B}_k$ is sufficiently weakly coupled its corresponding lead mode that it can be accurately modelled by a Lindblad dissipator. The dissipator on a lead mode then that injects and ejects fermions at rates which in isolation damp the mode into a thermal state.}
\label{fig:meso_lead_lindblad}
\end{figure}

To analyse the steady-state physics it is sufficient to focus on long times, such that $t \gg \gamma_k^{-1},\tau_{\rm rel}$. Here $\tau_{\rm rel}$ represents the characteristic relaxation timescale of ${\tt S}$ due to its coupling with the bath~\footnote{Note that some systems, such as glassy systems, may never relax when coupled to a bath. In such cases, our arguments regarding the equivalence of mesoscopic and infinite reservoirs do not hold. Indeed, one expects that for such systems the effect of a bath must be highly dependent on the microscopic details of the bath and its coupling to the system.}. In this limit, we find that the Heisenberg equations for the system variables in this configuration are identical to Eq.~\eqref{eq:Heisenberg_cj}, but the statistics of the noise operator are now determined by an effective spectral density
\begin{equation}
    \label{eq:spectral_density_effective}
    \mathcal{J}^{\rm eff}(\omega)  = \sum_{k=1}^L \frac{|\kappa_{kp}|^2 \gamma_k}{(\omega-\varepsilon_k)^2 + (\gamma_k/2)^2}.
\end{equation}
It follows that this damped mesoscopic lead configuration reproduces the correct steady state of $\tt S$, so long as the true spectral density $\mathcal{J}(\omega)$ can be well approximated by a sum of Lorentzians as above. This is depicted in Fig.~\ref{fig:meso_lead}. In particular, consider a given set of lead energies $\varepsilon_k$ that sample the spectral density and are arranged in ascending order, with energy spacing $e_k = \varepsilon_{k+1} - \varepsilon_k$. By taking $\kappa_{kp} = \sqrt{\mathcal{J}(\varepsilon_k)e_k/2\pi}$ and $\gamma_k = e_k$, we have $\gamma_k \sim L^{-1}$ so that Eq.~\eqref{eq:spectral_density_effective} reduces to Eq.~\eqref{eq:spectral_density_def} in the limit $L\to \infty$. We therefore obtain a controlled approximation of the bath spectral function as the lead size $L$ increases.

In order to obtain a tractable description of the augmented system-lead configuration, we use the fact that both the damping rates $\gamma_k$ and the coupling constants $\kappa_{kp}$ are small in the large-$L$ limit. Tracing out the baths, we derive a master equation describing the joint state of $\tt S$ and $\tt L$, valid  for times $ t\gg \gamma_k^{-1},\tau_{\rm rel}$ and up to second order in both the lead-bath and system-lead coupling (see Appendix~\ref{app:meso_equivalence}). We emphasise that the assumption that \textit{individual} modes of the lead couple weakly to the system does not imply that the overall system-bath coupling $\Gamma$ is weak. The quantum master equation is
\begin{align}
    \label{eq:Lindblad}
    \frac{{\rm d}\hat{\rho}}{{\rm d}t} & = \ii[\hat{\rho},\hat{H}] + \mathcal{L}_{\tt L}\{\hat{\rho}\},
\end{align}
where $\hat{H} = \hat{H}_{\tt S} + \hat{H}_{\tt L} + \hat{H}_{\tt SL}$ denotes the Hamiltonian of the system and lead, while thermalisation of the lead is described by the Lindblad dissipator
\begin{align}
\label{eq:dissipator}
\mathcal{L}_{\tt L}\{\hat{\rho}\} = &\sum_{k=1}^{L} \gamma_k(1 - f_k) \left[\hat{a}_k \hat{\rho} \hat{a}^{\dagger}_k - \tfrac{1}{2}\{ \hat{a}^{\dagger}_k \hat{a}_k, \hat{\rho} \} \right] \nonumber \\
& + \sum_{k=1}^{L} \gamma_k f_k \left[\hat{a}^{\dagger}_k \hat{\rho} \hat{a}_k - \tfrac{1}{2}\{ \hat{a}_k \hat{a}^{\dagger}_k, \hat{\rho} \} \right].
\end{align}
with $f_k = f(\varepsilon_k)$ denoting the sampling of the Fermi distribution by the lead modes. This master equation configuration is illustrated in Fig.~\ref{fig:meso_lead_lindblad}. 

The above representation does not simplify the problem a priori, since it is strictly valid only in the large-$L$ limit. However, a simplification may arise if the expectation values of operators converge with increasing $L$. We show numerically in later sections that this convergence occurs rapidly in several examples of interest for quantum thermodynamics. In such cases, a tractable number of lead sites $L$ can be used to obtain a good approximation of an infinite bath with a continuous spectral density. For this, it is crucial that $\gamma_k$ remains the smallest energy scale in the physical configuration, to both model the spectral function correctly and accurately approximate the baths via the Lindblad equation~\cite{Guimares2016,Reichental2018}. 

So far we have considered a single bath coupled to a particular site of the system. However, the above results are easily generalised to describe the situation of several sites connected to multiple baths at different temperatures and chemical potentials. The steps of the above analysis are carried out independently for each bath, leading to additive contributions to the master equation. 

\section{Superfermion representation of non-equilibrium dynamics}
\label{sec:superfermion}

In order to solve the dissipative dynamics under a master equation of the form in Eq.~\eqref{eq:Lindblad}, we use the superfermion formalism introduced in Ref.~\cite{Dzhioev2011}. For a non-interacting (quadratic) open system, this method provides numerically tractable analytical expressions for steady-state quantities. The superfermion representation is also central to our approach to simulating interacting systems, as discussed in Sec.~\ref{sec:tensor}. Here, we limit ourselves to a concise review of the formalism; for more details, see Appendix~\ref{ap:superf}.

The superfermion approach is akin to a purification or thermofield scheme for open systems. It doubles the system size by introducing a new fermionic ancilla mode for each of the modes present in the system and leads. To describe the formalism succinctly we stick for now to the single-lead setup of Eq.~\eqref{eq:Lindblad}. In order to distinguish clearly between the ancillary modes and the physical modes of the system and lead, we introduce a unified notation for the latter. In this single-lead setup the total number of system and lead modes is $M = D+L$ and so we define $M$ fermion mode operators
\begin{eqnarray}
    \hat{d}_k \defeq \left\{
\begin{array}{cl}
 \hat{a}_k & \quad k=1,\ldots,L   \\
 \hat{c}_k & \quad k=(L+1),\ldots,M 
\end{array}
\right. . \label{eq:unified_notation}
\end{eqnarray}
The ancillary modes are described by $M$ additional canonical creation and annihilation operators $\hat{s}^{\dagger}_k$ and $\hat{s}_k$. We use an interleaved ordering for the physical and ancillary operators, so that the Fock basis of the combined Hilbert space is defined by
\begin{align}
\label{eq:superfermion_Fock_state}
\ket{\underline{n}| \underline{m}} = ( \hat{d}_1^{\dagger} )^{n_1}& ( \hat{s}_1^{\dagger} )^{m_1} \cdots ( \hat{d}_{M}^{\dagger} )^{n_{M}} ( \hat{s}_{M}^{\dagger} )^{m_{M}} \ket{\textrm{vac}}.
\end{align}
Here $\underline{n}$ are $\underline{m}$ are binary strings of length $M$ that describe occupation numbers for the physical and ancillary modes, respectively. While the ordering used for the Fock basis is entirely arbitrary, we shall see shortly that interleaving has useful locality properties exploited later in Sec.~\ref{sec:tensor}. We now define a new (unnormalised) ket vector called the {\em left vacuum} as
\begin{align}
\ket{I} \defeq \sum_{\underline{n}} \ket{\underline{n} |\underline{n}},
\end{align}
where the sum runs over all $2^{M}$ binary strings $\underline{n}$. Using this ket, we can define a quantum state representing the system-lead density operator as
\begin{align}
\hat{\rho}(t)\ket{I} = \ket{\hat{\rho}(t)},
\end{align}
and the expectation values of any system or lead operator $\hat{A}$ as
\begin{align}
\label{eq:expec}
\braket{ I | \hat{A} | \hat{\rho}(t) } = \langle \hat{A}(t) \rangle.
\end{align}

A key aspect of this formalism are the conjugation relations allowing physical creation (annihilation) operators to be swapped for ancillary annihilation (creation) operators. For the interleaved Fock ordering these conjugation relations are given by
\begin{align}
\hat{d}^{\dagger}_j \ket{I} &= -\hat{s}_j \ket{I},\quad \bra{I}\hat{d}_j = -\bra{I}\hat{s}^{\dagger}_j, \nonumber \\
\hat{d}_j \ket{I} &= \phantom{-}\hat{s}^{\dagger}_j \ket{I},\quad \bra{I}\hat{d}^{\dagger}_j = \phantom{-}\bra{I}\hat{s}_j.
\end{align}
Acting the master equation Eq.~\eqref{eq:Lindblad} on $\ket{I}$ and using the conjugation relations yields a Schr\"odinger-type equation for the state,
\begin{align}
\frac{\rm d}{{\rm d}t} \ket{\hat{\rho}(t)} = -\textrm{i}\hat{L}\ket{\hat{\rho}(t)}, \label{eq:superfermion_evolve}
\end{align}
with the (non-Hermitian) generator of time evolution given by
\begin{align}
\label{eq:l_single}
\hat{L} &= \hat{H} - \hat{H}_{d\Leftrightarrow s} \nonumber - \textrm{i}\sum_{k=1}^{L} \gamma_k f_k\\
&\quad-\frac{\textrm{i}}{2}\sum_{k=1}^{L} \gamma_k (1 - 2f_k) \left( \hat{d}^{\dagger}_{k} \hat{d}_{k} + \hat{s}^{\dagger}_{k} \hat{s}_{k} \right)  \nonumber \\
&\quad+\textrm{i}\sum_{k=1}^{L} \gamma_k \left( f_k \hat{d}^{\dagger}_{k} \hat{s}^{\dagger}_{k} -(1-f_k) \hat{d}_{k} \hat{s}_{k}  \right),
\end{align}
where $\hat{H}_{d\Leftrightarrow s}$ is the same as the system-lead Hamiltonian $\hat{H}$ but with all physical operators replaced by their ancillary counterparts, $\hat{d}_k \to \hat{s}_k$. Crucially, dissipative processes are now described by non-Hermitian quadratic operators that, according to the interleaved mode ordering of Eq.~\eqref{eq:superfermion_Fock_state}, couple only nearest neighbours $\hat{d}_{k}$ and $\hat{s}_{k}$. The formalism generalises straightforwardly to multiple leads by introducing an additional ancilla mode needed for each additional lead mode.

So far the superfermion formalism is entirely general. In the special case where the system Hamiltonian $\hat{H}_{\tt S}$ is non-interacting the formalism provides a compact expression for the exact solution of the NESS. In this case the system-lead Hamiltonian is quadratic with the form  
\begin{equation}
    \label{eq:quadratic_Hamiltonian}
    \hat{H} = \sum_{i,j = 1}^M [{\bf H}]_{ij} \hat{d}^\dagger_i \hat{d}_j,
\end{equation}
where ${\bf H}$ is an Hermitian $M\times M$ matrix. Next we define $M\times M$ diagonal matrices $\mathbf{\Gamma}_+$ and $\mathbf{\Gamma}_-$ containing the injection and ejection rates of fermions for each site. Specifically, for the single-lead setup the first $L$ follow the thermal damping rates contained in the dissipator Eq.~\eqref{eq:dissipator}, while the last $D$ entries corresponding to the system modes are zero, giving
\begin{eqnarray}
  \mathbf{\Gamma}_+ &=& {\rm diag}\Big(\gamma_1 f_1,\dots,\gamma_L f_L,0,\dots,0\Big), \nonumber \\
  \mathbf{\Gamma}_- &=& {\rm diag}\Big(\gamma_1 (1-f_1),\dots,\gamma_L (1-f_L),0,\dots,0\Big). \nonumber
\end{eqnarray}
Using these we define two additional diagonal matrices $\mathbf{\Lambda} = (\mathbf{\Gamma}_- + \mathbf{\Gamma}_+)/2$ and $\mathbf{\Omega} = (\mathbf{\Gamma}_- - \mathbf{\Gamma}_+)/2$. Consequently, for the case of a non-interacting system the generator $\hat{L}$ is quadratic with the form
\begin{align}
    \label{eq:L_generator_quadratic}
\hat{L} &= \hat{\mathbf{f}}^{\dagger} \begin{pmatrix} \mathbf{H} - \textrm{i}\mathbf{\Omega} & \textrm{i}\mathbf{\Gamma}_+ \\ \textrm{i}\mathbf{\Gamma}_- & \mathbf{H} + \textrm{i}\mathbf{\Omega} \end{pmatrix} \hat{\mathbf{f}} - \textrm{Tr}\left( \mathbf{H} + \textrm{i}\mathbf{\Lambda} \right) \nonumber \\
&= \hat{\mathbf{f}}^{\dagger}\, \mathbf{L}\, \hat{\mathbf{f}} - \eta,
\end{align}
where $\mathbf{\hat{f}}=(\hat{d}_1,\ldots,\hat{d}_M,\hat{s}^\dagger_1,\ldots,\hat{s}^\dagger_M)^{\rm T}$ is the full $2M$-dimensional column vector of all physical and ancillary operators~\footnote{Note that the ordering of operators in this vector is completely unrelated to that used to define the Fock basis.}. 

To determine the NESS we diagonalise $\hat{L}$ by a similarity transformation, $\mathbf{L} = \mathbf{V} \,\bm{\epsilon}\, \mathbf{V}^{-1}$, to find the complex eigenvalues $\bm{\epsilon} = \mathrm{diag}(\epsilon_1,\ldots,\epsilon_{2M})$ and the matrix of right eigenvectors $\mathbf{V}$ of $\mathbf{L}$. As shown in Appendix~\ref{ap:superf}, the many-body NESS is a Fermi-sea-like state in which only modes with $\Im(\epsilon_\mu)>0$ are occupied, furnishing us with a complete solution of the problem. In particular, two-point correlation functions of physical modes in the NESS are found to be
\begin{equation}
    \label{eq:RDM_SF_quadratic}
    \langle \hat{d}_i^\dagger \hat{d}_j\rangle = [\mathbf{V}\,\mathbf{D}\,\mathbf{V}^{-1}]_{ji},
\end{equation}
where $\mathbf{D}_{\mu\nu} = \delta_{\mu\nu} \Theta(\Im\{\epsilon_\mu\})$, with $\Theta(x)$ the Heaviside step function. This gives an efficient prescription to find steady state observables such as currents for non-interacting systems, while higher-order correlation functions follow from Wick's theorem.

\section{Non-equilibrium thermodynamics with mesoscopic leads}
\label{sec:thermo_meso}

\begin{figure}
\fontsize{13}{10}\selectfont 
\centering
\includegraphics[width=0.8\columnwidth]{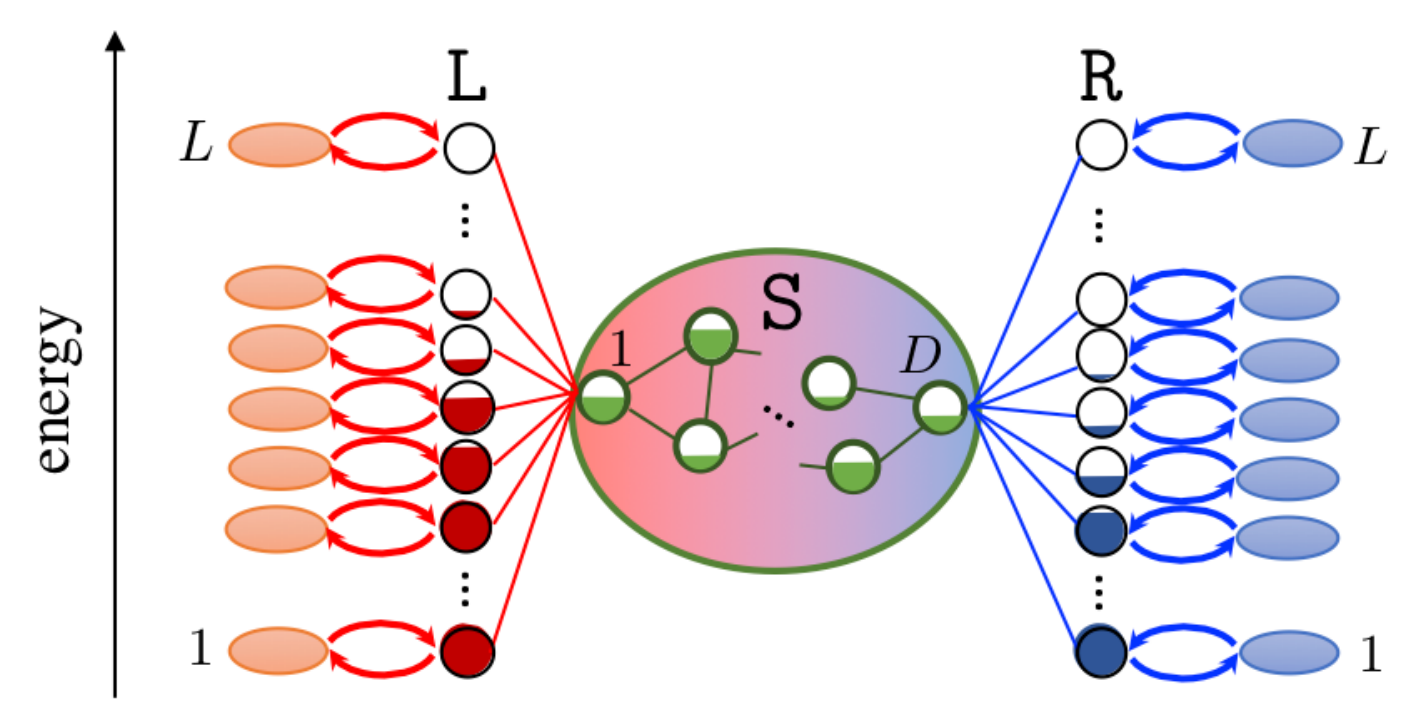}
\caption{The Lindblad mesoscopic lead approximation of the simple thermal machine setup shown in Fig.~\ref{fig:thermal_machine} where some generic system $\tt S$ is coupled to two reservoirs with differing chemical potentials and temperatures.}
\label{fig:meso_leads_lindblad}
\end{figure}

The central focus of our work is autonomous thermal machines in the two-lead configuration illustrated in Fig.~\ref{fig:meso_leads_lindblad}, with mesoscopic reservoirs labelled by $\alpha = {\tt L},{\tt R}$. These two leads of size $L$ are described by Hamiltonians of the form Eq.~\eqref{eq:H_lead} and Eq.~\eqref{eq:H_lead_sys}, where the left lead couples to the first system site, $p=1$, and the right lead to the last system site, $p=D$. Each lead is also acted on by a dissipator of the form given in Eq.~\eqref{eq:dissipator}. The master equation for this set-up thus reads as
\begin{equation}
    \label{eq:two_terminal_ME}
    \frac{{\rm d}\hat{\rho}}{{\rm d}t} = \ii[\hat{\rho},\hat{H}] + \mathcal{L}_{\tt L}\{\hat{\rho}\}+ \mathcal{L}_{\tt R}\{\hat{\rho}\},
\end{equation}
where $\hat{H} = \hat{H}_{\tt S} + \hat{H}_{\tt L} + \hat{H}_{\tt R} + \hat{H}_{\tt S L}+ \hat{H}_{\tt SR}$.

To find expressions for the particle and energy currents, we need to consider the continuity equations for the total particle-number operator $\hat{N} =  \hat{N}_{\tt S} + \hat{N}_{\tt L} + \hat{N}_{\tt R}$ and total energy operator $\hat{H}$ for the system and the leads. Since $[\hat{H},\hat{N}] = 0$, we derive
\begin{align}
\label{eq:continuity_equations}
\frac{{\rm d}\langle\hat{N}\rangle}{{\rm d}t}  =  J^{\rm P}_{\tt L} + J^{\rm P}_{\tt R}, \quad
\frac{{\rm d}\langle\hat{H}\rangle}{{\rm d}t}  =  J^{\rm E}_{\tt L} + J^{\rm E}_{\tt R},
\end{align}
where $J^{\rm P}_\alpha$ and $J^{\rm E}_\alpha$ are respectively the particle and energy currents flowing into the entire configuration via lead $\alpha$, given by
\begin{align}
    \label{eq:currents_def}
    J^{\rm P}_\alpha = \Tr \left[ \hat{N} \mathcal{L}_\alpha \{\rho\}\right], \quad {\rm and} \quad  J^{\rm E}_\alpha = \Tr \left[ \hat{H} \mathcal{L}_\alpha \{\rho\}\right].
\end{align}
In the NESS, the time derivatives in Eqs.~\eqref{eq:continuity_equations} vanish. Defining positive currents to flow across the system from left to right, we thus take $J^{\rm P} = J^{\rm P}_{\tt L} = -J^{\rm P}_{\tt R}$ and similarly $J^{\rm E} = J^{\rm E}_{\tt L} = - J^{\rm E}_{\tt R}$. Explicitly, we show in Appendix~\ref{ap:conteqs} that
\begin{align}
\label{eq:partsf_explicit}
J^{\rm P} & = \sum_{k=1}^L \gamma_k  \left\langle  f_{{\tt L},k} - \hat{a}_k^\dagger \hat{a}_k\right\rangle, \\
\label{eq:enersf_explicit}
J^{\rm E}  & = \sum_{k=1}^L \gamma_k \varepsilon_k  \left \langle f_{{\tt L},k} - \hat{a}_{k}^\dagger \hat{a}_{k} \right \rangle \notag \\
& \qquad -  \frac{1}{2}\sum_{k=1}^L \gamma_k  \left \langle \kappa_{k1}\hat{c}_1^\dagger \hat{a}_{k}  + \kappa_{k1}^* \hat{a}^\dagger_{k} \hat{c}_1 \right\rangle,
\end{align}
where the sum runs over only the modes of the left lead with $f_{\tt L}(\varepsilon) = (\ee^{\beta_{\tt L}(\varepsilon-\mu_{\tt L})} +1 )^{-1}$ being its corresponding equilibrium distribution and $f_{{\tt L},k} = f_{\tt L}(\varepsilon_k)$. 

For sufficiently large systems with short-range interactions~\footnote{For nearest-neighbour interactions $D \geq 3$ is sufficient.}, it is possible to define current operators $\hat{J}_{\tt S}^{{\rm P},{\rm E}}$ supported only on ${\tt S}$. In this case, we show in Appendix~\ref{ap:conteqs} that the expected values of these operators agree with the formulae given above, i.e. $\langle \hat{J}_{\tt S}^{{\rm P},{\rm E}}\rangle = J^{{\rm P},{\rm E}}$. However, in some cases, e.g.\ if ${\tt S}$ comprises just a single lattice site, no system operator for the currents can be defined. Nevertheless, whether or not such a system operator exists, we show in Appendix~\ref{app:meso_equivalence} that the average currents computed from Eqs.~\eqref{eq:partsf_explicit} and~~\eqref{eq:enersf_explicit} converge to the infinite-reservoir prediction when $L\to \infty$.

\section{Non-interacting example: the resonant-level heat engine}
\label{sec:non_interacting_example}

\begin{figure*}[t]
\fontsize{13}{10}\selectfont 
\centering
\includegraphics[width=1.8\columnwidth]{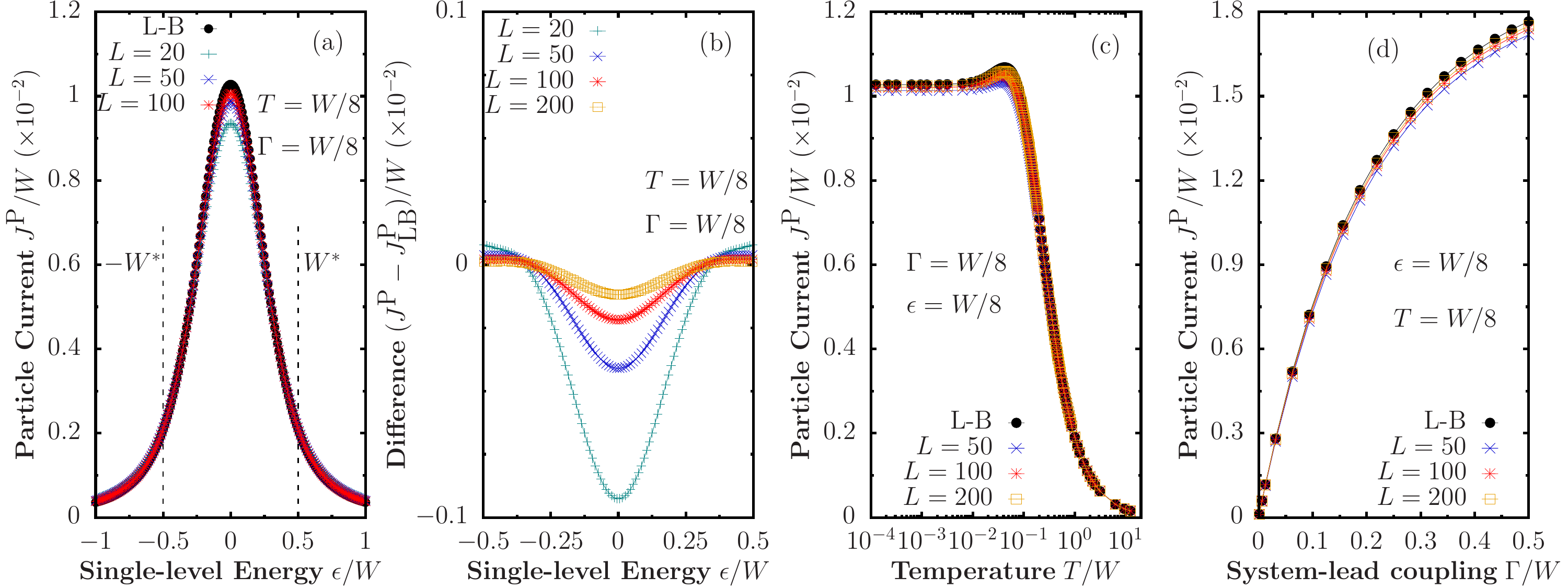}
\caption{Comparison between L-B predictions and the mesoscopic configuration of the expectation value of the total particle current flowing from the left lead through a single level, (a) as a function of the energy of the level, (b) absolute difference in the predictions from both scenarios with increasing number of modes in the leads $L$, (c) as a function of temperature and (d) as a function of the system-lead coupling strength $\Gamma$. In these calculations we used $\mu_{\texttt L} = -\mu_{\texttt R} = W / 16$, $T_{\texttt L} = T_{\texttt R}$, $L_{\textrm{log}} / L = 0.2$ and $W^* = W / 2$.}
\label{fig:6}
\end{figure*}

In this section, we apply our methods to analyse the performance of an autonomous thermal machine with a non-interacting working medium. Since exact results are available here for the $L\to\infty$ limit, this serves as a benchmark to evaluate the performance of the mesoscopic-reservoir formalism which can also be solved numerically exactly using the superfermion formalism. Using this we estimate the number of lead modes needed to accurately reproduce the continuum limit of infinite baths. 
We take a single resonant level as our working medium, described by the Hamiltonian
\begin{align}
\label{eq:h_s_d}
\hat{H}_{\tt S} = \epsilon\,\hat{c}^{\dagger}\hat{c},
\end{align}
where $\hat{c}^{\dagger}$ and $\hat{c}$ are the fermionic creation and annihilation operators in the system, respectively, and $\epsilon$ is the energy of the level. This models a single quantum dot in the spin-polarised regime running as a heat engine between two baths~\cite{Linke2010}. We note that a quantum-dot heat engine was recently realised experimentally~\cite{Linke2018}.

In principle, our methods can handle structured spectral densities that are different for each bath. For simplicity, however, we take both reservoirs to be characterised by identical, flat spectral densities within a finite energy band, given by
\begin{align}
\label{eq:wideband}
\mathcal{J}(\omega) = \begin{cases} \Gamma,\; \forall\, \omega \in [-W, W] \\ 0,\; \textrm{otherwise} \end{cases}
\end{align}
where $\Gamma$ is the coupling strength between the system and the leads. In the continuum limit of macroscopic baths, the particle and energy currents for a non-interacting system can be computed from the Landauer-B\"uttiker (L-B) formulae
\begin{align}
\label{eq:partlb}
J^{\textrm{P}}_{\textrm{LB}} & = \frac{1}{2\pi}\int_{-W}^{W} {\rm d} \omega\, \tau(\omega) [ f_{\tt L}(\omega) - f_{\tt R}(\omega) ],\\
\label{eq:enerlb}
J^{\textrm{E}}_{\textrm{LB}} & = \frac{1}{2\pi}\int_{-W}^{W} {\rm d} \omega\,  \omega\tau(\omega) [ f_{\tt L}(\omega) - f_{\tt R}(\omega) ],
\end{align}
where $f_\alpha(\omega)$ denotes the Fermi-Dirac distribution for lead $\alpha = {\tt L,R}$ and $\tau(\omega)$ is the transmission function. The latter is computed using the formalism described in Appendix \ref{ap:transmission}.

In the mesoscopic-reservoir approach, the spectral density is sampled by a finite number $L$ of lead modes, as in Eq.~\eqref{eq:spectral_density_effective}. Taking the distribution of lead mode energies $\{\varepsilon_k\}$, widths $\{\gamma_k\}$ and couplings $\{\kappa_{kp}\}$ to be identical for each lead, there remains significant freedom to choose these distributions in order to well approximate the continuum limit using moderate values of $L$. In particular, we use the logarithmic-linear discretisation scheme proposed in Refs.~\cite{Schwarz2016,vondelft2009}. Here, $L_{\textrm{lin}}$ modes are placed in the energy window $[-W^*, W^*]$, with equally spaced frequencies, i.e.\ $e_k = \varepsilon_{k+1} - \varepsilon_k = 2W^* / L_{\textrm{lin}}.$ Energies outside of this range are sampled by a smaller set of modes $L_{\textrm{log}}$, with frequencies logarithmically spaced from $W^*$ ($-W^*$) to $W$ ($-W$), with energy intervals $[\varepsilon_{n-1}, \varepsilon_n] = [\pm \Lambda^{-(n-1)}, \pm \Lambda^{-n}]$ for $n = 1, \cdots, L_{\textrm{log}}$ and $\Lambda^{-L_{\rm log}} = W^*$. The dissipation rates are taken equal to these spacings, $\gamma_k = e_k$, while the coupling constants $\kappa_{kp}$ ($p=1,D$) are determined by the equation $\Gamma = 2\pi \kappa_{kp}^2/e_k$~\cite{Dzhioev2011}, in accordance with the considerations of Sec.~\ref{sec:mesoscopic_leads}. For a given number of modes $L = L_{\rm log} + L_{\rm lin}$, this discretisation scheme gives better resolution within a smaller energy window $[-W^*, W^*]$ that includes the most relevant energy scales for the problem at hand. We remark that this discretisation scheme was chosen due to the featureless nature of $\mathcal{J}(\omega)$ in Eq.~\eqref{eq:wideband} to contain more energy modes in a given transport window. If $\mathcal{J}(\omega)$ was structured a different discretisation scheme to resolve its features could provide a better approximation of the spectral function. With respect to smooth spectral functions, however, we expect the chosen discretisation scheme to yield accurate results as the number of modes is increased. In our calculations, we henceforth set $W = 8$ and use this parameter as the overall energy scale, while $W^* = W / 2$. Moreover, we choose $L_{\textrm{log}} / L = 0.2$.

\begin{figure*}[t]
\fontsize{13}{10}\selectfont 
\centering
\includegraphics[width=1.8\columnwidth]{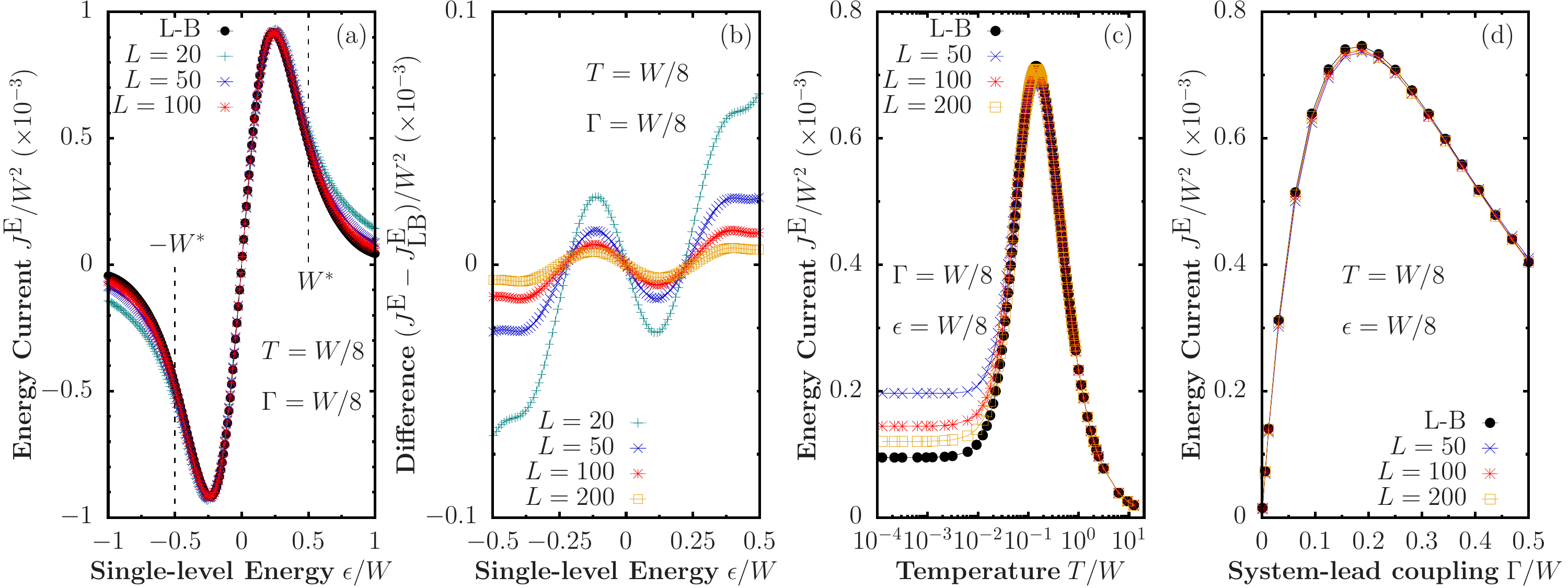}
\caption{Comparison between L-B predictions and the mesoscopic configuration of the expectation value of the total energy current flowing from the left lead through a single level, (a) as a function of the energy of the level, (b) absolute difference in the predictions from both scenarios with increasing number of modes in the leads $L$, (c) as a function of temperature and (d) as a function of the system-lead coupling strength $\Gamma$. In these calculations we used $\mu_{\texttt L} = -\mu_{\texttt R} = W / 16$, $T_{\texttt L} = T_{\texttt R}$, $L_{\textrm{log}} / L = 0.2$ and $W^* = W / 2$.}
\label{fig:7}
\end{figure*}

Under these conditions, we show in Fig.~\ref{fig:6} the behaviour of the particle current, where we have set equal temperatures in the leads $T_{\tt L} = T_{\tt R} = W / 8$ but used different chemical potentials $\mu_{\tt L} = -\mu_{\tt R} = W / 16$. In Fig.~\ref{fig:6}(a) we show the results for the particle current as a function of the system energy $\epsilon$ for different numbers of modes $L$ in the leads and compare it with L-B theory. From both Fig.~\ref{fig:6}(a) and Fig.~\ref{fig:6}(b), it can be observed that a good agreement is obtained, the biggest difference observed as $\epsilon \to 0$, when the current reaches its maximum value. As expected, the agreement is improved with increasing $L$, although even moderate values of $L \sim O(10)$ approximately reproduce the continuum. In our calculations, we fixed the bath parameters as we varied the self-energy of the single-level $\epsilon$, however, the approximation could be improved by adapting the mode distribution around the relevant transport window dictated by $\epsilon$. Furthermore, in Fig.~\ref{fig:6}(c) we fix the energy $\epsilon$ of the level to study the behaviour with increasing $L$ as a function of temperature $T_{\tt L} = T_{\tt R} = T$ with system-lead coupling strength $\Gamma$ fixed, and in Fig.~\ref{fig:6}(d) the behaviour with $\Gamma$ for fixed $T$. For this particular choice of parameters we find the particle currents are robust to a wide range of $T$ and $\Gamma$. Either low or high temperatures and weak or strong coupling yield similar results in both continuum or mesoscopic scenarios, even for a moderate number of modes in the mesoscopic leads.

\begin{figure*}[t]
\fontsize{13}{10}\selectfont 
\centering
\includegraphics[width=1.8\columnwidth]{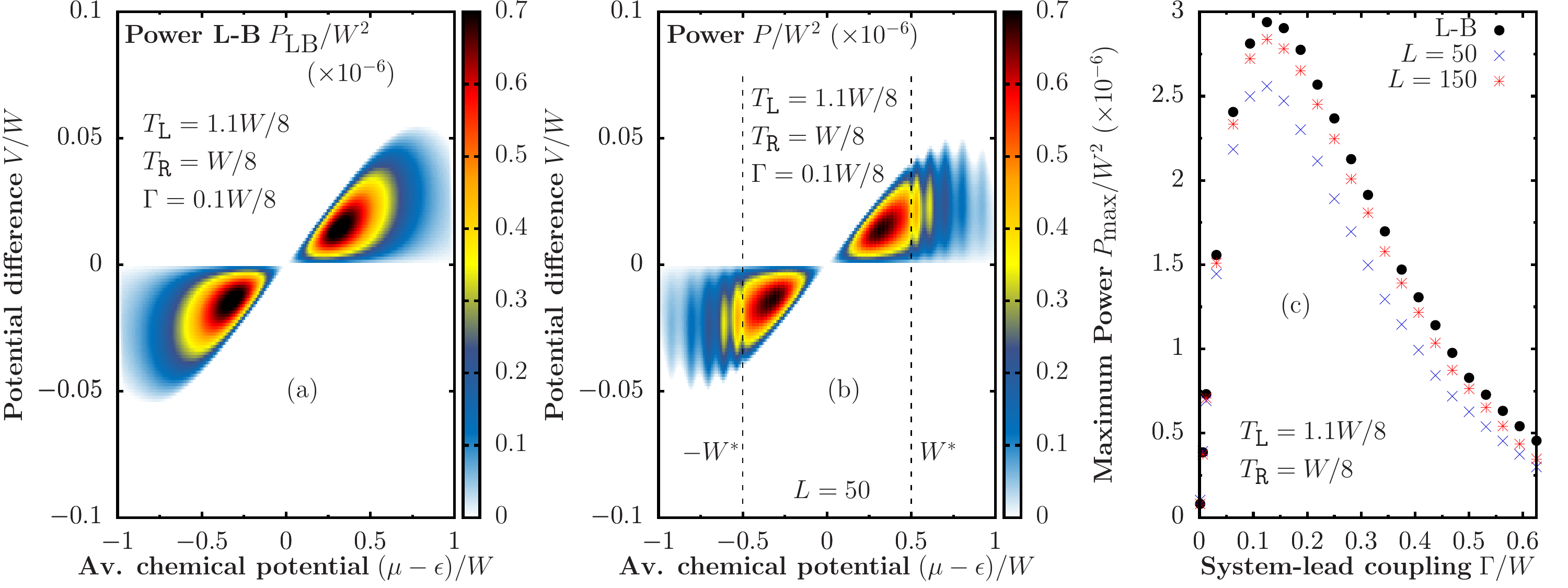}
\caption{Power as a function of potential difference $V$ and average chemical potential $\mu - \epsilon$ for the single-level system using (a) continuum leads and (b) mesoscopic reservoirs. In (c) we present the maximum power as a function of the system-lead coupling for both configurations. In (b) and (c) we used $L_{\textrm{log}} / L = 0.2$ and $W^* = W / 2$.}
\label{fig:8}
\end{figure*}

\begin{figure*}[t]
\fontsize{13}{10}\selectfont 
\centering
\includegraphics[width=1.8\columnwidth]{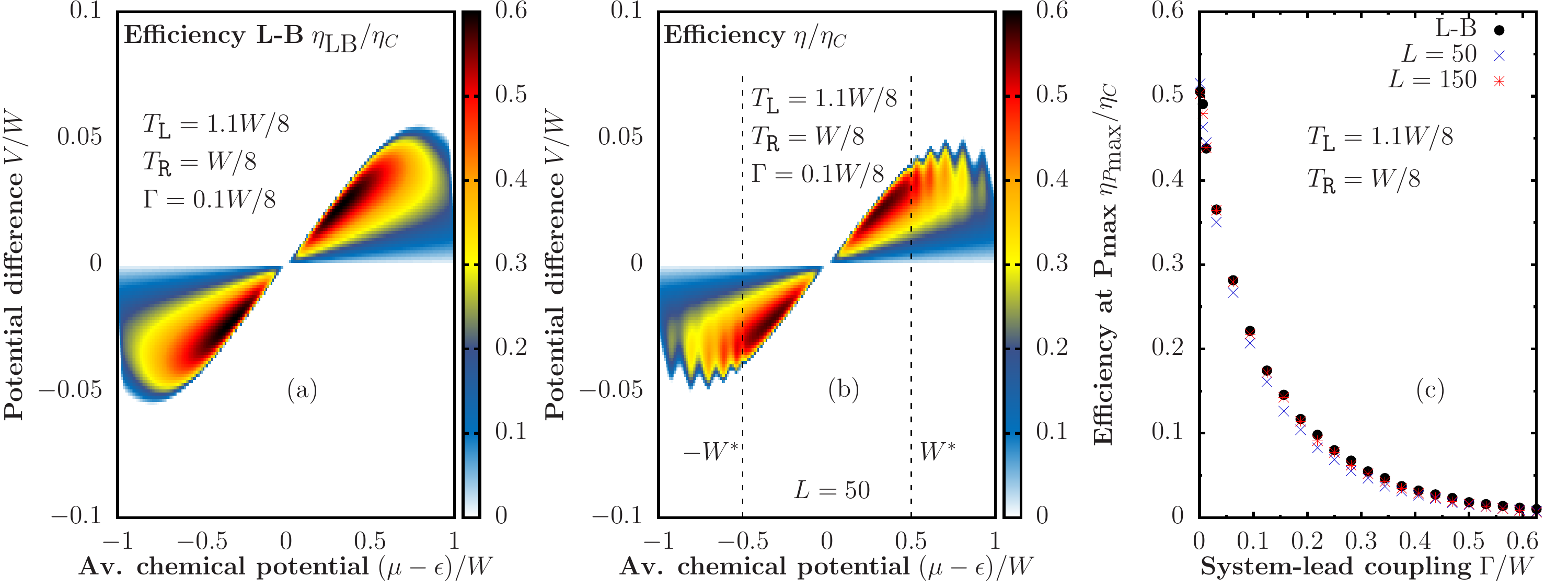}
\caption{Efficiency (normalised by the Carnot efficiency $\eta_C$) as a function of potential difference $V$ and average chemical potential $\mu - \epsilon$ for the single-level system using (a) continuum leads and (b) mesoscopic reservoirs. In (c) we present the efficiency at maximum power as a function of the system-lead coupling for both configurations. In (b) and (c) we used $L_{\textrm{log}} / L = 0.2$ and $W^* = W / 2$.}
\label{fig:9}
\end{figure*}

In Fig.~\ref{fig:7} we show the corresponding results for energy current. From Fig.~\ref{fig:7}(a) it can be observed that a better approximation is obtained when the number of modes in each lead is increased for a fixed set of parameters, with the absolute difference decreasing as a function of $L$, as can be concluded from Fig.~\ref{fig:7}(b). In Fig.~\ref{fig:7}(c) a key difference can be observed from the results obtained for particle current. The mesoscopic lead configuration is a good approximation as long as $T$ is kept above a given threshold. This threshold is dictated by the smallest energy spacing in the leads $e_k$ and can be understood as follows. The effective spectral function of the mesoscopic leads is a sum of Lorentzian peaks, as in Eq.~\eqref{eq:spectral_density_effective}. When the temperature is smaller than the minimum energy spacing $e_k$ in the mesoscopic lead, these peaks are too far apart to properly resolve the variation of the Fermi-Dirac distribution. In this regime, the noise statistics given by Eq.~\eqref{eq:noise_correlations} are significantly modified and the approximation is not reliable. It can be observed from Fig.~\ref{fig:7}(c) that the approximation at lower temperatures is much better for larger leads~\footnote{One method that can be used to obtain a better approximation at lower temperatures, that reduces the value of $e_k$ in the leads and without increasing the number of modes, is to change the width and position of the window $[-W^*, W^*]$ depending on the region of the parameter space that needs to be resolved in greater detail.}. 

In Fig.~\ref{fig:7}(d) we analyse the energy current as a function of the system-lead coupling $\Gamma$. We observe that the approximation for energy current in the mesoscopic lead configuration is quite robust to a wide range of couplings. This provides further evidence that the accuracy of the approximation is primarily determined by the size of $\gamma_k$ and $e_k$ relative to the temperature and voltage bias of the reservoirs~\cite{Gruss2016}.
 
Next we evaluate the power and efficiency given by Eqs.~\eqref{eq:power_def} and \eqref{eq:efficiency_def}. In Fig.~\ref{fig:8}(a) we show the power output as a function of average chemical potential $\mu = (\mu_{\tt L}+\mu_{\tt R})/2$ and the potential difference $V = \mu_{\tt R} - \mu_{\tt L}$ using the L-B prediction for continuum leads. In our calculations we set $T_{\tt L} = 1.1W/8$ and $T_{\tt R} = W/8$ and show the power output results only for the values of $\mu - \epsilon$ and $V$ for which the system acts as a power generator. It can be observed that the power output reaches a maximum value depending on bias and average chemical potential. In Fig.~\ref{fig:8}(b) we show the results for the same calculation, but instead we substitute the continuum leads with our mesoscopic lead configuration. The results are in good agreement up to the point where $\mu - \epsilon$ reaches the boundary of linearly discretised and logarithmically discretised lead modes. Beyond $W^*$ and $-W^*$, the spectral function is not sampled as finely and the power output results get distorted. We note that the window can be increased to resolve a bigger set of the parameter space, however, this would require more lead modes to resolve the maximum power output with the same accuracy. Alternatively, the range of linearly discretised modes could be adapted for each value of $\epsilon$ to ensure that the relevant energy range for transport is always included within this window. In Fig.~\ref{fig:8}(c) we show the maximum power output $P_{\textrm{max}}$ as a function of the system-lead coupling for both the L-B and mesoscopic lead predictions, which in turn reveals the value of $\Gamma_\textrm{max}$ for which $P_{\textrm{max}}$ reaches its maximum value. With our choice of parameters, $\Gamma_{\textrm{max}}$ lies very close in both configurations, as well as the overall behaviour as a function of system-lead coupling. The absolute value of the maximum power is better approximated, following the expected behaviour from Fig.~\ref{fig:6}(a), as the number of lead modes is increased.

In Fig.~\ref{fig:9}(a) we show the efficiency obtained using continuum leads, normalised by the Carnot efficiency. It can be observed that the points of maximum efficiency lie close to the boundary where the system stops operating as an engine, i.e., where the potential difference becomes too large for the temperature gradient to drive electrons in the opposite direction of the bias. In Fig.~\ref{fig:9}(b) we present the results for the mesoscopic lead configuration. As before, we find that both predictions are quantitatively similar up to the point where the boundary of $W^*$ is reached. In Fig.~\ref{fig:9}(c) we show the efficiency at the point where the maximum power is obtained from the configuration as a function of $\Gamma$, where we observe that both the continuum and mesoscopic lead configurations predict very similar results, even with a moderate number of lead modes. As expected, the approximation becomes more accurate as the lead size is increased. Furthermore, not only is the strong system-lead coupling behaviour well-captured, but so is the Curzon-Ahlborn efficiency limit (approximately given by $\eta_C / 2$) at weak coupling \cite{Curzon1975}.

\section{Tensor network approach}
\label{sec:tensor}

Having established that relatively modest sized mesoscopic leads can capture the continuum behaviour of a non-interacting system we now move on to consider the highly non-trivial problem of interacting systems. To do this we introduce in this section a tensor network based numerical method that can efficiently and accurately compute the interacting NESS of the the two-reservoir problem illustrated in Fig.~\ref{fig:meso_leads_lindblad}. To describe the method we will return briefly to the single-lead configuration shown in Fig.~\ref{fig:meso_lead_lindblad} in which the first site $p=1$ of the system $\tt S$ is coupled to the mesoscopic lead. Since we will exploit the superfermion formalism we continue to use the unified notation for modes $\hat{d}_k$ given in Eq.~\eqref{eq:unified_notation}.

\subsection{Spin-1/2 representation}
\label{sec:s12_rep}
\begin{figure}[b]
\fontsize{13}{10}\selectfont 
\centering
\includegraphics[width=0.6\columnwidth]{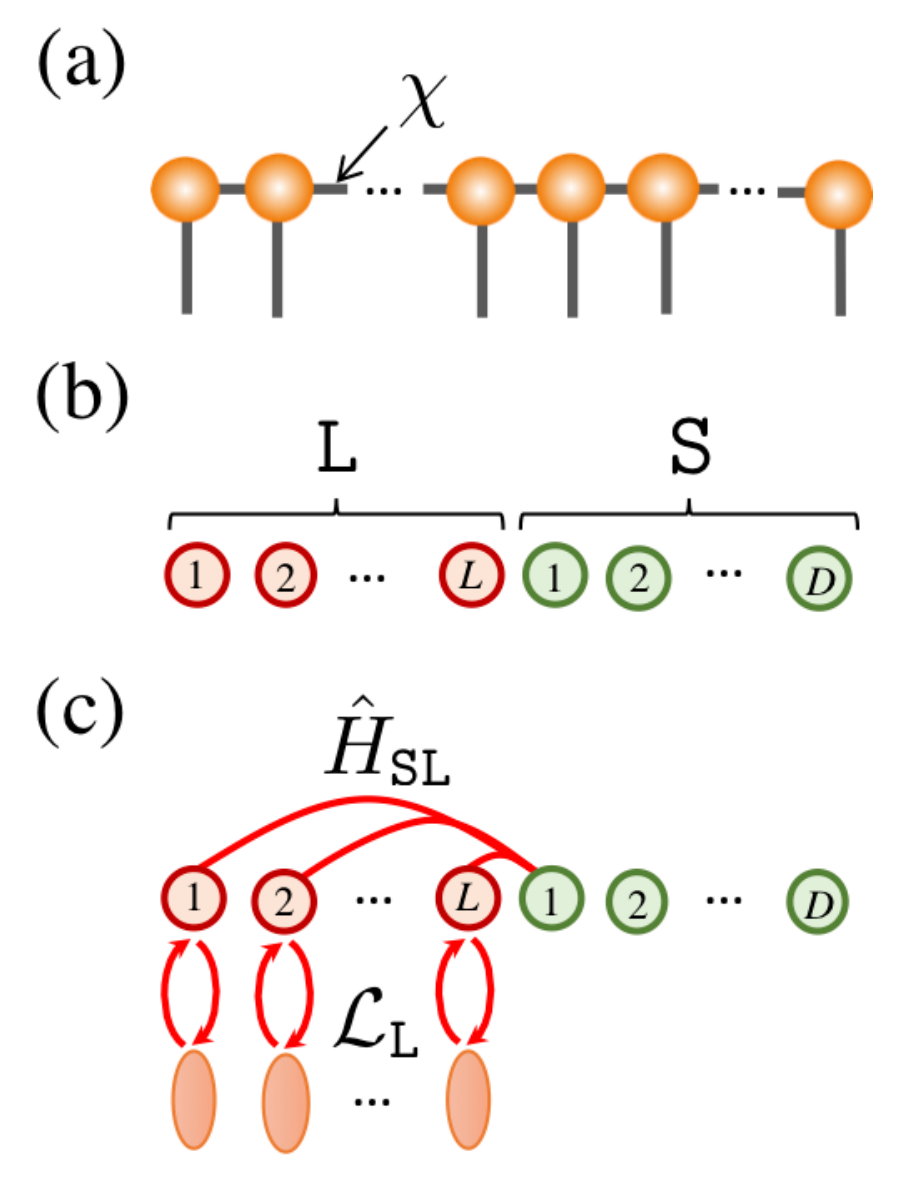}
\caption{(a) A MPS tensor network in which every site (except the boundaries) have an order-3 tensor associated to it. The vertical dangling legs are the physical indices of the system of dimension 2 in our case, the horizontal contracted legs are the internal bonds of the MPS of dimension $\chi$. (b) The lead and system modes are ordered into a one-dimensional geometry to match the MPS. (c) With this ordering the star geometry system-lead coupling $\hat{H}_{\tt SL}$ is long-ranged and the local fermionic dissipators $\mathcal{L}_{\tt L}$ on the lead also become long-ranged due to JW strings.}
\label{fig:mps_star}
\end{figure}

Our approach uses the matrix product state (MPS) decomposition that is a tensor network with a one-dimensional chain-like geometry~\cite{schollwock:2011}, as shown in Fig.~\ref{fig:mps_star}(a). To apply this powerful ansatz to our setup we first map the lead and system modes into a one-dimensional chain. In doing so the coherent coupling between the lead modes and the system become long-ranged within this chain since they corresponding to a so-called ``star geometry''. Fundamentally this is because we use the energy eigenbasis of the lead.

Additionally, since MPS apply to systems built from a tensor product of local Hilbert spaces, to describe a spinless fermionic system requires that we transform it into a spin-1/2 representation. Our starting point is to introduce Fock states constructed from the unified physical modes with occupation-number vector $\underline{n}$ as
\begin{align} \label{fock_states}
\ket{\underline{n}} = \left( \hat{d}_1^{\dagger} \right)^{n_1} \cdots \left( \hat{d}_{M}^{\dagger} \right)^{n_{M}}  \ket{\textrm{vac}},
\end{align}
which in the single-lead case has $M=L+D$ and is ordered with lead modes first, as shown in Fig.~\ref{fig:mps_star}(b). A spin-1/2 representation is then obtain via the well-known Jordan-Wigner (JW) transformation involving $M$ spins \cite{Jordan1928,Coleman2015}   
\begin{align}
\hat{d}^{\dagger}_j &= \left( \prod_{q=1}^{j-1} \hat{\sigma}^z_q \right) \hat{\sigma}^{-}_j,
\end{align}
where $\hat{\sigma}^z_q$ is the Pauli spin matrix in the $z$ direction and $\hat{\sigma}^{\pm}_q$ are the spin raising/lowering operators for the $q$-th spin. Correspondingly, the Fock states of Eq.~\eqref{fock_states} are equivalent to the spin states
\begin{align}
\ket{\underline{n}} = \left( \hat{\sigma}^-_1 \right)^{n_1} \cdots  \left( \hat{\sigma}^-_{M} \right)^{n_{M}} \ket{\uparrow \cdots \uparrow},
\end{align}
since each JW string vanishes on polarised spins it is applied to. Transforming the total Hamiltonian $\hat{H} =  \hat{H}_{\tt S} + \hat{H}_{\tt L} + \hat{H}_{\tt SL}$ [from Eqs.~\eqref{eq:H_lead} and \eqref{eq:H_lead_sys}] to this representation gives
\begin{align}
\label{eq:h_jw}
\hat{H} &= \hat{H}_{\tt S} + \sum_{k=1}^L \left\{ \kappa_{k1} \hat{\sigma}^+_{k}\left( \prod_{q=k+1}^{L} \hat{\sigma}^z_q \right)\hat{\sigma}^-_{L+1} \right. \\
&\left. \quad+ \kappa^*_{k1}\hat{\sigma}^-_{k} \left( \prod_{q=k+1}^{L} \hat{\sigma}^z_q \right)\hat{\sigma}^+_{L+1} \right\} + \sum_{k=1}^L \varepsilon_k \hat{\sigma}^-_{k} \hat{\sigma}^+_{k}.\notag
\end{align}
The star geometry, shown in Fig.~\ref{fig:mps_star}(c), thus introduces JW strings to the lead-system coupling terms making them long-ranged multi-body spin operators. Similarly, the Lindblad dissipator of Eq.~\eqref{eq:dissipator} becomes
\begin{align}
\label{eq:dissipator_jw}
\mathcal{L}_{\tt L}&\{\hat{\rho}\}  = \sum_{k=1}^{L} \gamma_k(1 - f_k) \Biggl[- \tfrac{1}{2}\{ \hat{\sigma}^-_{k} \hat{\sigma}^+_{k}, \hat{\rho} \} \nonumber \\ &\qquad\qquad\qquad + \hat{\sigma}^+_{k} \left( \prod_{q=1}^{k-1} \hat{\sigma}^z_q \right) \hat{\rho} \left( \prod_{q=1}^{k-1} \hat{\sigma}^z_q \right) \hat{\sigma}^-_{k} \Biggl] \nonumber \\
&\qquad\qquad + \sum_{k=1}^{L} \gamma_k f_k \Biggl[- \tfrac{1}{2}\{ \hat{\sigma}^+_{k} \hat{\sigma}^-_{k}, \hat{\rho} \} \nonumber \\ &\qquad\qquad\qquad + \hat{\sigma}^-_{k} \left( \prod_{q=1}^{k-1} \hat{\sigma}^z_q \right) \hat{\rho} \left( \prod_{q=1}^{k-1} \hat{\sigma}^z_q \right) \hat{\sigma}^+_{k} \Biggl],
\end{align}
showing that the jump operators are now also non-local due to the JW strings.

\subsection{Superfermion representation}
\label{sec:superfermion_spin}

By using the energy eigenbasis of the lead we have arrived at a master equation with a highly non-local multi-body Hamiltonian and dissipator. The JW strings therefore appear to severely frustrate the use of MPS algorithms in this setup. Typically those arising from the star geometry of the Hamiltonian in Eq.~\eqref{eq:h_jw} are dealt with by tridiagonalising the lead Hamiltonian, transforming it into a chain geometry and localising its coupling to the system. However, it is clear that this procedure profoundly complicates the dissipator in Eq.~\eqref{eq:dissipator_jw}. The thermal damping of the lead induced by the dissipator is most naturally described in the lead's energy eigenbasis. 

In the lead energy eigenbasis, the JW strings of the dissipators can be eliminated by exploiting the superfermion representation of the open system introduced in Sec.~\ref{sec:superfermion}. There, an {\em interleaved} physical and ancillary mode ordering was used, resulting in the dissipative processes becoming nearest-neighbour non-Hermitian Hamiltonian terms, as shown in Eq.~\eqref{eq:l_single}. In this form, when moving to a spin-1/2 representation, the JW string of each system or lead site cancels with that of the corresponding ancillary mode, rendering the dissipator terms local. 

To observe this explicitly, first note that the Fock basis of the combined Hilbert space of the physical and ancilla sites, namely Eq.~\eqref{eq:superfermion_Fock_state}, can be written in the spin-1/2 basis as 
\begin{align}
\label{eq:superfermion_Fock_state_spin}
\ket{\underline{n}| \underline{m}} & = \left( \hat{\sigma}^-_1 \right)^{n_1} \left( \hat{\sigma}^-_2 \right)^{m_1} \cdots \\ & ~\quad\qquad \cdots \left( \hat{\sigma}^-_{2M} \right)^{n_{M}} \left( \hat{\sigma}^-_{2M} \right)^{m_{M}} \ket{\uparrow \uparrow \cdots \uparrow \uparrow}. \nonumber
\end{align}
The non-Hermitian generator of the superfermion time evolution thus becomes
\begin{align}
\label{eq:l_single_spin}
&\hat{L} = \hat{H} - \hat{H}_{d\Leftrightarrow s} \nonumber +\textrm{i}\sum_{k=1}^{L} \gamma_k  (1-f_k) \hat{\sigma}^+_{2k-1} \hat{\sigma}^+_{2k}   \nonumber \\
&\quad+\textrm{i}\sum_{k=1}^{L} \gamma_k  f_k \hat{\sigma}^{-}_{2k-1} \hat{\sigma}^{-}_{2k} - \textrm{i}\sum_{k=1}^{L} \gamma_k f_k \\
&\quad-\frac{\textrm{i}}{2}\sum_{k=1}^{L} \Bigl[\gamma_k (1 - 2f_k)\left( \hat{\sigma}^{-}_{2k-1} \hat{\sigma}^+_{2k-1} + \hat{\sigma}^{-}_{2k}  \hat{\sigma}^+_{2k} \right)\Bigr], \nonumber
\end{align}
showing that the dissipator contribution consists of on-site and nearest-neighbour terms.  

\begin{figure}[t]
\fontsize{13}{10}\selectfont 
\centering
\includegraphics[width=0.5\columnwidth]{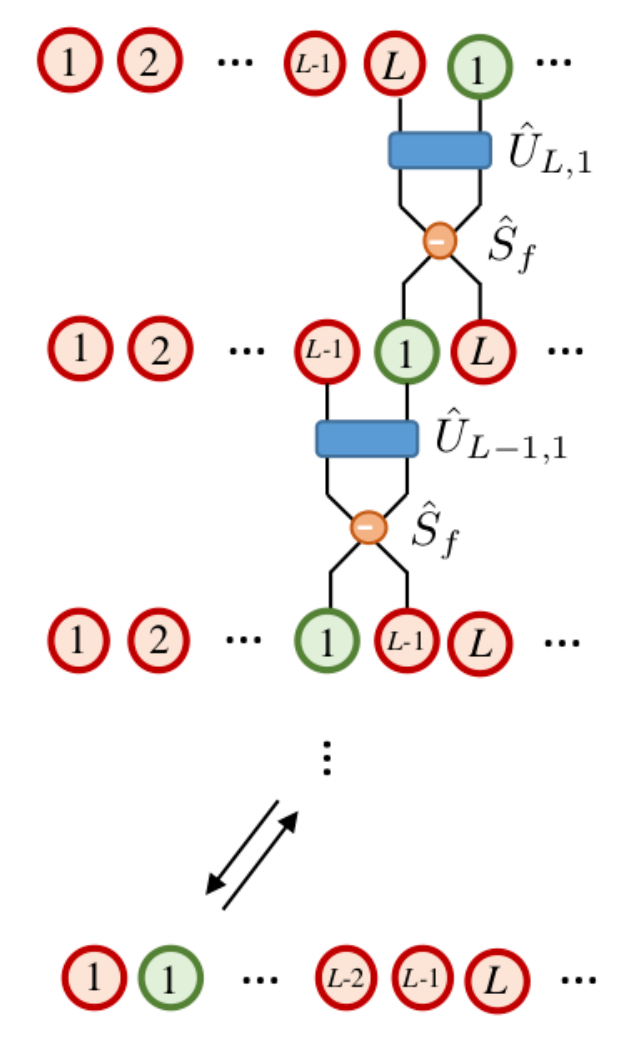}
\caption{The sweeping sequence of two-site gates $\hat{U}_{k,1}$ between the $k$-th lead mode and the first system site along with the fermionic SWAPs $\hat{S}_f$ needed to implement a Trotter step for the star geometry couplings shown in Fig.~\ref{fig:mps_star}(c).}
\label{fig:swap_seq}
\end{figure}

\subsection{Time evolving block decimation with swaps}
\label{sec:tebd}
To efficiently simulate the time evolution of the correlated system described by Eq.~\eqref{eq:l_single_spin}, we use one of the most well-known algorithms within the tensor network family, namely, the time-evolving block decimation (TEBD)~\cite{VidalTEBD2004,VerstraeteTEBD2004}. Given some system governed by a Hamiltonian $\hat{H}_{\rm loc} = \sum_i \hat{h}_{i,i+1}$, comprising a sum of 2-site terms $\hat{h}_{i,i+1}$ along a chain of length $M$, the standard formulation of TEBD computes the MPS approximation of the propagation $\ket{\psi(t)} = \exp(-\ii \hat{H}_{\rm loc}t)\ket{\psi(0)}$. This is done by first breaking up the evolution into many small time-steps $\delta t$ and then performing a second-order Trotter expansion as
\begin{align}
\label{eq:time_step}
e^{-\ii \hat{H}_{\rm loc} \delta t}\approx\left(\prod_{i=1}^{M-1}\hat{U}_{i,i+1}\right)\left(\prod_{i=M-1}^{1}\hat{U}_{i,i+1}\right),
\end{align}
where $\hat{U}_{i,i+1} = \exp(-\frac{\ii}{2} \hat{h}_{i,i+1}\delta t)$. In this way, a time step of propagation is implemented by a staircase circuit of two-site gates sweeping right-to-left and then left-to-right. Each two-site gate can be applied to the MPS and, via a singular value decomposition, the result can be re-factorised and truncated back into MPS form. 

Here, we use a simple modification of TEBD that allows us to compute the time-evolution under fermionic star-geometry Hamiltonians $\hat{H}_{\rm star} = \sum_i \hat{h}_{i,M}$, where all sites $i<M$ interact with the last site $M$. The key ingredient is the fermionic SWAP gate $\hat{S}_f$, which is a conventional SWAP gate between spins $j$ and $j+1$ that exchanges their spin configurations, but also incorporates the application of the local $\hat{\sigma}_j^z$ operator from the JW string of Eq.~\eqref{eq:h_jw}. For two sites, the gate is given by 
\begin{align} \label{swap2sites}
\hat{S}_f = \begin{pmatrix*}[r] 1 & 0 & 0 & 0 \\ 0 & 0 & \phantom{-}1 & 0 \\ 0 & \phantom{-}1 & 0 & 0 \\ 0 & 0 & 0 & -1 \end{pmatrix*},
\end{align}
where the negative sign accounts for the anticommutation relation between two fermionic creation operators when both sites $j$ and $j+1$ are occupied. By interspersing fermionic SWAP gates within the Trotter expansion, as shown in Fig.~\ref{fig:swap_seq}, distant sites are temporarily made adjacent, allowing the standard nearest-neighbour two-gate gate update to be applied. 

Time-evolution under a long-ranged Hamiltonian is generally considered impractical for tensor network calculations, due to very fast growth of entanglement across the system. This conjecture has been challenged in recent studies of fermionic impurity models, where efficient tensor network calculations have been performed using a star-like geometry~\cite{Wolf:2014,Mendoza:2017}. The proliferation of correlations in these models is curtailed by Pauli exclusion within the majority of the modes of the lead, limiting them to the range of modes around the Fermi energy. This favourable situation persists in the mesoscopic thermal lead setup considered here. Furthermore, it has been recently shown that using a suitable order of the lead modes can significantly enhance the efficiency of tensor network simulations~\cite{Rams2020}. 

\begin{figure}[b]
\fontsize{13}{10}\selectfont 
\centering
\includegraphics[width=0.8\columnwidth]{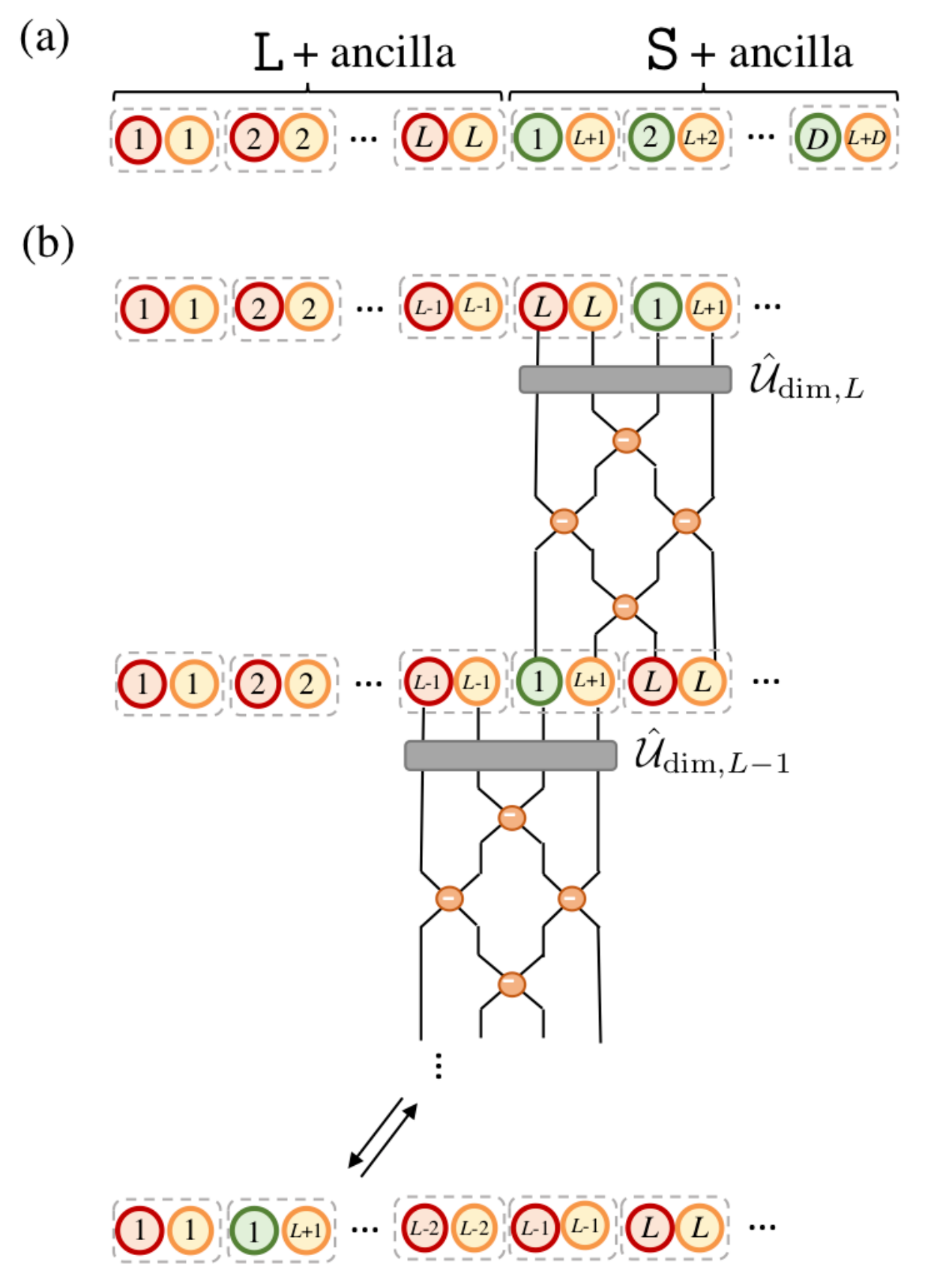}
\caption{(a) Ancilla modes are interleaved with the system and lead modes they are associated to. Computationally the system or lead site and its ancilla are bundled together as a dimer site. (b) A two-dimer site gate $\hat{\mathcal U}_{{\rm dim},k}$ is applied between the $k$-th lead mode dimer, and the first system site dimer. This is followed by four fermionic SWAPs $\hat{S}_f$ to shuffle the system site and its ancilla through the lead and its ancilla making the next lead mode adjacent. This is repeated all the way along the chain and back to complete one time-step.}
\label{fig:ancilla_modes}
\end{figure}

\begin{figure}[b]
\fontsize{13}{10}\selectfont 
\centering
\includegraphics[width=0.75\columnwidth]{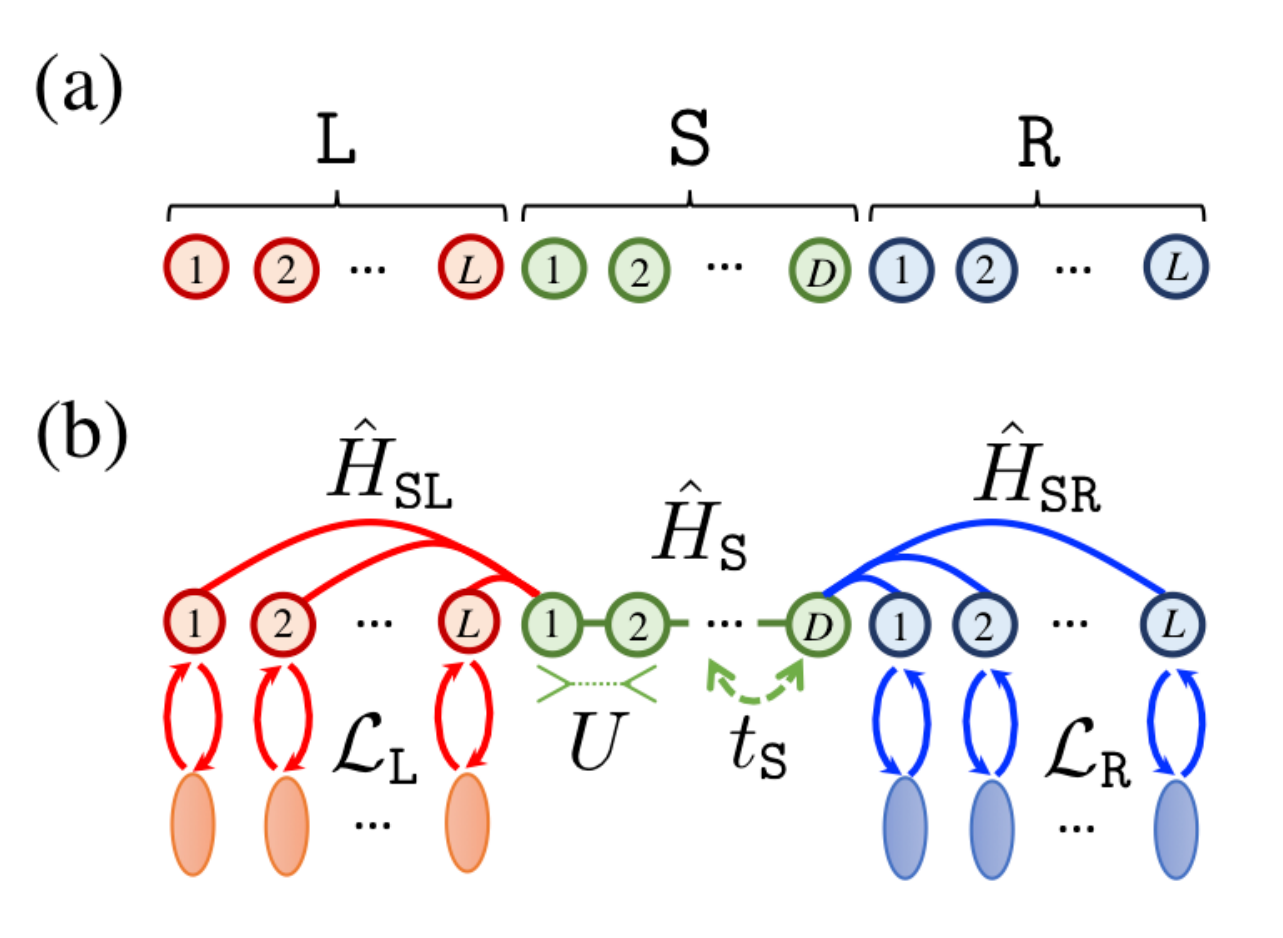}
\caption{(a) The lead and system mode ordering for a two lead setup. (b) The configuration used for the interacting system examples. Here the system $\tt S$ is a fermionic chain with hopping amplitude $t_{\tt S}$ and nearest-neighbour interaction $U$.}
\label{fig:chain_setup}
\end{figure}

\begin{figure*}[t]
\fontsize{13}{10}\selectfont 
\centering
\includegraphics[width=1.7\columnwidth]{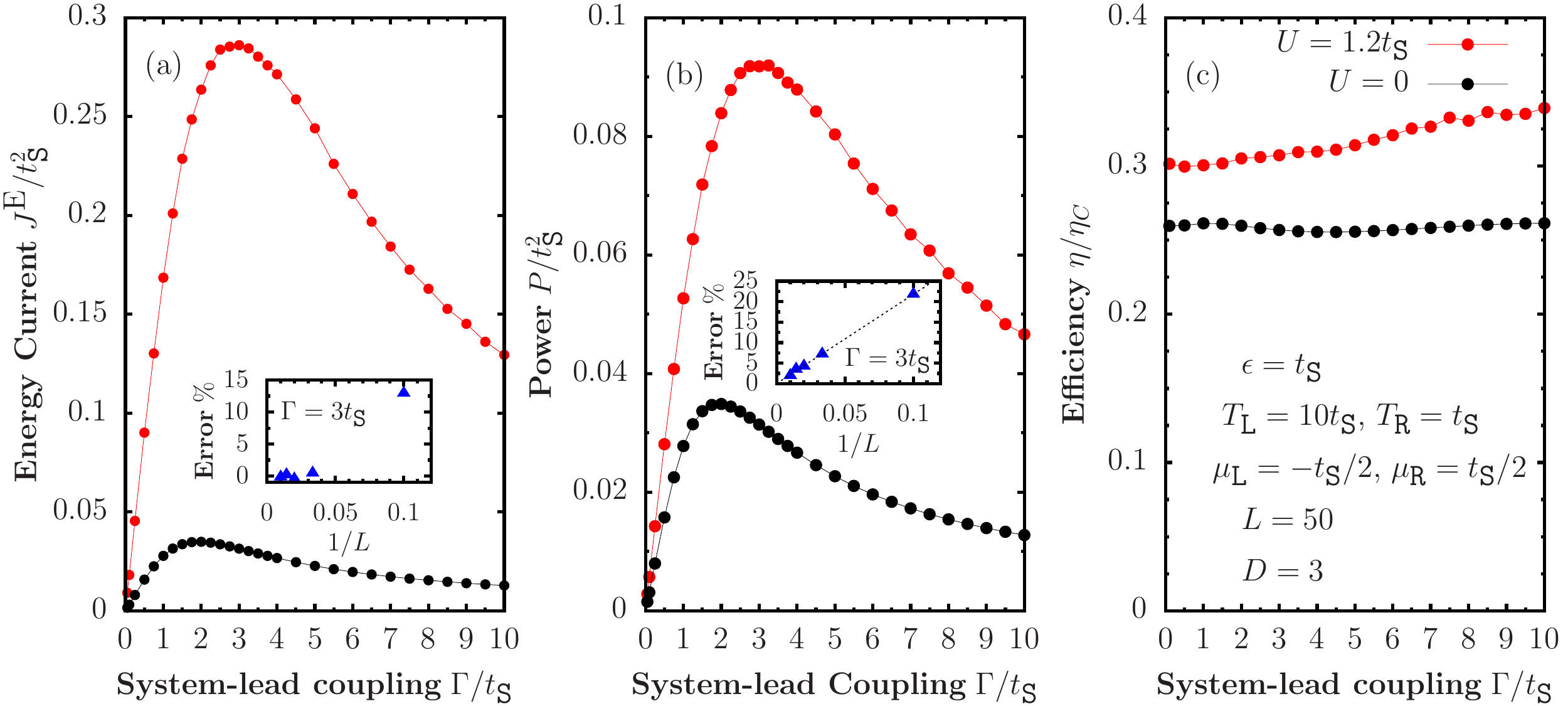}
\caption{(a) Energy current, (b) power and (c) efficiency of the interacting three-site system as a function of the system-lead coupling strength $\Gamma$. The insets in (a) and (b) show the error associated to the finite number of modes in the leads $L$ (up to $L=100$), estimated from extrapolated values of the currents at the point in which the maximum is observed ($\Gamma \approx 3t_{\texttt S}$). In these calculations we used $L_{\textrm{log}}/L = 0.2$, $W^* = W/2$ and $W = 8t_{\texttt S}$.}
\label{fig:14}
\end{figure*}

\subsection{Non-equilibrium steady state solver}
\label{sec:ness_solve}
The TEBD algorithm works equally well for non-Hermitian Hamiltonians generating non-unitary propagation. Indeed, it has been widely used to study the NESS of incoherently driven chains where the coupling to the reservoirs is localised to one~\cite{Benenti:2009,Znidaric:2010,Znidaric:2010b,ZnidaricXXZspintransport,Mendoza:2013a,Mendoza:2013b,Mendoza:2014,vznidarivc2016diffusive,vznidarivc2017dephasing,Brenes2018} or two sites~\cite{Znidaric:2011c,Mendoza:2015,Schulz2018,Mendoza2019} at the boundaries. We have now introduced all the elements required to extend the capabilities of TEBD to simulate the open-system governed by the Hamiltonian Eq.~\eqref{eq:h_jw} and the dissipator Eq.~\eqref{eq:dissipator_jw}.

First, we move to the superfermion representation where the generator $\hat{L}$ is given by Eq.~\eqref{eq:l_single_spin}. We define dimer sites composed of a physical (system or lead) site and its corresponding ancilla, as shown in Fig.~\ref{fig:ancilla_modes}(a). This procedure squares the dimension of the local basis. The left vacuum state $\ket{I}$ in this representation is a product state of dimers, with each dimer local to a given site being an equal superposition of $\ket{\uparrow \uparrow}$ and $\ket{\downarrow \downarrow}$.

Next, we identify all the terms in $\hat{L}$ that correlate the dimers located at lead site $k$ and system site $p=1$. Assuming these sites are adjacent to each other through SWAP operations, we express 
\begin{align}
\label{eq:l_dimer}
\hat{L}_{{\rm dim},k} &= \varepsilon_k \left( \hat{\sigma}_1^- \hat{\sigma}_1^+ - \hat{\sigma}_2^- \hat{\sigma}_2^+ \right) + \frac{\epsilon}{L} \left( \hat{\sigma}_3^- \hat{\sigma}_3^+ - \hat{\sigma}_4^- \hat{\sigma}_4^+ \right) \nonumber \\
&+ \kappa_{kL} \hat{\sigma}^-_1 \hat{\sigma}^z_2 \hat{\sigma}^+_3 + \kappa^*_{kL} \hat{\sigma}^+_1 \hat{\sigma}^z_2 \hat{\sigma}^-_3
\nonumber \\
& \qquad -\kappa_{kL} \hat{\sigma}^+_2 \hat{\sigma}^z_3 \hat{\sigma}^-_4 - \kappa^*_{kL} \hat{\sigma}^-_2 \hat{\sigma}^z_3 \hat{\sigma}^+_4  \nonumber \\
& -\frac{\textrm{i}}{2} \gamma_k (1 - 2f_k) \left( \hat{\sigma}^-_1 \hat{\sigma}^+_1 + \hat{\sigma}^-_2 \hat{\sigma}^+_2 \right) -\textrm{i} \gamma_k f_k \nonumber \\
& +\textrm{i}\gamma_k(1 - f_k) \hat{\sigma}^+_1 \hat{\sigma}^+_2 + \textrm{i}\gamma_k f_k \hat{\sigma}^-_1 \hat{\sigma}^-_2.
\end{align}
We identify spin 1 as the $k$-th lead eigenmode with spin 2 being its corresponding ancilla mode. On the other hand, spin 3 is the system site coupled to the lead with spin 4 its corresponding ancilla mode. A JW string appears between interacting spins that are not adjacent, however, they remain local to the dimer pair. The exponential of this operator, $\hat{\mathcal U}_{{\rm dim},k} = \textrm{exp} (-\textrm{i} \hat{L}_{{\rm dim},k} \delta t / 2)$, defines a non-unitary gate for a half time step $\delta t$. This operator accounts for all the coherent interactions and the non-Hermitian terms, describing the dissipation between the lead mode and the system site. We have assumed a Hamiltonian of the form Eq.~\eqref{eq:h_s_d} in Eq.~\eqref{eq:l_dimer}.

Finally, the non-unitary gates $\hat{\mathcal U}_{{\rm dim},k}$ are then applied along with fermionic SWAP gates that shuffle the system dimer along the chain, as shown in Fig.~\ref{fig:ancilla_modes}(b). The latter can be defined from the two-site SWAP gates of Eq.~\eqref{swap2sites} in the following way: naming $\hat{A} = I_2 \otimes S_f \otimes I_2$, with $I_2$ the $2\times2$ identity matrix, and $\hat{B} = S_f \otimes S_f$, the two-dimer SWAP gate depicted in Fig.~\ref{fig:ancilla_modes}(b) is given by $ABA$. Altogether, this sequence of gates computes the action of the propagator $\exp(-\ii\hat{L}\delta t)$ and formally solves Eq.~\eqref{eq:superfermion_evolve} for a single time-step. We take the initial state to be $\ket{\rho(0)} = \ket{I}$, and find the steady state $\ket{\rho(\infty)}$ by evolving towards the long-time limit. Expectation values and the trace of the density operator follow from the inner product with $\ket{I}$ as given in Eq.~\eqref{eq:expec}.

The same simulation scheme can be readily extended to the two-lead configuration, as shown in Fig.~\ref{fig:chain_setup}(a), with the long-time limit now giving rise to a NESS. The approach to the stationary state is assessed by evaluating the convergence of observables such as the particle and energy currents. In practice, we used a dynamically-increasing truncation parameter $\chi$ for different time-step parameters $\delta t$. In the standard MPS language~\cite{VidalTEBD2004,VerstraeteTEBD2004}, $\chi$ refers to the maximum MPS bond dimension in between each pair of neighbouring nodes in the network, where each node represents a dimer. To perform the simulation, we chose an initial value of $\chi$ and $\delta t$, and evolved the system up to an intermediate time. The resulting state was then further evolved in time with a larger $\chi$ and an appropriately reduced $\delta t$. This procedure is repeated until the currents obtained converged up to a small tolerance of $1-2 \%$. The largest bond dimension used in our calculations was $\chi_{\textrm{max}} = 220$, showing that a moderate computational effort was required to access the NESS (see Appendix \ref{ap:converge} for further details). All MPS calculations in this work were performed using the open-source Tensor Network Theory (TNT) library~\cite{tnt,tnt_review1}.

\section{Interacting examples}
\label{sec:interacting}

In this section, we employ the tensor network algorithm from Sec.~\ref{sec:tensor} to study an autonomous thermal machine with an interacting working medium, as depicted in Fig.~\ref{fig:chain_setup}(b). Our methods enable us to consider the challenging problem of simultaneously strong interactions and system-bath coupling, far beyond the linear-response regime. 

\subsection{Interacting three-site engine}

Our first example is an autonomous quantum heat engine with a three-site interacting working medium, which is described by the Hamiltonian
\begin{align}
\label{eq:h_s_i}
\hat{H}_{\tt S} = \sum_{j = 1}^{D} \epsilon_j \hat{n}_j - \sum_{j = 1}^{D - 1} t_{\tt S} \left( \hat{c}^{\dagger}_{j+1} \hat{c}_{j} + \textrm{H.c.} \right) + \sum_{j=1}^{D} U \hat{n}_j \hat{n}_{j+1},
\end{align}
where $\hat{n}_j = \hat{c}^{\dagger}_j \hat{c}_j$ is the density operator for site $j$ and $U$ is the interaction strength. The last term in the equation above corresponds to a density-density interaction of neighbouring particles. A small central system composed of $D=3$ interacting fermionic sites can be interpreted as a three-site version of the interacting resonant level model \cite{Kennes2012}.

We set the system hopping $t_{\tt S} = W/8$ and focus on the regime in which the temperature gradient and the difference in chemical potential between the mesoscopic reservoirs is strong. We set $T_{\tt L} = 10t_{\tt S}$, $T_{\tt R} = t_{\tt S}$, $\mu_{\tt L} = -t_{\tt S}/2$, $\mu_{\tt R} = t_{\tt S}/2$ and $\epsilon_j = \epsilon = t_{\tt S}$. With these parameters, the system operates as a heat engine, i.e. particle current flows from the left reservoir to the right reservoir, driven by the temperature gradient against a chemical potential gradient. As in Sec.~\ref{sec:non_interacting_example}, both leads are assumed to have identical, flat spectral densities given by Eq.~\eqref{eq:wideband} and we use the logarithmic-linear discretisation scheme with $W^* = W / 2$ and $L_{\textrm{log}}/L = 0.2$. We remark that the chosen Hamiltonian parameters are far apart from the energy scale dictated by $W$, such that the effect of the finite bandwidth is expected to be negligible. This choice of parameters is thus a useful representative example for exposing the efficacy of the proposed methodology. 

We first focus on the dependence of the currents on the system-lead coupling $\Gamma$, as shown in Fig.~\ref{fig:14}. In Fig.~\ref{fig:14}(a), the energy current for a particular value of the interaction strength $U=1.2t_{\tt S}$ is shown as a function of $\Gamma$. Remarkably, a density-density interaction yields a larger energy current flowing through the system compared to the non-interacting case in the chosen regime. The same observation holds for the particle current in Fig.~\ref{fig:14}(b), since for our choice of parameters the particle current and the power output are equivalent [see Eq.~\eqref{eq:power_def}]. The efficiency shown in Fig.~\ref{fig:14}(c), remains approximately constant as a function of system-lead coupling strength just like the non-interacting case. Future work will investigate a larger range of parameters to identify a maximum power output for a given interaction strength.  

The insets in Figs.~\ref{fig:14}(a) and \ref{fig:14}(b) show the error associated to employing a finite number of modes in each reservoir for a specific value of $\Gamma = 3t_{\tt S}$, where the currents in the interacting case reach the maximum value. The error is computed from an extrapolated value of the currents to the $L \to \infty$ limit, based on the currents for finite $L$, for each respective case. We define $\textrm{Error \%} \defeq |K(L \to \infty) - K(L)| \cdot 100 / K(L \to \infty)$, where $K = J^{\textrm{E}}, P$ for energy current and power, respectively. The value $K(L \to \infty)$ is taken from an extrapolation following the trend of $K(L)$. A linear extrapolation was made for the power as shown in the inset in Fig.~\ref{fig:14}(b), while no extrapolation is required for the energy current in Fig.~\ref{fig:14}(a), as the current has converged for $L$ smaller than the final value of $L = 100$. It can be observed that for the specific choice of parameters, a good approximation can be obtained to a few percent accuracy using $L=50$, compared to larger reservoirs. The energy current converges faster than the particle current (power) in this case. This behaviour is expected, as observing Figs.~\ref{fig:18} and \ref{fig:19} for the non-interacting case in Appendix \ref{sec:noninteracting}, the largest deviation for the particle current occurs where the maximum value is obtained, while the largest deviation for the energy current is observed near the edges of the band. 

\subsection{High-temperature transport}
The transport properties of spin chains have been studied extensively using standard open-system MPS approaches based on a boundary driving Lindblad master equation. This approach has been successful in accurately describing the high-temperature spin/particle transport behaviour of the integrable anisotropic XXZ Heisenberg model \cite{Znidaric:2010,Znidaric:2010b,Prosen:2011,ZnidaricXXZspintransport} as well as non-integrable versions of the model when integrability-breaking perturbations are introduced, such as magnetic impurities \cite{Brenes2018} or disorder~\cite{vznidarivc2016diffusive,vznidarivc2017dephasing,Schulz2018,Mendoza2019}. However, driving on the boundary spins is formally equivalent to infinite temperature baths. Modelling energy currents therefore requires more elaborate multi-site boundary driving to mimic finite temperature differences. While this approach has proven successful for the very high temperature limit, its reliability as the temperature is lowered is questionable. The mesoscopic leads construction introduced here overcomes this deficiency. 

\begin{figure}[t]
\fontsize{13}{10}\selectfont 
\centering
\includegraphics[width=0.9\columnwidth]{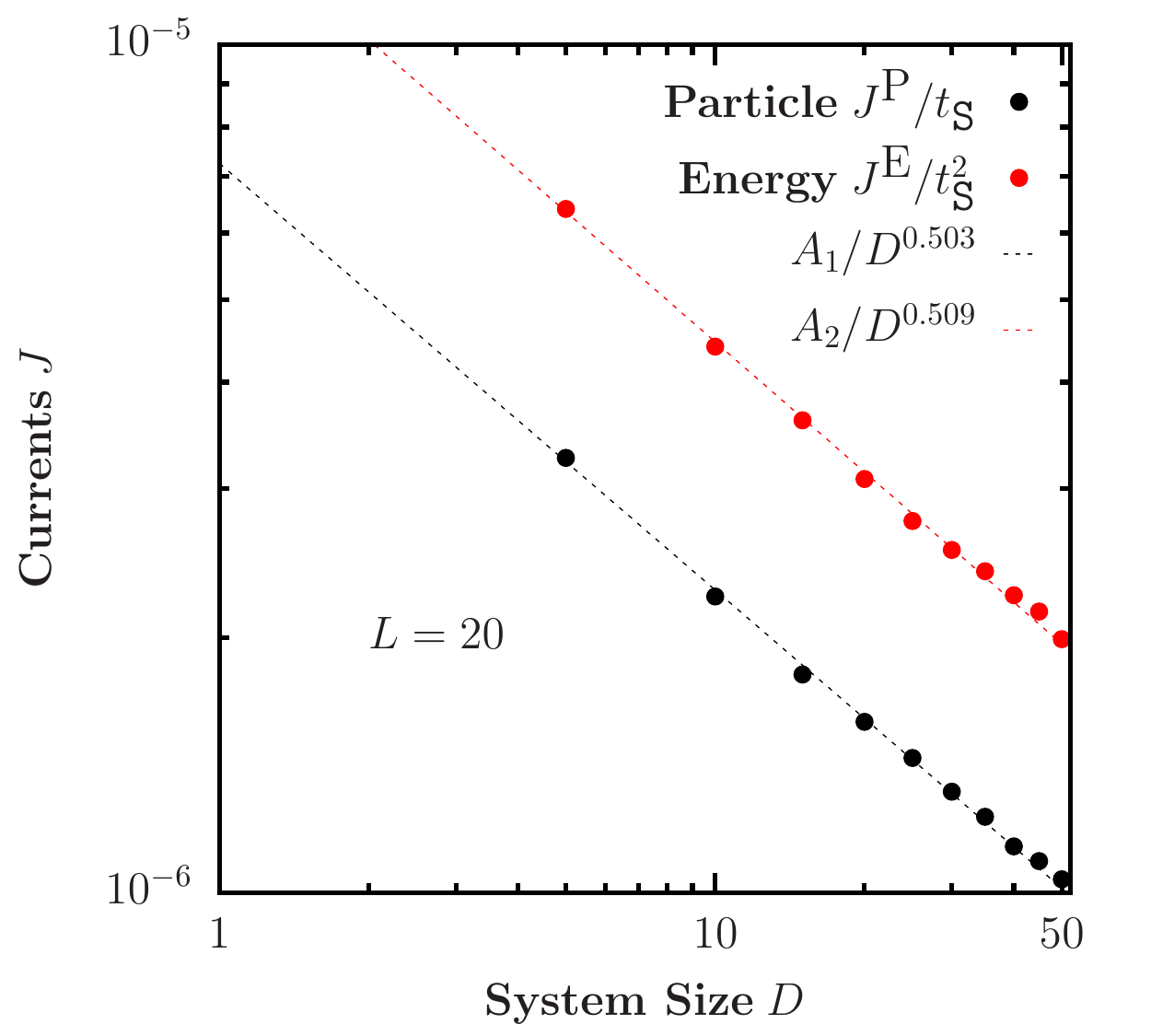}
\caption{Particle and energy currents as a function of system size $D$ for the isotropic Heisenberg model $U / t_{\tt S} = 2$ [see Eq.\eqref{eq:h_s_i}]. The results shown correspond to a very high temperature $T_{\tt L} = T_{\tt R} = 1000t_{\tt S}$ and a small chemical potential bias $\mu_{\tt L} = -\mu_{\tt R} = 0.025t_{\tt S}$, where the system is expected to be in linear response regime.  In these calculations we used $L_{\textrm{log}}/L=0.2$, $W^* = W/2$, $W = 8t_{\texttt S}$ and $\Gamma = \epsilon = t_{\texttt S}$.}
\label{fig:15}
\end{figure}

The system Hamiltonian introduced in Eq.~\eqref{eq:h_s_i} is the spinless fermion equivalent of the anisotropic XXZ Heisenberg model. This model exhibits a range of distinct linear response particle and energy transport behaviour as a function of the anisotropy $U$. Specifically, these include ballistic transport which is characterised by a constant value of the current as a function of system size $D$, as well as diffusive transport, where $J^{\textrm{P}} \propto 1/D^{\nu}$ with $\nu = 1$ \cite{Brenes2018}. Anomalous diffusion is signalled by $0 < \nu < 1$ and $\nu > 1$, corresponding to superdiffusion and subdiffusion, respectively. A sharp transition in the system's transport properties is known to occur at the isotropic point $U / t_{\tt S} = 2$, with the system displaying ballistic transport for $U / t_{\tt S} < 2$, while for $U / t_{\tt S} > 2$ transport becomes diffusive. Furthermore, precisely at the isotropic point $U / t_{\tt S} = 2$, boundary driving calculations have shown that transport is superdiffusive with $\nu = 1/2$ \cite{ZnidaricXXZspintransport}. These results are expected to hold only in the linear-response regime at high temperatures, where the structure of the thermal baths becomes irrelevant. We now corroborate these results using our mesoscopic reservoir formalism.

\begin{figure}[b]
\fontsize{13}{10}\selectfont 
\centering
\includegraphics[width=0.9\columnwidth]{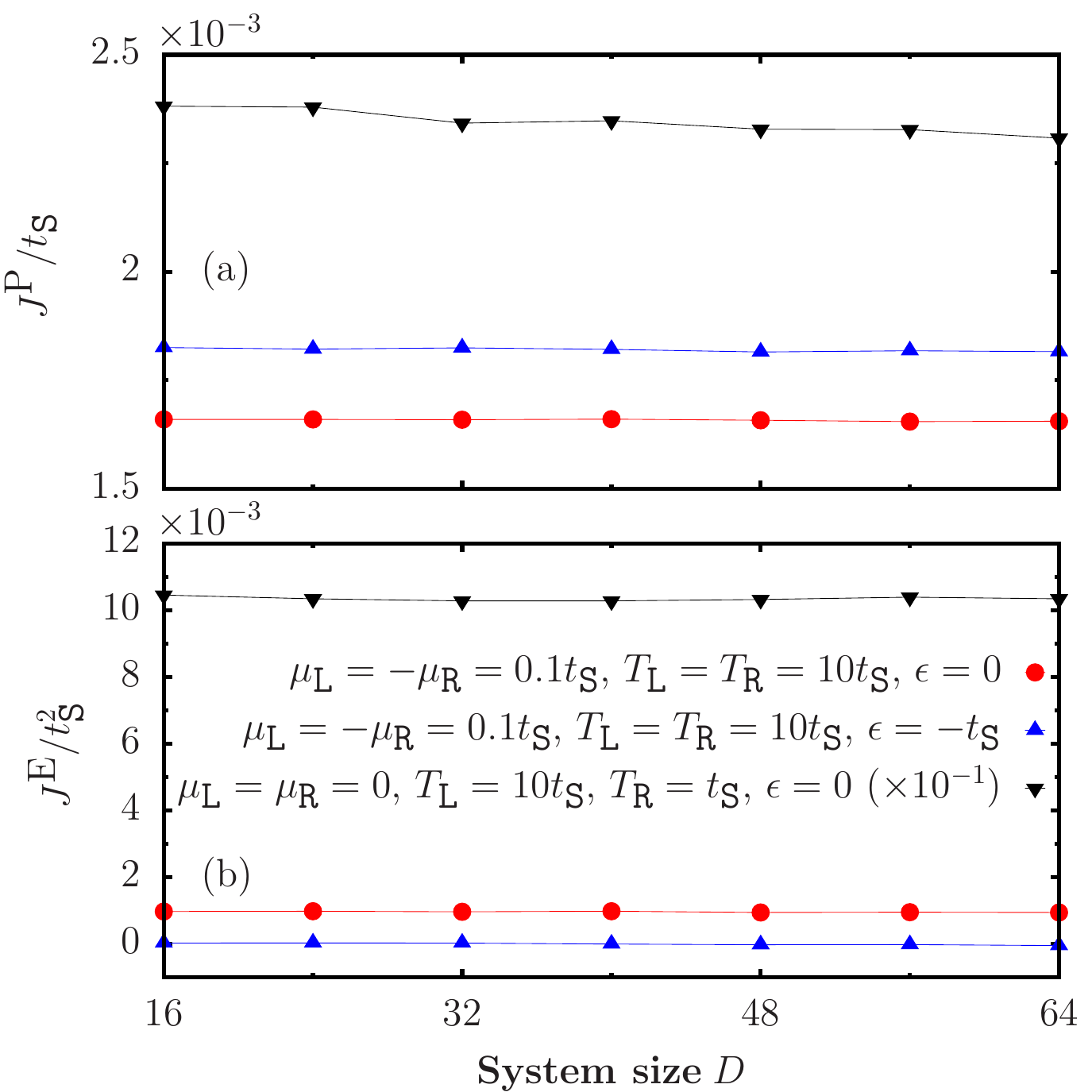}
\caption{ Finite-size scaling of (a)~particle current and (b)~energy current for the anisotropic Heisenberg model in Eq.~\eqref{eq:h_s_i} with ${U=t_{\tt S}}$. Size-independent currents imply ballistic particle and energy transport under chemical-potential or temperature bias. Data shown by the black triangles are rescaled by a factor of $10^{-1}$ to be visible on the same scale. In these calculations we used $L = 20$, $L_{\textrm{log}}/L=0.2$, $W^* = W/2$, $W = 8t_{\texttt S}$ and $\Gamma = U = t_{\texttt S}$.}
\label{fig:16}
\end{figure}

As before, we choose the same discretisation scheme and bath structure parameters. We focus on the isotropic point $U / t_{\tt S} = 2$ and set $\epsilon_j / t_{\tt S} = \epsilon / t_{\tt S} = 1$. We set the temperature on each reservoir to a high value of $T_{\tt L} = T_{\tt R} = 1000t_{\tt S}$ and choose a small chemical potential gradient $\mu_{\tt L} = -\mu_{\tt R} = 0.025t_{\tt S}$, where we expect the system to be in linear response regime. In Fig.~\ref{fig:15} we show both the particle and energy currents as a function of system size $D$. We have used $L = 20$ modes for both left and right reservoirs. As can be observed, the currents fit a power law scaling with an exponent very close to $\nu = 1/2$ in clear indication of super-diffusive behaviour. We remark that at high temperature, fewer reservoir modes can be used to obtain the correct transport exponent, as observed from boundary driving calculations \cite{ZnidaricXXZspintransport}. 

\subsection{Finite-temperature transport and CP symmetry}

We now test the capabilities of our method to extract transport properties outside of the high-temperature limit. As a benchmark, we focus on the anisotropic Heisenberg Hamiltonian given by Eq.~\eqref{eq:h_s_i} with $U = t_{\tt S}$ and homogeneous on-site energies, $\epsilon_j = \epsilon$. 

In this regime, the Hamiltonian is integrable and the total energy current is conserved, implying ballistic energy transport at all temperatures under linear-response conditions~\cite{Zotos1997,Bertini2020}. Ballistic particle conduction is also expected for $U<2t_{\tt S}$, as indicated by extensive numerical calculations~\cite{Bertini2020} and arguments based on quasilocal conservation laws~\cite{prosen:2011-2,prosen:2013}. We confirm the ballistic nature of transport at finite temperature by a scaling analysis with the system size $D$ of the particle and energy currents, as shown in Fig.~\ref{fig:16}. We drive the system out of equilibrium either by applying a chemical-potential bias at fixed temperature, or by a temperature gradient applied at fixed chemical potential. In each case we find that the particle and energy currents are essentially independent of system size, as expected. We note that our method can be applied far outside linear response, for example with a large temperature bias $T_{\tt L} - T_{\tt R} \gg T_{\tt R}$, as shown by the black triangles in Fig.~\ref{fig:16}.

The magnitudes of the currents strongly depend on the bulk Hamiltonian and the thermodynamic potentials of the baths. Configurations that are invariant under a charge conjugation-parity (CP) transformation, i.e.~a combined reflection and particle-hole symmetry, are found to exhibit vanishing energy current. More precisely, CP symmetry requires equal bath temperatures, $T_{\tt L}=T_{\tt R}$, opposite chemical potentials, $\mu_{\tt L} = -\mu_{\tt R}$, and bulk Hamiltonian parameters $\epsilon = -U$. As shown by the blue triangles in Fig.~\ref{fig:16}(b), the energy current is zero in this case, in agreement with exact analytical calculations detailed in Appendix~\ref{app:CP}. A finite energy current emerges whenever the on-site energies of $\hat{H}_{\tt S}$ are moved away from the CP-symmetric point, even when the forcing from the baths remains CP-symmetric (red circles in Fig.~\ref{fig:16}). This is in stark contrast with the predictions of single-site boundary driving transport calculations on the Heisenberg model, where symmetric driving leads to vanishing energy current independent of the bulk Hamiltonian parameters~\cite{Popkov2013}. This ultimately stems from the fact that boundary driving simulates white noise and thus does not capture the energy dependence of true thermal fluctuations.

\begin{figure}[t]
\fontsize{13}{10}\selectfont 
\centering
\includegraphics[width=0.9\columnwidth]{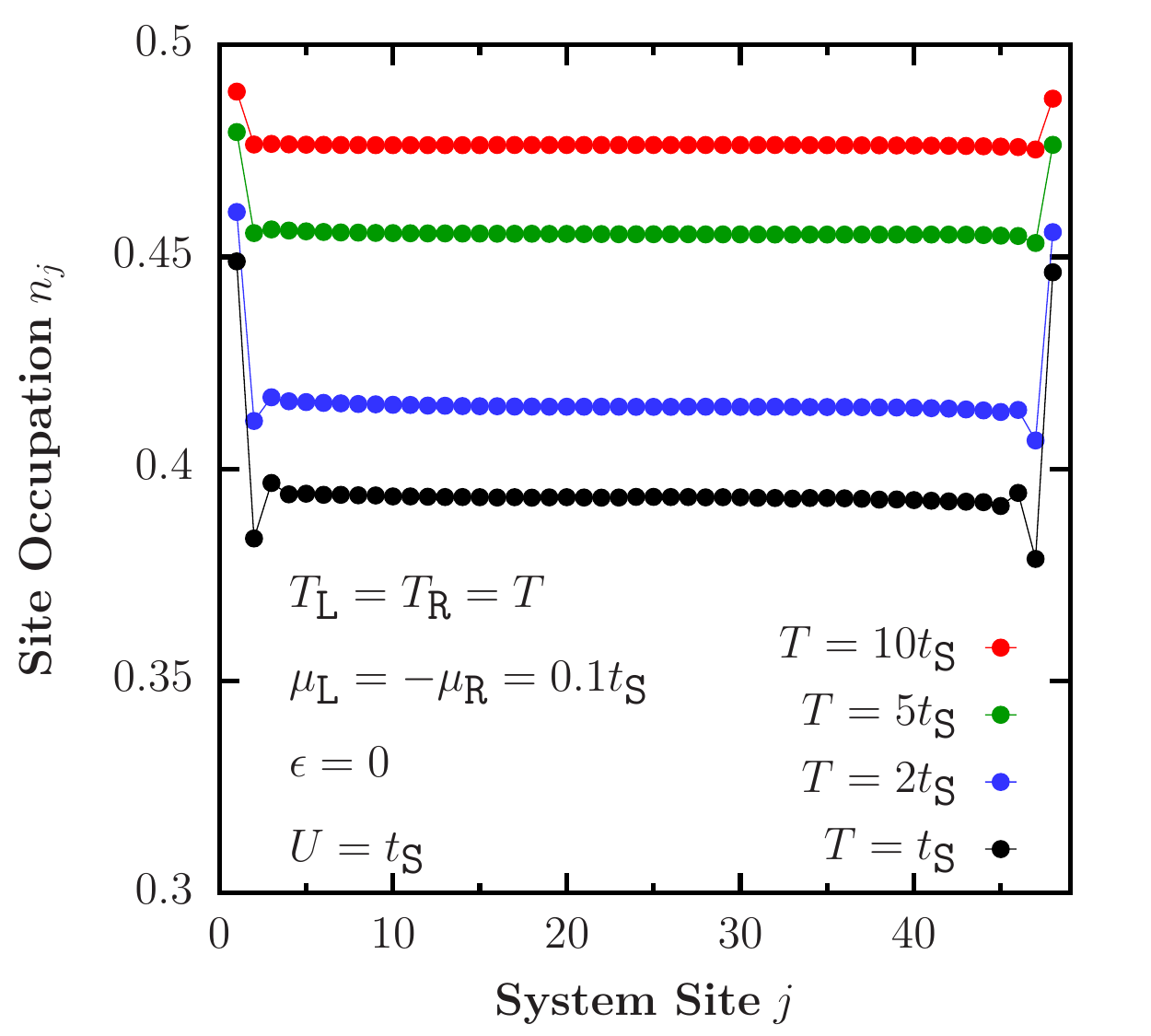}
\caption{Non-equilibrium density profile in the anisotropic Heisenberg model, Eq.~\eqref{eq:h_s_i}, under symmetric chemical-potential bias and various temperatures. In these calculations we used $D = 48$, $L = 40, $ $L_{\textrm{log}}/L=0.2$, $W^* = W/2$, $W = 8t_{\texttt S}$ and $\Gamma = t_{\texttt S}$. We checked that the site occupation results are robust towards changes from $L = 20$ to $L = 40$ in all cases shown.}
\label{fig:17}
\end{figure}

We further explore the effect of temperature by examining the non-equilibrium density profile of the system in Fig.~\ref{fig:17}. We consider equal reservoir temperatures, $T_{\tt L} = T_{\tt R} = T$, fixed system ($D = 48$) and lead ($L = 40$) sizes, and a symmetric chemical potential bias, $\mu_{\tt L}= -\mu_{\tt R}$. We also take $\epsilon \neq -U$, to break CP-symmetry. Away from the boundaries, we find the flat profile characteristic of ballistic transport, with a density that depends on temperature. Lower temperatures correspond to lower densities and larger currents. As the temperature is increased, the bulk density tends to the CP-symmetric value $\langle \hat{n}_j\rangle \to 0.5$. This shows that the CP symmetry enforced by the single-site boundary driving configuration is indeed recovered in the high-temperature limit.

\section{Conclusions and outlook}
\label{sec:conclusions}
In this work we introduced a novel methodology to simulate the heat and particle currents in thermal machines which comprise a complex working medium coupled to fermionic leads at fixed temperatures and chemical potentials. The method is based on the concept of mesoscopic reservoirs whose energy modes are damped in order the replicate the continuum. The method allows for calculations in highly non-equilibrium scenarios such as strong system-lead coupling and large biases. In order to cope with non-quadratic interactions in the working medium, we implemented a novel tensor network algorithm directly in the star geometry using auxiliary modes. 

For the purpose of expounding the method, in this paper we considered only autonomous thermal machines where the working medium is time independent. In order to benchmark our technique we first focused on replicating the steady-state thermodynamics of the resonant-level heat engine. The simplicity of this quadratic model allows for direct comparison with the Landauer-B\"uttiker theory for quantum transport. We observed excellent agreement across a wide parameter regime. We then explored efficiency and power in a strongly interacting three-qubit machine in a parameter regime where other methods are known to struggle. In doing this we observed that, remarkably, the efficiency is enhanced as a function of the system-lead coupling in the presence of non-quadratic interactions. Furthermore, we demonstrated that our technique is capable of highly non-trivial heat and particle transport calculations in strongly correlated many-body systems by performing a scaling analysis at the isotropic point of the paradigmatic Heisenberg model. Finally, we analysed the current scaling and non-equilibrium density profile  in the integrable regime of the anisotropic Heisenberg model, confirming the ballistic nature of transport at finite temperature and well beyond linear response.

Due to the flexibility of our technique we expect that the method is extendable further in the direction of steady-state thermodynamics of complex interacting quantum systems. Beyond strong coupling and far-from-equilibrium scenarios, our technique may also find useful applications in the study of time-dependent working media, bulk noise effects and non-trivial spectral densities, thus taking quantum thermodynamics to unexplored horizons.  

\textit{Note added in proof.} During the preparation of this manuscript, several articles have appeared that propose different yet related tensor-network algorithms to study transport with mesoscopic reservoirs~\cite{Rams2020,Wojtowicz2020,Lotem2020}. 

\section*{Acknowledgements}

M.B. and J.G. acknowledge the DJEI/DES/SFI/HEA Irish Centre for High-End Computing (ICHEC) for the provision of computational facilities and support, project TCPHY104B, and the Trinity Centre for High-Performance Computing. This work was supported by a SFI-Royal Society University Research Fellowship (J.G.) and the Royal Society (M.B.). J.G. acknowledges funding from European Research Council Starting Grant ODYSSEY (Grant Agreement No. 758403). S.R.C. gratefully acknowledges support from the UK's Engineering and Physical Sciences Research Council (EPSRC) under grant No. EP/P025110/2. J.J.M.-A. is thankful for the support of Ministerio de Ciencia, Tecnolog\'ia e Innovaci\'on (MINCIENCIAS), through the project Producci\'on y Caracterizaci\'on de Nuevos Materiales Cu\'anticos
de Baja Dimensionalidad: Criticalidad Cu\'antica y Transiciones de Fase Electr\'onicas (Grant No. 120480863414).

\appendix

\section{Connection between mesoscopic and macroscopic reservoirs}
\label{app:meso_equivalence}

In this appendix we give further mathematical details of the connection between mesoscopic and infinite reservoirs described in Sec.~\ref{sec:mesoscopic_leads}. 

\subsection{Infinite-bath configuration}
\label{app:infinite_bath}

We begin by discussing the equations of motion assuming that the system is in contact with an infinite thermal reservoir. The total Hamiltonian is thus $\hat{H} = \hat{H}_{\tt S} + \hat{H}_{\tt B} + \hat{H}_{\tt SB}$, where $\hat{H}_{\tt B}$ and $\hat{H}_{\tt SB}$ are respectively given by
\begin{align}
\label{eq:H_B_infinite_1}
&\hat{H}_{\tt B} = \sum_{m=1}^\infty \omega_m \hat{b}^{\dagger}_m \hat{b}_m,\\
&\label{eq:H_SB_infinite_1}
\hat{H}_{\tt SB} = \sum_{m=1}^{\infty} \left( \lambda_{m} \hat{c}^{\dagger}_p \hat{b}_m + \lambda^*_{m} \hat{b}^{\dagger}_m \hat{c}_p \right),
\end{align}
while $\hat{H}_{\tt S}$ is arbitrary. In the Heisenberg picture, the equations of motion read as
\begin{align}
    \label{b_EOM}
    \frac{\rm d}{{\rm d} t}\hat{b}_m(t) & = -{\rm i}\omega_m \hat{b}_m(t) -{\rm i} \lambda_m^*\hat{c}_p(t),\\
    \label{c_EOM}
\frac{\rm d}{{\rm d} t}\hat{c}_j(t) & = {\rm i}[\hat{H}_{\tt S},\hat{c}_j(t)] - {\rm i}\delta_{jp} \sum_m \lambda_m \hat{b}_m(t),
\end{align}
where $p$ denotes the system site connected to the bath. The formal solution of Eq.~\eqref{b_EOM} reads as
\begin{equation}
    \label{b_formal_solution}
    \hat{b}_m(t) = {\rm e}^{-{\rm i}\omega_m t} \hat{b}_m(0) -{\rm i} \lambda_m^*\int_0^t{\rm d}t'\, {\rm e}^{-{\rm i}\omega_m (t-t')}\hat{c}_p(t'). 
    \end{equation}
Substituting this back into Eq.~\eqref{c_EOM} yields the quantum Langevin equation
\begin{equation}
    \label{QLE}
        \frac{{\rm d}}{{\rm d} t}\hat{c}_j(t) = \ii [\hat{H}_{\tt S}, \hat{c}_j(t)] + \delta_{jp} \left[\hat{\xi}(t) - \int_0^t {\rm d}t'\, \chi(t-t') \hat{c}_p(t')\right].
\end{equation}
Here, the noise operator is $\hat{\xi}(t) = -{\rm i}\sum_m{\rm e}^{-{\rm i}\omega_m t}\lambda_m \hat{b}_m(0)$ and the memory kernel is $\chi(t-t') = \langle \{\hat{\xi}(t),\hat{\xi}^\dagger(t')\}\rangle$.

The solution of Eq.~\eqref{QLE} at time $t$ depends in principle on the entire past history of the noise operator $\hat{\xi}(s)$ for $s<t$. Once found, the solution for $\hat{c}_j(t)$ is sufficient to reconstruct all $n$-point correlation functions of $\tt S$, which together uniquely specify the quantum state (amongst other information). Since the initial bath state is Gaussian, these correlation functions depend on the noise only via its two-time correlations
\begin{align}
    \label{chi_t}
    \langle \{\hat{\xi}(t),\hat{\xi}^\dagger(t')\}\rangle & = \int \frac{{\rm d}\omega}{2\pi} \mathcal{J}(\omega) \ee^{-\ii \omega(t-t')},\\
    \label{phi_t}
    \langle \hat{\xi}^\dagger(t)\hat{\xi}(t')\rangle & = \int \frac{{\rm d}\omega}{2\pi} \mathcal{J}(\omega) f(\omega) \ee^{\ii \omega(t-t')}.
\end{align}

In some cases, like for a single site system, the particle and energy currents from the bath also become important. The particle and energy currents from the bath are given by
\begin{align}
&J^{\textrm{P}} = i\left \langle\sum_{m=1}^{\infty} \left( \lambda_{m} \hat{c}^{\dagger}_p \hat{b}_m  - \lambda^*_{m} \hat{b}^{\dagger}_m \hat{c}_p  \right)\right\rangle, \\
&J^{\textrm{E}} = i\left \langle\sum_{m=1}^{\infty} \omega_m\left( \lambda_{m} \hat{c}^{\dagger}_p \hat{b}_m  - \lambda^*_{m} \hat{b}^{\dagger}_m \hat{c}_p  \right)\right\rangle.
\end{align}
This requires evaluation of the operators $\langle\sum_{m=1}^{\infty} \lambda_{m} \hat{c}^{\dagger}_p \hat{b}_m\rangle$ and $\langle\sum_{m=1}^{\infty}\omega_m\lambda_{m} \hat{c}^{\dagger}_p \hat{b}_m\rangle$. The evolution of these operators can be written down from Eq.~\ref{b_formal_solution} and are given by
\begin{align}
\label{curr_op}
&\langle\sum_{m=1}^{\infty} \lambda_{m} \hat{c}^{\dagger}_p(t) \hat{b}_m(t)\rangle \nonumber \\
&= i\langle \hat{c}^{\dagger}_p(t) \hat{\xi}(t) \rangle -i\int_0^t {\rm d}t'\, \chi(t-t') \langle\hat{c}^{\dagger}_p(t)\hat{c}_p(t')\rangle, \\
\label{energy_curr_op}
&\langle\sum_{m=1}^{\infty} \omega_m\lambda_{m} \hat{c}^{\dagger}_p(t) \hat{b}_m(t)\rangle \nonumber \\
&=i\langle \hat{c}^{\dagger}_p(t) \hat{\tilde{\xi}}(t) \rangle -i\int_0^t {\rm d}t'\, \tilde{\chi}(t-t') \langle\hat{c}^{\dagger}_p(t)\hat{c}_p(t')\rangle,
\end{align} 
where we have additionally defined  
\begin{align}
\label{xi_tilde_t}
&\hat{\tilde{\xi}}(t)~=-{\rm i}\sum_m{\rm e}^{-{\rm i}\omega_m t}\omega_m\lambda_m \hat{b}_m(0), \\
    \label{chi_tilde_t}
   & \tilde{\chi}(t-t')= \int \frac{{\rm d}\omega}{2\pi} \omega\mathcal{J}(\omega) \ee^{-\ii \omega(t-t')}.
\end{align}
The operator $\hat{\tilde{\xi}}(t)$ satisfies
\begin{align}
    \label{phi_tilde_t1}
    &\langle \hat{\tilde{\xi}}^\dagger(t)\hat{\tilde{\xi}}(t')\rangle & = \int \frac{{\rm d}\omega}{2\pi} \omega^2\mathcal{J}(\omega) f(\omega) \ee^{\ii \omega(t-t')}, \\
    \label{phi_tilde_t2}
    &\langle \hat{\tilde{\xi}}^\dagger(t)\hat{\xi}(t')\rangle & = \int \frac{{\rm d}\omega}{2\pi} \omega\mathcal{J}(\omega) f(\omega) \ee^{\ii \omega(t-t')}.
\end{align}
Eqs.~(\ref{QLE}), (\ref{chi_tilde_t}), (\ref{phi_t}), (\ref{curr_op}), (\ref{energy_curr_op}), (\ref{phi_tilde_t1}), (\ref{phi_tilde_t2}) completely define time evolution of any operator of the system, as well as that of the energy and particle currents from the baths. In the following, we show that the same equations can be recovered in the mesoscopic-lead configuration, thereby showing their equivalence.

\subsection{Mesoscopic-lead configuration}
\label{app:meso_reservoir}
We now turn to the mesoscopic-reservoir configuration, with total Hamiltonian $\hat{H} = \hat{H}_{\tt S} +  \hat{H}_{\tt SL} + \hat{H}_{\tt L} + \hat{H}_{\tt LB} + \hat{H}_{\tt B}$. Here $\hat{H}_{\tt L}$ and $\hat{H}_{\tt SL}$ describe the lead and its coupling to the system and are given explicitly by
\begin{align}
\label{eq:H_lead_1}
    \hat{H}_{\tt L} & = \sum_{k=1}^L \varepsilon_k \hat{a}^\dagger_k \hat{a}_k,\\
    \label{eq:H_lead_sys_1}
    \hat{H}_{\tt SL} & = \sum_{k=1}^{L} \left( \kappa_{kp} \hat{c}^{\dagger}_p \hat{a}_k + \kappa^*_{kp} \hat{a}^{\dagger}_k\hat{c}_p \right).
\end{align}
Each mode of the lead is further coupled to an infinite reservoir according to 
\begin{align}
    \label{H_bath}
    \hat{H}_{\tt B} &= \sum_{k=1}^L \sum_{q=1}^\infty \Omega_{kq}\hat{b}_{kq}^\dagger \hat{b}_{kq}, \\
    \label{H_lead_bath}
     \hat{H}_{\tt LB} &= \sum_{k=1}^L \sum_{q=1}^\infty \left(\zeta_{kq}\hat{a}_{k}^\dagger \hat{b}_{kq} + \zeta_{kq}^*\hat{b}_{kq}^\dagger \hat{a}_{k}\right),
\end{align}
where $\hat{a}_k$ describes mode $k$ of the lead, while the ladder operators $\hat{b}_{kq}$ describe the bath connected to mode $k$. Each bath is described by the flat spectral density
\begin{equation}\label{flat_spectral_density}
    \mathcal{J}_k(\omega) = 2\pi \sum_q |\zeta_{kq}|^2 \delta(\omega - \Omega_{kq}) = \gamma_k.
\end{equation}
We are interested in the evolution of the joint system-lead state $\rho(t)$ starting from the initial product state Eq.~\eqref{eq:product_state}, where all baths are initialised at the same temperature and chemical potential.

As in Eq~\eqref{b_formal_solution}, we formally solve the Heisenberg equation of motion for the bath variables to find 
\begin{equation}
    \hat{b}_{kq}(t) = {\rm e}^{-{\rm i}\Omega_{kq}t} \hat{b}_{kq}(0) - {\rm i}\zeta_{kq}^*\int_0^t{\rm d}t' \, {\rm e}^{-{\rm i}\Omega_{kq}(t-t')}\hat{a}_k(t').
\end{equation}
Substituting this into the equation of motion for $\hat{a}_k(t)$, we obtain
\begin{align}
\label{a_k_Langevin}
    \frac{\rm d}{{\rm d} t}\hat{a}_k(t) & = -{\rm i}\varepsilon_k \hat{a}_k(t) - {\rm i}\kappa_{kp}^* \hat{c}_p(t) \notag \\ & \quad + \hat{\xi}_k(t) - \int_0^t{\rm d}t'\, \chi_k(t-t')\hat{a}_k(t').
\end{align}
Here, we defined the noise operators
\begin{equation}
    \label{noise_lead_site_k}
    \hat{\xi}_k(t) = -{\rm i}\sum_q \zeta_{kq} {\rm e}^{-{\rm i}\Omega_{kq}t} \hat{b}_{kq}(0),
\end{equation}
and the memory kernels $\chi_k(t-t') = \langle \{\hat{\xi}_k(t),\hat{\xi}_{k}^\dagger(t')\}\rangle$. For the flat spectral density in Eq.~\eqref{flat_spectral_density}, the noise correlations are given by
\begin{align}
    \label{noise_k_memory}
     \langle \{\hat{\xi}_k(t),\hat{\xi}_{k'}^\dagger(t')\}\rangle & = \delta_{kk'}\gamma_k \delta(t-t'), \\
     \label{noise_k_correlation}
     \langle \hat{\xi}_k^\dagger(t)\hat{\xi}_{k'}(t')\rangle & = \delta_{kk'} \gamma_k \int\frac{{\rm d}\omega}{2\pi}\, f(\omega) {\rm e}^{{\rm i}\omega(t-t')}.
\end{align}

Next we formally solve Eq.~\eqref{a_k_Langevin} to find
\begin{align}
\label{lead_formal_solution}
    \hat{a}_k(t) & = {\rm e}^{-{\rm i}\varepsilon_k t - \gamma_k t/2} \hat{a}_k(0) \\ &\quad + \int_0^t{\rm d}t'\, {\rm e}^{(-{\rm i}\varepsilon_k - \gamma_k/2)(t-t')}\left[\hat{\xi}_k(t') - {\rm i} \kappa_{kp}^* \hat{c}_p(t') \right].\notag
\end{align}
Considering long times, such that $\gamma_k t \gg 1$, the first term above is negligible and will be ignored in the following. Substituting this solution into the equations of motion for the system variables, we finally obtain an effective quantum Langevin equation
\begin{align}
    \label{Langevin_eff}
        \frac{{\rm d}}{{\rm d} t}\hat{c}_j(t) & = \ii [\hat{H}_{\tt S}, \hat{c}_j(t)]  \\ 
        &\quad + \delta_{jp} \left[\hat{\xi}_{\rm eff}(t) - \int_0^t {\rm d}t'\, \chi_{\rm eff}(t-t') \hat{c}_p(t')\right].\notag
\end{align}
This is of the same form as Eq.~\eqref{QLE}, but with the noise operator
\begin{equation}\label{noise_eff}
    \hat{\xi}_{\rm eff}(t) = -{\rm i} \sum_{k=1}^L \kappa_{kp}\int_0^t{\rm d}t'\,{\rm e}^{(-{\rm i}\varepsilon_k - \gamma_k/2)(t-t')} \hat{\xi}_k(t'),
\end{equation}
and the memory kernel
\begin{align}
    \chi_{\rm eff}(t-t') & = \sum_{k=1}^L|\kappa_{kp}|^2{\rm e}^{(-{\rm i}\varepsilon_k - \gamma_k/2)(t-t')} \notag \\
    & = \int \frac{{\rm d}\omega}{2\pi} \mathcal{J}^{\rm eff}(\omega) \ee^{-\ii \omega(t-t')},
\end{align}
where the effective spectral density $\mathcal{J}^{\rm eff}(\omega)$ is the sum of Lorentzian functions 
\begin{equation}
    \label{eq:spectral_density_eff}
    \mathcal{J}^{\rm eff}(\omega)  = \sum_{k=1}^L \frac{|\kappa_{kp}|^2 \gamma_k}{(\omega-\varepsilon_k)^2 + (\gamma_k/2)^2}.
\end{equation}
The second equality above follows via an identity which can be proved by contour integration:
\begin{equation}
    \label{exp_identity}
    {\rm e}^{-{\rm i}\varepsilon_k t - \gamma_k t/2} = \int\frac{{\rm d}\omega}{2\pi}\, \frac{\gamma_k{\rm e}^{-{\rm i}\omega t}}{(\omega-\varepsilon_k)^2 + (\gamma_k/2)^2}.
\end{equation}

It remains to check the effective noise correlations. We have, using Eqs.~\eqref{noise_k_memory},~\eqref{noise_k_correlation} and~\eqref{exp_identity}, 
\begin{align}\label{memory_eff}
    \langle \{\hat{\xi}_{\rm eff}(s),\hat{\xi}_{\rm eff}^\dagger(s')\}\rangle &  \approx \int \frac{{\rm d}\omega}{2\pi} \mathcal{J}^{\rm eff}(\omega) \ee^{-\ii \omega(s-s')},\\ 
        \label{noise_eff_corr}
    \langle \hat{\xi}_{\rm eff}^\dagger(s)\hat{\xi}_{\rm eff}(s')\rangle & \approx \int \frac{{\rm d}\omega}{2\pi} \mathcal{J}^{\rm eff}(\omega) f(\omega)\ee^{\ii \omega(s-s')},
\end{align}
where we have neglected all terms proportional to ${\rm e}^{-\gamma_k s}$ or ${\rm e}^{-\gamma_k s'}$. This approximation is valid at long times, so long as the solution of Eq.~\eqref{Langevin_eff} depends only on the past history of $\hat{\xi}_{\rm eff}(s)$ within a time window that is essentially finite. This will generically be the case for any system that relaxes to a steady state when coupled to a bath, since any memory of environmental fluctuations in the far past is eventually lost. In particular, if $\tau_{\rm rel}$ is the (slowest) characteristic timescale of relaxation of ${\tt S}$, then we need consider only arguments of $\hat{\xi}_{\rm eff}(s)$ in the range $t - \tau_{\rm rel}\lesssim s < t$. Hence, the approximations leading to Eqs.~\eqref{memory_eff} and \eqref{noise_eff_corr} are valid for all times such that
\begin{equation}
    \label{noise_eff_approx}
    t \gg \gamma_k^{-1}, \tau_{\rm rel}.
\end{equation}
If this holds, we have shown that the effective noise generated by the mesoscopic lead is equivalent to an infinite bath with a spectral density given by Eq.~\eqref{eq:spectral_density_eff}, giving rise to an identical equation of motion for the system, Eq.~\eqref{Langevin_eff}.

Under this condition, the currents from the mesoscopic leads also become the same as the currents obtained in the infinite bath case. To see this, we write down the expressions for particle and energy currents from the lead,
\begin{align}
\label{curr_op_lead_def}
&J^{\textrm{P}} = i\left \langle \sum_{k=1}^{L} \left( \kappa_{kp} \hat{c}^{\dagger}_p \hat{a}_k - \kappa^*_{kp} \hat{a}^{\dagger}_k\hat{c}_p \right)\right \rangle,  \\
\label{ecurr_op_lead_def}
&J^{\textrm{E}} = i\left \langle\sum_{k=1}^{L} \varepsilon_k\left( \kappa_{kp} \hat{c}^{\dagger}_p \hat{a}_k - \kappa^*_{kp} \hat{a}^{\dagger}_k\hat{c}_p \right)\right \rangle.
\end{align}
This requires evaluation of the operators $\langle \sum_{k=1}^{L} \kappa_{kp} \hat{c}^{\dagger}_p \hat{a}_k \rangle$ and $\langle \sum_{k=1}^{L} \varepsilon_k \kappa_{kp} \hat{c}^{\dagger}_p \hat{a}_k \rangle$. From Eq.~(\ref{lead_formal_solution}), and considering the time regime in Eq.~(\ref{noise_eff_approx}), we have the following equations for evolution of these operators,
\begin{align}
\label{curr_op_lead}
&\langle \sum_{k=1}^{L} \kappa_{kp} \hat{c}^{\dagger}_p \hat{a}_k \rangle \nonumber \\
&= i\langle \hat{c}^{\dagger}_p(t) \hat{\xi}_{\rm eff}(t) \rangle -i\int_0^t {\rm d}t'\, \chi_{\rm eff}(t-t') \langle\hat{c}^{\dagger}_p(t)\hat{c}_p(t')\rangle, \\
\label{energy_curr_op_lead}
&\langle \sum_{k=1}^{L} \varepsilon_k \kappa_{kp} \hat{c}^{\dagger}_p \hat{a}_k \rangle \nonumber \\
&=i\langle \hat{c}^{\dagger}_p(t) \hat{\tilde{\xi}}_{\rm eff}(t) \rangle -i\int_0^t {\rm d}t'\, \tilde{\chi}_{\rm eff}(t-t') \langle\hat{c}^{\dagger}_p(t)\hat{c}_p(t')\rangle,
\end{align} 
where  
\begin{align}
\label{xi_tilde_t_mesoscopic}
&\hat{\tilde{\xi}}_{\rm eff}(t) = -{\rm i} \sum_{k=1}^L \varepsilon_k \kappa_{kp}\int_0^t{\rm d}t'\,{\rm e}^{(-{\rm i}\varepsilon_k - \gamma_k/2)(t-t')} \hat{\xi}_k(t'), \\
    \label{chi_tilde_t_mesoscopic}
    &\tilde{\chi}_{\rm eff}(t-t')= \int \frac{{\rm d}\omega}{2\pi} \omega\mathcal{J}_{\rm eff}(\omega) \ee^{-\ii \omega(t-t')}.
\end{align}
The operator $\hat{\tilde{\xi}}(t)$ satisfies
\begin{align}
    \label{phi_tilde_t_mesoscopic1}
\langle \hat{\tilde{\xi}}_{\rm eff}^\dagger(t)\hat{\tilde{\xi}}_{\rm eff}(t')\rangle & = \int \frac{{\rm d}\omega}{2\pi} \omega^2\mathcal{J}^{\rm eff}(\omega) f(\omega) \ee^{\ii \omega(t-t')},  \\
\label{phi_tilde_t_mesoscopic2}
\langle \hat{\tilde{\xi}}_{\rm eff}^\dagger(t)\hat{\xi}_{\rm eff}(t')\rangle & = \int \frac{{\rm d}\omega}{2\pi} \omega\mathcal{J}^{\rm eff}(\omega) f(\omega) \ee^{\ii \omega(t-t')}.
\end{align}
Here we have neglected terms proportional to $\ee^{-\gamma_k t}$ and $\ee^{-\gamma_k t'}$, following the same arguments that led to Eqs.~\eqref{memory_eff} and \eqref{noise_eff_corr}. In addition, we have made the approximation $\sum_k \varepsilon_k^n|\kappa_{kp}|^2 \gamma_k/[(\omega - \varepsilon_k)^2+ (\gamma_k/2)^2] \approx \omega^n \mathcal{J}^{\rm eff}(\omega)$, which holds so long as $\gamma_k$ is sufficiently small that the replacement $\varepsilon_k \to \omega$ in the numerator is valid. In this limit, $\mathcal{J}^{\rm eff}(\omega)$ reproduces $\mathcal{J}(\omega)$ faithfully and therefore the above equations become equivalent to Eqs.~\eqref{curr_op}--\eqref{phi_tilde_t2}. 

We note that in Eq.~\eqref{ecurr_op_lead_def} we have considered only the contribution to the current associated with the change in the lead energy, i.e. $J^E = -\langle \dd \hat{H}_{\tt L} /\dd t\rangle$. However, due to the Lindblad damping, there is an additional term associated with the change in $\hat{H}_{\tt SL}$, i.e. the second term in Eq.~\eqref{eq:enersf_explicit}. This term is of order $O(\gamma_k\kappa_{kp})$ and therefore becomes negligible in comparison to the first term in the limit $L\to \infty$. Thus, currents from the baths in the infinite bath configuration also become the same as currents from the mesoscopic lead in this regime. 
\subsection{Quantum master equation}

Finally, we briefly discuss the derivation of the quantum master equation. In the limit of large lead size, $L\to \infty$, the energy spacing $e_k=\varepsilon_{k+1}-\varepsilon_k \to 0$. So both the lead-bath couplings $\kappa_{kp} \propto \sqrt{e_k}$ and the system-lead coupling $\gamma_k=e_k$ must tend to zero in order to recover the continuum spectral density $\mathcal{J}(\omega)$ (see the discussion below Eq.~\eqref{eq:spectral_density_effective}). In this limit, we derive a quantum master equation using perturbation theory correct to $O(e_k)$. Following the standard procedure~\cite{BreuerPetruccione}, and working in an interaction picture with respect to the free Hamiltonian $\hat{H}_0 = \hat{H}_{\tt S} + \hat{H}_{\tt SL} + \hat{H}_{\tt L} + \hat{H}_{\tt B}$, we obtain
\begin{equation}
    \label{QME_app}
     \frac{\rm d}{{\rm d} t}\hat{\rho}(t) = -\int_0^\infty {\rm d}t' \,{\rm Tr}_{\tt B}\, [\hat{H}_{\tt LB}(t),[ \hat{H}_{\tt LB}(t-t'),\hat{\rho}(t)\hat{\rho}_{\tt B}]].
\end{equation}
Here, the upper limit of the $t'$ integration is taken to infinity because we consider the long-time limit, i.e. only the Born approximation and not the Markov approximation is invoked in Eq.~\eqref{QME_app}. In the interaction picture, the free evolution of the lead operators is given by
\begin{equation}\label{a_k_interaction}
    \hat{a}_k(t) = {\rm e}^{{\rm i}\hat{H}_0 t}\hat{a}_k{\rm e}^{-{\rm i}\hat{H}_0 t} = {\rm e}^{-{\rm i}\varepsilon_k t}\hat{a}_k + O(\kappa_{kp}).
\end{equation}
Since Eq.~\eqref{QME_app} is already of order $O(\gamma_k)$, we keep only the leading-order term in Eq.~\eqref{a_k_interaction}. Straightforward manipulations then lead to the master equation given by Eq.~\eqref{eq:Lindblad}. Note that the usual Lamb-shift Hamiltonian does not appear here due to the flat spectral densities in Eq.~\eqref{flat_spectral_density}.

The quantum master derived up to $O(e_k)$ is of the form
\begin{align}
\frac{\rm d}{{\rm d} t}\hat{\rho}(t) = \mathcal{L}^{(0)}\hat{\rho} + \mathcal{L}^{(1)}\hat{\rho},
\end{align}
where $\mathcal{L}^{(0)}$ is the $O(1)$ term of the Liouvillian, and $\mathcal{L}^{(1)}$ is the $O(e_k)$ term of the Liouvillian. The solution of this equation is 
\begin{align}
\hat{\rho}(t) = e^{(\mathcal{L}^{(0)} + \mathcal{L}^{(1)})t} \hat{\rho}(0),
\end{align}
which has all orders of $O(e_k)$. Clearly, all orders of $O(e_k)$ are not accurate. Following Ref.~\cite{Fleming2011}, it can be shown that the diagonal elements of $\hat{\rho}(t)$ in the eigenbasis of the system Hamiltonian $\hat{H}_{\tt S}$ are correct to $O(1)$ and error occurs at $O(e_k)$, whereas the off-diagonal elements are correct to $O(e_k)$ and the error occurs at $O(e_k^{3/2})$. Thus, by reducing $e_k$, i.e., by increasing the number of lead modes, it is possible to make results from the quantum master equation arbitrarily close to those obtained from the infinite-bath configuration.  

\section{Super-fermion formalism for non-equilibrium steady states}
\label{ap:superf}

In this appendix we give further details the superfermion \cite{Dzhioev2011} steady state solution of the master equation in Eq.~\eqref{eq:Lindblad} for a non-interacting system of size $D$ coupled a single mesoscopic lead of size $L$.

This open system has a quadratic generator $\hat{L} = \hat{\mathbf{f}}^{\dagger}\, \mathbf{L}\, \hat{\mathbf{f}} - \eta$ defined by the $2M \times 2M$ non-Hermitian matrix $\bf L$ where $M=D+L$. To compute its NESS we proceed to diagonalise this matrix as ${\bf L} = {\bf V} \, \bm{\varepsilon} \, {\bf V}^{-1}$ to give a diagonal matrix $\bm{\varepsilon}$ of complex eigenvalues $\varepsilon_\mu$. These eigenvalues come in conjugate pairs and we shall denote the half with ${\rm Im}\{\epsilon_\mu\}>0$ as set $\Xi^+$ and the other half with ${\rm Im}\{\epsilon_\mu\}<0$ as $\Xi^-$. 

We identify the corresponding normal mode operators as $\hat{\bm{\xi}}^{\dagger} = \hat{\mathbf{f}}^{\dagger}\mathbf{V}$ and $\hat{\bm{\chi}} = \mathbf{V}^{-1}\hat{\mathbf{f}}$.
Although $\hat{\chi}_{\mu}$ and $\hat{\xi}_{\mu}$ mix physical $\hat{d}_k$ and ancillary modes $\hat{s}_k$ via a similarity transformation, and so are not Hermitian conjugates of one another, they still obey canonical anticonmmutation relations \cite{Dorda2014}, e.g.
\begin{align}
\{ \hat{\chi}_{\mu}, \hat{\xi}^{\dagger}_{\nu} \} = \delta_{\mu \nu} \mathds{1}.  \label{eq:canonical_normal_modes}
\end{align}
The equations of motion for the normal mode operators follow from the commutator with $\hat{L}$ giving 
\begin{align}
[\hat{L}, \hat{\chi}_{\mu}] &= -\epsilon_{\mu} \hat{\chi}_{\mu}, \quad {\rm and} \quad [\hat{L}, \hat{\xi}^{\dagger}_{\mu}] = \epsilon_{\mu} \hat{\xi}^{\dagger}_{\mu},
\end{align}
so in vector form the time-evolved mode operators are
\begin{align}
\hat{\bm{\xi}}^{\dagger} (t) = \hat{\bm{\xi}}^{\dagger} e^{i \bm{\epsilon} t} \quad {\rm and} \quad
\hat{\bm{\chi}} (t) = e^{-\ii \bm{\epsilon} t} \hat{\bm{\chi}}.
\end{align} 

A defining property of the NESS is $\hat{L}\ket{\rho(\infty)} = 0$. Using this we compute the time-evolution of the NESS when acted upon by a normal mode operator to obtain
\begin{align}
e^{-\ii \hat{L} t}\hat{\xi}^{\dagger}_\mu\ket{\rho(\infty)} = e^{-\ii \epsilon_\mu t}\hat{\xi}^\dagger_\mu\ket{\rho(\infty)},
\end{align} 
and also
\begin{align}
e^{-\ii \hat{L} t}\hat{\chi}_\nu\ket{\rho(\infty)} = e^{\ii \epsilon_\nu t}\hat{\chi}_\nu\ket{\rho(\infty)}.
\end{align} 
For these time-evolved states not to diverge in time we require that $\hat{\xi}^\dagger_\mu\ket{\rho(\infty)} = 0$ when $\mu \in \Xi^+$ and $\hat{\chi}_\nu\ket{\rho(\infty)} = 0$ when $\nu \in \Xi^-$. This pair of constraints is analogous to those of a Fermi sea state $\ket{\rm FS}$ where $\hat{c}^\dagger_j\ket{\rm FS} = 0$ when mode $j$ is occupied, and $\hat{c}_j\ket{\rm FS} = 0$ when it is empty. Similarly for the left vacuum state $\ket{I}$ we get
\begin{align}
\bra{I} \hat{\xi}^{\dagger}_\mu e^{-\ii \hat{L} t} = e^{\ii \varepsilon_\mu t}\bra{I}\hat{\xi}^\dagger_\mu \quad {\rm and} ~\, \bra{I} \hat{\chi}_\nu e^{-\ii \hat{L} t} = e^{-\ii \varepsilon_\mu t}\bra{I}\hat{\chi}_\nu, \nonumber
\end{align} 
implying the complementary constraints $\bra{I}\hat{\xi}^\dagger_\mu = 0$ when $\mu \in \Xi^-$ and $\bra{I}\hat{\chi}_\nu = 0$ when $\nu \in \Xi^+$. Together these relations fully define the $2M \times 2M$ matrix $\mathbf{D}$ of normal mode two-point correlations of the NESS with elements
\begin{align}
D_{\mu\nu} = \bra{I} \hat{\xi}^{\dagger}_\mu\hat{\chi}_\nu\ket{\rho(\infty)}.
\end{align} 
We immediately see that $D_{\mu\nu} = 0$ whenever $\mu \in \Xi^-$ and/or $\nu \in \Xi^-$. The case $\mu,\nu \in \Xi^+$ is then determined using Eq.~\eqref{eq:canonical_normal_modes} to find that $D_{\mu\nu} = \delta_{\mu\nu}$. Hence in general we have 
\begin{eqnarray}
D_{\mu\nu} &=& \delta_{\mu\nu}\Theta({\rm Im}\{\epsilon_\mu\}>0).
\end{eqnarray} 
indicating that the set $\Xi^+$ of normal modes are the unit filled Fermi sea of the NESS.

Using this result we can evaluate physical quantities such as the single-particle Green function $G_{ij}(t,t^{\prime}) = \langle \hat{c}^{\dagger}_i(t) \hat{c}_j(t^{\prime}) \rangle = \braket{I | \hat{c}^{\dagger}_i(t) \hat{c}_j(t^{\prime}) | \rho(\infty)}$ for the system $\tt S$. Transforming back from the normal modes we have 
\begin{align}
\hat{\mathbf{f}}^{\dagger}(t) &= \hat{\bm{\xi}}^{\dagger} e^{i \bm{\epsilon} t} \mathbf{V}^{-1}, \quad {\rm and} \quad
\hat{\mathbf{f}}(t) &= \mathbf{V} e^{-i \bm{\epsilon} t} \hat{\bm{\chi}},
\end{align}
and thus the Green function follows as
\begin{align}
G_{ij}(t, t^{\prime}) &= \braket{I | \left[\hat{\mathbf{f}}^{\dagger}(t)\right]_i \left[\hat{\mathbf{f}}(t^{\prime})\right]_j | \rho(\infty) } \nonumber \\
&= \sum_{\mu , \nu} \left[ e^{i \bm{\epsilon} t} \mathbf{V}^{-1} \right]_{\mu i} \left[ \mathbf{V} e^{-i \bm{\epsilon} t^{\prime}} \right]_{j\nu} \braket{I | \hat{\xi}^{\dagger}_{\mu} \hat{\chi}_{\nu} | \rho(\infty)},  \nonumber \\
&= \sum_{\mu , \nu} \left[ \mathbf{V} e^{-i \bm{\epsilon} t^{\prime}} \right]_{j\nu} D_{\mu\nu} \left[ e^{i \bm{\epsilon} t} \mathbf{V}^{-1} \right]_{\mu i},  \nonumber \\
&= \left[ \mathbf{V} e^{-i \bm{\epsilon} t^{\prime}} \, \mathbf{D} \, e^{i \bm{\epsilon} t} \mathbf{V}^{-1} \right]_{j i},
\end{align}
where we have used that $\mathbf{D}$ is diagonal and the indices $i,j = (L+1),\dots,M$ give the physical system $\tt S$ modes. This reduces to the NESS expectation value in Eq.~\eqref{eq:RDM_SF_quadratic} once $t=t'=0$. The Fermi sea structure of the NESS allows Wick's theorem to be applied to breakup expectation values for high-order correlations into two-point ones, for example
\begin{align}
\bra{I} \hat{\xi}^{\dagger}_\mu\hat{\chi}_\nu \hat{\xi}^{\dagger}_\tau\hat{\chi}_\sigma \ket{\rho(\infty)} = \bra{I} \hat{\xi}^{\dagger}_\mu\hat{\chi}_\nu \ket{\rho(\infty)}\bra{I} \hat{\xi}^{\dagger}_\tau\hat{\chi}_\sigma \ket{\rho(\infty)}\nonumber \\
+ \bra{I} \hat{\xi}^{\dagger}_\mu\hat{\chi}_\sigma \ket{\rho(\infty)}\bra{I} \hat{\chi}_\nu\hat{\xi}^{\dagger}_\tau \ket{\rho(\infty)}\nonumber \\
- \bra{I} \hat{\xi}^{\dagger}_\mu\hat{\xi}^{\dagger}_\tau \ket{\rho(\infty)}\bra{I} \hat{\chi}_\nu\hat{\chi}_\sigma \nonumber \ket{\rho(\infty)},
\end{align}
leaving products of terms that can be readily evaluated using the NESS normal mode constraints determined above.

\section{Transmission functions in Landauer-B\"uttiker theory}
\label{ap:transmission}

In this appendix we briefly introduce the methodology to compute the transmission functions $\tau(\omega)$ from Eqs.~\eqref{eq:partlb} and \eqref{eq:enerlb}. As remarked before, these functions are required to compute the currents in Landauer-B\"uttiker theory which correspond to our point of comparison for non-interacting systems [Secs.~\ref{sec:non_interacting_example} and \ref{sec:noninteracting}].

The transmission function can be obtained in terms of the non-equilibrium Green's function \cite{ryndyk2016nano,Purkayastha2019}
\begin{align}
\mathbf{G}(\omega) = \mathbf{M}^{-1}(\omega).
\end{align}
For the specific case of a system composed of $D$ fermionic sites connected to leads on sites $j = 1$ and $j = D$, $\mathbf{M}(\varepsilon)$ can be expressed as
\begin{align}
\mathbf{M}(\omega) = \omega \mathds{1} - \mathbf{H}_{\tt S} - \mathbf{\Sigma}^{(1)}(\omega) - \mathbf{\Sigma}^{(D)}(\omega),
\end{align}
where $\mathbf{H}_S$ is the Hamiltonian matrix of the system and $\mathbf{\Sigma}(\omega)$ corresponding to self-energy matrices of the leads. The only non-zero elements of the latter are given by
\begin{align}
[\mathbf{\Sigma}^{(j)}]_{jj} (\omega) = \frac{1}{2\pi} \textrm{P.V.}\int d \omega^{\prime} \frac{\mathcal{J}(\omega^{\prime})}{(\omega^{\prime} - \omega)} - \frac{i}{2}\mathcal{J}(\omega),\; \forall j = 1, D;
\end{align}
where P.V. denotes principal value and $\mathcal{J}(\omega)$ is the spectral function of the leads. In our configuration, both leads are of equivalent form. For the sake of comparison between L-B theory and mesoscopic reservoirs, we employ the wide-band approximation in which 
\begin{align}
\mathcal{J}(\omega) = \begin{cases} \Gamma,\; \forall\, \omega \in [-W, W] \\ 0,\; \textrm{otherwise} \end{cases}
\end{align}
where $\Gamma$ is the coupling strength between the system and the leads. Under these considerations, the transmission function for a system composed of $D$ fermionic sites with $\hat{H}_{\tt S}$ from Eq.~\eqref{eq:h_s_m} is given by
\begin{align}
\label{eq:transm}
\tau(\varepsilon) = \mathcal{J}^2(\varepsilon) | [\mathbf{G}(\varepsilon)]_{1D} | ^2 = \frac{\mathcal{J}^2(\varepsilon)}{| \textrm{det} [\mathbf{M}] |^2} \prod_{i=1}^{D-1}t_{S,i}^2.
\end{align}
When the central system is a single-level with $\hat{H}_{\tt S}$ from Eq.~\eqref{eq:h_s_d}, the transmission function can be proven to be of Lorentzian form and equivalent to
\begin{align}
\label{eq:lorentzian_transmission}
\tau_{\rm SL}(\varepsilon) = \frac{\mathcal{J}^2(\varepsilon)}{| \textrm{det} [\mathbf{M}] |^2},
\end{align}
while a central system composed of $D$ fermionic sites with $\hat{H}_S$ from Eq.~\eqref{eq:h_s_m} has a transmission function which corresponds to a convolution of Lorentzian functions whose form depends on the site energies $\epsilon$ and hopping amplitudes $t_{\tt S}$, as observed from Eq.~\eqref{eq:transm}. With the previous expressions for $\tau(\varepsilon)$, Eqs.~\eqref{eq:partlb} and \eqref{eq:enerlb} can then be evaluated numerically to obtain particle and energy currents for a given system. 

\section{Definitions of currents}
\label{ap:conteqs}
 
We discuss here the energy and particle current in more detail. In the mesoscopic-lead configuration, the currents are found from the continuity equation and given by Eq.~\eqref{eq:currents_def}. The currents are straightforward to evaluate using the adjoint dissipator $\mathcal{L}_{\alpha}^\dagger$, for $\alpha = L,R$, which satisfies ${\rm Tr}[\hat{A} \mathcal{L}_\alpha\{\hat{B}\}] = {\rm Tr}[\mathcal{L}_\alpha^\dagger\{\hat{B}\} \hat{A}]$ for an arbitrary operator $\hat{A}$. For the Lindblad dissipator in Eq.~\eqref{eq:Lindblad}, we have
\begin{align}
    \mathcal{L}_{\tt L}^\dagger \{ \bullet\} & = \sum_{k=1}^{L} \gamma_k(1 - f(\varepsilon_k)) \left[\hat{a}_k^{\dagger} \bullet \hat{a}_k - \tfrac{1}{2}\{ \hat{a}^{\dagger}_k \hat{a}_k, \bullet \} \right] \nonumber \\
& + \sum_{k=1}^{L} \gamma_k f(\varepsilon_k) \left[\hat{a}_k \bullet \hat{a}^{\dagger}_k - \tfrac{1}{2}\{ \hat{a}_k \hat{a}^{\dagger}_k, \bullet\} \right].
\end{align}
Since this superoperator acts only on the lead degrees of freedom, we find the explicit expressions quoted in Eqs.~\eqref{eq:partsf_explicit} and ~\eqref{eq:enersf_explicit} with straightforward algebra.

In sufficiently large central systems, an alternative definition of the currents can be derived from the continuity equations within the system itself. Let us focus on 1D systems with two-body interactions coupled to two baths at the first and final sites $j=1,D$, as considered in the examples of Secs.~\ref{sec:non_interacting_example} and \ref{sec:interacting}. In this case, the fermion number and Hamiltonian can be written as
\begin{equation}
    \label{N_H_1D}
    \hat{N}_{\tt S} = \sum_{j=1}^{D} \hat{n}_j, \qquad \hat{H}_{\tt S} = \sum_{j=1}^{D-1} \hat{h}_{j,j+1},
\end{equation}
where $ \hat{n}_j = \hat{c}_j^\dagger \hat{c}_j$ is the local fermion density on site $j$ and $\hat{h}_{j,j+1}$ denotes a local energy density operator. Since $\hat{h}_{j,j+1}$ has support only on sites $j$ and $j+1$, we derive the continuity equation for number density from the Heisenberg equation for $\hat{n}_j$:
\begin{equation}
    \label{cont_n_sys}
    \frac{\rm d}{{\rm d} t} \hat{n}_j = \hat{J}^{\textrm{P}}_{j-1\to j} -  \hat{J}^{\textrm{P}}_{j\to j+1},
\end{equation}
where we defined the particle current operator 
\begin{equation}
\hat{J}^{\textrm{P}}_{j-1\to j} = {\rm i} [\hat{h}_{j-1,j}, \hat{n}_j],    
\end{equation}
 which clearly depends only on system variables. In the steady state, the time derivatives of all expectation values vanish and we find that the current is homogeneous, i.e.\ $\langle \hat{J}^{\textrm{P}}_{j-1\to j}\rangle = \langle \hat{J}^{\textrm{P}}_{j\to j+1}\rangle$. 
 
\begin{figure*}[t]
\fontsize{13}{10}\selectfont 
\centering
\includegraphics[width=1.8\columnwidth]{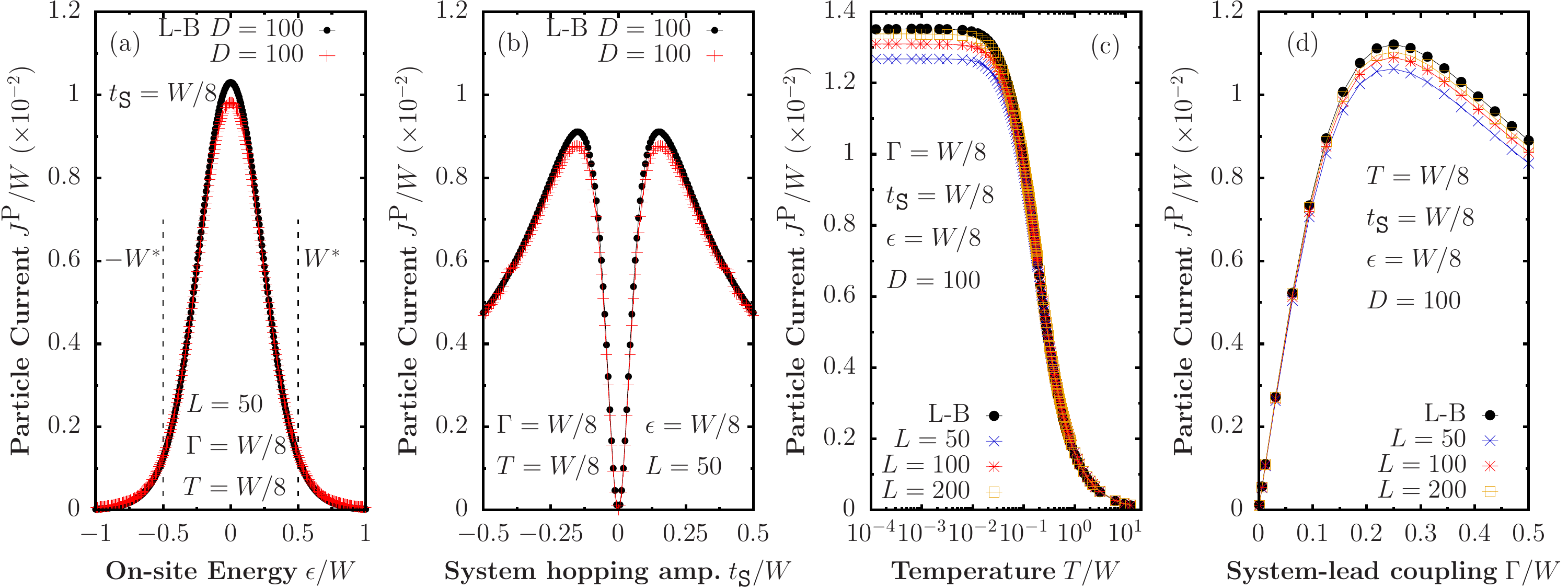}
\caption{Particle current from L-B and mesoscopic reservoir predictions flowing from the left lead and into the system (a) as a function of the on-site energy (same parameter for every site) for a central system with $D = 100$ sites and a fixed number of modes in the leads $L = 50$, and (b) as a function of the hopping amplitude $t_{\tt S}$ (same parameter for every site). In panels (c) and (d) we fix every parameter and study the particle current as a function of temperature and system-lead coupling, respectively. In these calculations we used $\mu_{\texttt L} = -\mu_{\texttt R} = W / 16$, $T_{\texttt L} = T_{\texttt R}$, $L_{\textrm{log}} / L = 0.2$ and $W^* = W / 2$.}
\label{fig:18}
\end{figure*}
 
Eq.~\eqref{cont_n_sys} holds only for $j\neq 1,D$. For $j=1$, for example, we have instead that \begin{equation}
    \label{cont_n_site1}
    \frac{\rm d}{{\rm d} t} \hat{n}_1 =  {\rm i}[\hat{H}_{\tt SL}, \hat{n}_{1}] - \hat{J}^{\textrm{P}}_{1\to 2}.
\end{equation}
Meanwhile, the mean number of particles in the left reservoir obeys the equation
\begin{equation}
    \label{cont_N_L}
    \frac{\rm d}{{\rm d} t} \left\langle \hat{N}_{\tt L} \right \rangle = J^{\textrm{P}}_{\tt L} + \left\langle {\rm i}[\hat{H}_{\tt SL}, \hat{n}_{1}]\right\rangle.
\end{equation}
Here we used the fact that $[\hat{H}_{\tt SL}, \hat{N}_{\tt L} + \hat{n}_1] = 0$, which merely reflects the overall conservation of fermion number and the fact that ${\tt L}$ couples only to site $j=1$. Combining Eqs.~\eqref{cont_n_site1} and \eqref{cont_N_L} and assuming steady-state conditions we deduce that
\begin{equation}
    \label{J_P_equiv}
    J^{\textrm{P}}_{\tt L} = \left\langle \hat{J}^{\textrm{P}}_{1\to 2}\right \rangle.
\end{equation}
Therefore, so long as the system comprises $D\geq 2$ sites, the current computed via Eq.~\eqref{eq:partsf_explicit} coincides with the expectation value of a system operator.

For the energy current, one similarly finds in the bulk of the system 
\begin{equation}
    \label{cont_h_sys}
    \frac{\rm d}{{\rm d} t} \hat{h}_{j,j+1} = \hat{J}^{\textrm{E}}_{j-1\to j+1} -  \hat{J}^{\textrm{E}}_{j\to j+2},
\end{equation}
where
\begin{equation}
\hat{J}^{\textrm{E}}_{j-1\to j+1} = {\rm i} [\hat{h}_{j-1,j},\hat{h}_{j,j+1}].
\end{equation}
Considering the leftmost site, on the other hand, 
\begin{equation}
    \label{cont_h1}
    \frac{\rm d}{{\rm d} t} \hat{h}_{1,2} = {\rm i}[\hat{H}_{\tt SL},\hat{h}_{1,2}] - \hat{J}^{\textrm{E}}_{1\to 3}.
\end{equation}
Now, considering the Heisenberg equations for both $\hat{H}_{\tt SL}$ and $\hat{H}_{\tt L}$ and assuming steady-state conditions, we conclude that
\begin{equation}
    \label{J_E_equiv}
    J^{\textrm{E}}_{\tt L} = \left\langle \hat{J}^{\textrm{E}}_{1\to 3}\right \rangle.
\end{equation}
Therefore, the energy current computed from Eq.~\eqref{eq:enersf_explicit} also coincides with the expected value of a system operator, so long as $D\geq 3$. 

The above arguments, although developed for the specific case of two-body interactions in one dimension, are based only on conservation laws and the locality of interactions, which are general principles. Similar arguments can thus be developed for more general $n$-body interacting systems in higher-dimensional geometries, so long as a sufficiently large region of the central system is not directly connected to the baths.

\begin{figure*}[t]
\fontsize{13}{10}\selectfont 
\centering
\includegraphics[width=1.8\columnwidth]{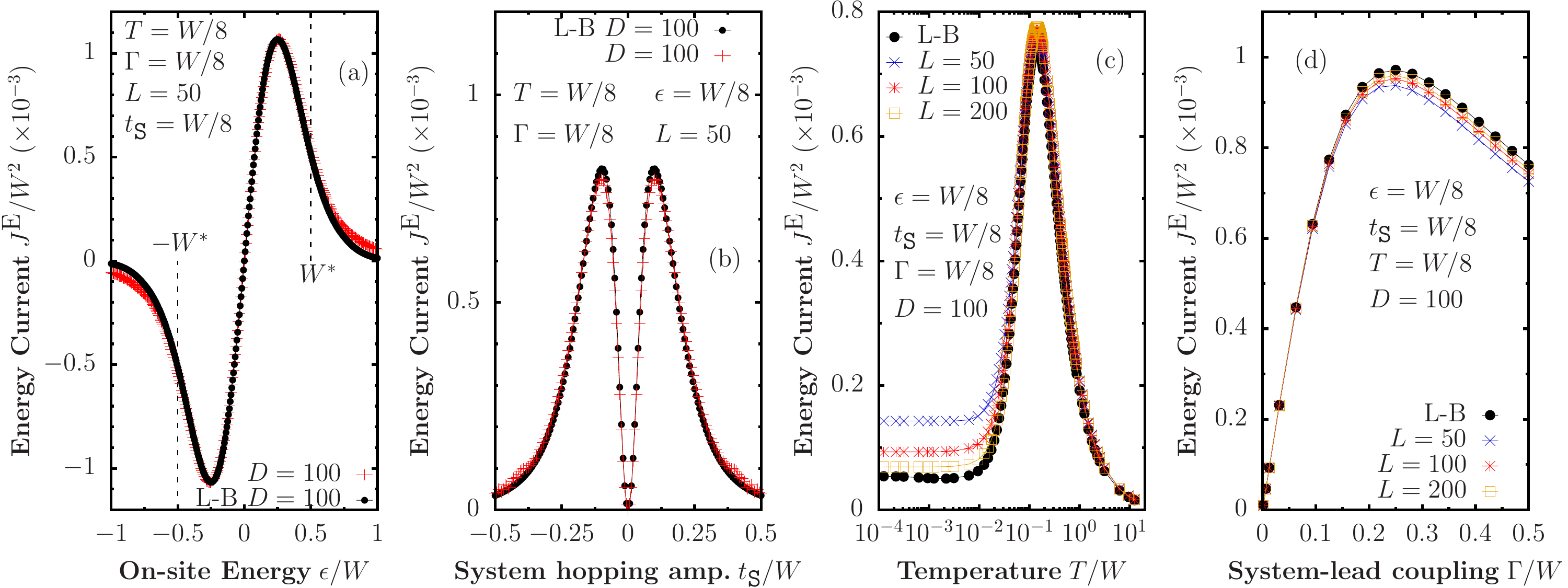}
\caption{Energy current from L-B and mesoscopic reservoir predictions flowing from the left lead and into the system (a) as a function of the on-site energy (same parameter for every site) for a central system with $D = 100$ sites and a fixed number of modes in the leads $L=50$, and (b) as a function of the hopping amplitude $t_{\tt S}$ (same parameter for every site). In panels (c) and (d) we fix every parameter and study the energy current as a function of temperature and system-lead coupling, respectively. In these calculations we used $\mu_{\texttt L} = -\mu_{\texttt R} = W / 16$, $T_{\texttt L} = T_{\texttt R}$, $L_{\textrm{log}} / L = 0.2$ and $W^* = W / 2$.}
\label{fig:19}
\end{figure*}

\section{Many fermionic sites}
\label{sec:noninteracting}

Another configuration of interest is a system composed of many fermionic sites, one for which we can express the Hamiltonian as
\begin{align}
\label{eq:h_s_m}
\hat{H}_{\tt S} = \sum_{j = 1}^{D} \epsilon_j \hat{c}^{\dagger}_j \hat{c}_j - \sum_{j = 1}^{D - 1} t_{\tt S} \left( \hat{c}^{\dagger}_{j+1} \hat{c}_{j} + \textrm{H.c.} \right),
\end{align}
where $\hat{c}^{\dagger}_j$ and $\hat{c}_j$ are fermionic creation and destruction operators and $D$ is the number of sites in the system. We couple the leftmost and rightmost sites of this system to mesoscopic reservoirs, as shown in Fig.~\ref{fig:meso_leads_lindblad}. 

Given that our expressions for particle and energy currents in Eqs.~\eqref{eq:partsf_explicit} and \eqref{eq:enersf_explicit} are defined in terms of canonical operators in the leads, the corresponding expressions for the case of a many-fermionic central system are equivalent to those of a single-level system. 
For a sufficiently large amount of sites in the central system, these operators can be defined in terms of just system operators. Here, however, we will use the expressions in Eqs.~\eqref{eq:partsf_explicit} and \eqref{eq:enersf_explicit} which are general for any number of sites $D$.

We now evaluate whether the mesoscopic lead configuration can provide a good approximation of the continuum even if the central system is composed of many fermionic sites. In a similar fashion as for the single-level system, in Fig.~\ref{fig:18}(a) we present the particle current flowing from the left lead and into system as a function of the on-site energy $\epsilon = \epsilon_j$ for every site $j$. In our calculations we use the same macroscopic parameters as before, given by $T_{\tt L} = T_{\tt R} = W / 8$ and $\mu_{\tt L} = -\mu_{\tt R} = W/16$. We fix the number of energy modes in each lead to $L = 50$ and the number of sites in the central system to $D = 100$. The Landauer-B\"uttiker calculations are done by evaluating Eq.~\eqref{eq:partlb} using the transmission function obtained as described in Appendix~\ref{ap:transmission}. It can be observed that for a fixed number of modes in the leads $L$ and a fixed number of sites in the central system $D$ the approximation to the continuum limit using mesoscopic reservoirs is robust to a wide range of on-site energies. The small oscillations that can be observed near the band edges at $|\epsilon| \gtrapprox |W^*|$ are due to the logarithmic spacing of modes. Furthermore, from Fig.~\ref{fig:18}(b), the same can be said when $\epsilon$ is fixed and $t_{\tt S}$ is changed to different values. Given that the energies in the central system are bounded by $-2t_{\tt S}$ and $2t_{\tt S}$, the oscillations due to logarithmic discretisation are observed close to $t_{\tt S} \approx W / 2$. The same observations hold for energy current in Figs.~\ref{fig:19}(a) and \ref{fig:19}(b)

As a function of temperature, a similar behaviour as for the single-level system can be observed. In particular, for particle current and energy current in Figs.~\ref{fig:18}(c) and \ref{fig:19}(c), respectively, the continuum is properly approximated with the exception of the values of temperature that are lower than the minimum energy spacing of the modes in the leads. For these small temperatures, the Fermi-Dirac distributions of the leads resemble a Heaviside step function and the discontinuity can no longer be well-captured by discrete and broadened energy modes. Following from our previous discussion for the single-level system, to obtain a better approximation at lower temperatures one can either increase the number of total energy modes or decrease the width of the window $[-W^*, W^*]$. The former choice comes with the cost of a larger computational complexity, while with the latter one can then only provide a good approximation of the continuum for a smaller range in the parameter space of $\epsilon$, $t_{\tt S}$, $\mu_{\tt L}$ and $\mu_{\tt R}$. If these values are fixed, a good choice of $[-W^*, W^*]$ can be used to obtain better approximations at lower temperatures with its limit, as discussed for the single-level system, related to the minimum value of $e_k$ in the linearly-discretised region. 

As a function of the system-lead coupling, the results are very robust to a wide range of values as observed from Figs.~\ref{fig:18}(d) and \ref{fig:19}(d). Because of the ballistic (coherent) nature of transport in the central system, currents become independent of $D$ in the asymptotic regime.

\begin{figure}[b]
\fontsize{13}{10}\selectfont 
\centering
\includegraphics[width=0.9\columnwidth]{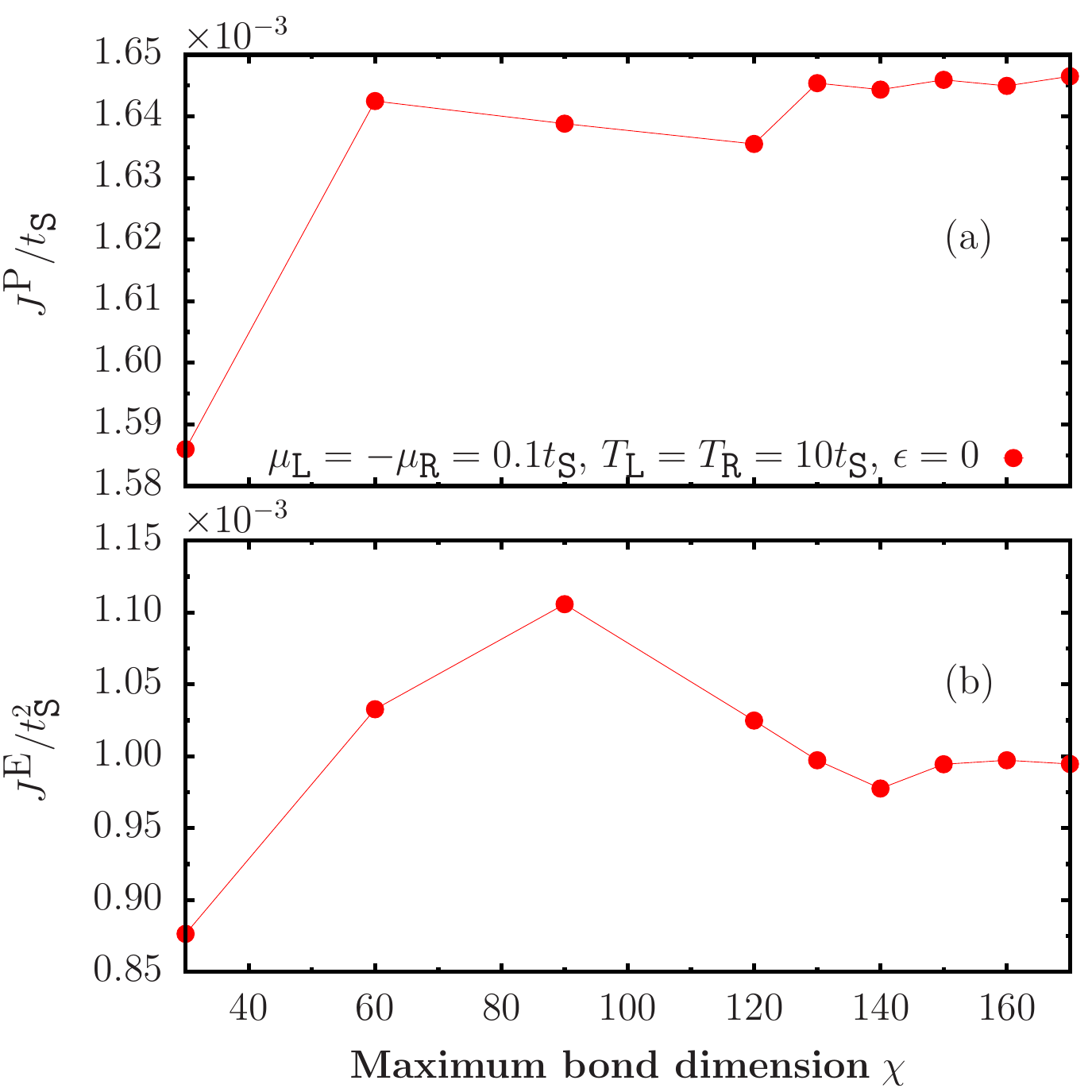}
\caption{ Convergence of (a)~particle current and (b)~energy current as a function of the maximal bond dimension $\chi$ across the system, for a particular driving configuration of the anisotropic Heisenberg model in Eq.~\eqref{eq:h_s_i}. The symbols correspond to the converged currents for a fixed $\chi$. From $\chi \approx 150$ both currents remain essentially unchanged. In these calculations we used $D = 40$, $L = 20$, $L_{\textrm{log}} / L = 0.2$, $W^* = W / 2$, $W = 8t_{\tt S}$ and $\Gamma = U = t_{\tt S}$.}
\label{fig:20}
\end{figure}

\section{Convergence and computation time}
\label{ap:converge}

The bond dimension $\chi$, discussed in Sec.~\ref{sec:ness_solve}, is a relevant parameter that is inherently associated to the fidelity with which a tensor network mathematically represents a quantum object. The complexity of finding the long-time solution to Eq.~\eqref{eq:superfermion_evolve} grows exponentially with system size using a full representation of the quantum state $\ket{\hat{\rho}(t)}$. However, such state can be described by a tensor network, with its maximum bond dimension directly connected to how accurately the state is represented \cite{Brenes2018}. The purpose of this Appendix is to exemplify how the NESS can be accurately represented with a bond dimension $\chi$ that keeps calculations tractable. 

Starting from our non-equilibrium configuration depicted in Fig.~\ref{fig:thermal_machine}, we set $L=20$ lead modes for both left and right reservoirs, $D=40$ system sites, $T_{\tt L} = T_{\tt R} = 10t_{\tt S}$, $\mu_{\tt L} = -\mu_{\tt R} = 0.1t_{\tt S}$, and $\epsilon_j / t_{\tt S} = \epsilon / t_{\tt S} = 0$, for the anisotropic Heisenberg model with $U = t_{\tt S}$; this configuration is thus away from CP symmetry. We proceed to evaluate both the particle and energy currents in the NESS by employing the algorithm described in Sec.~\ref{sec:tensor} as a function of the maximum bond dimension $\chi$. The results are shown in Fig.~\ref{fig:20}. It can be observed that as the bond dimension is increased, both currents converge to a given value within a few percent of accuracy.

\begin{figure}[t]
\fontsize{13}{10}\selectfont 
\centering
\includegraphics[width=0.9\columnwidth]{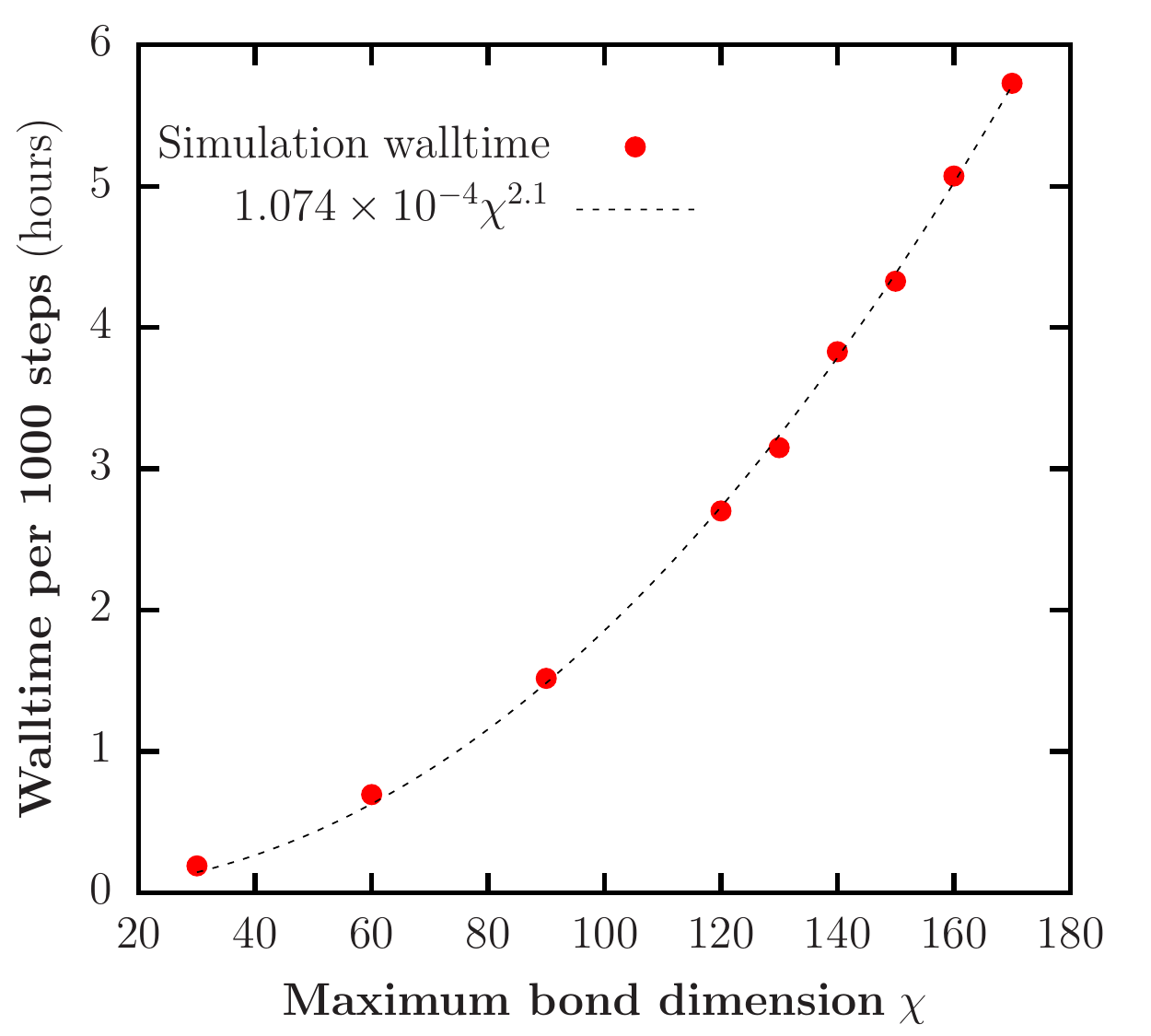}
\caption{Average simulation walltime for 1000 evolution steps as a function of the maximal bond dimension $\chi$ across the system, for a particular driving configuration of the anisotropic Heisenberg model in Eq.~\eqref{eq:h_s_i}. The symbols correspond to the converged currents for a fixed $\chi$, and the dashed line to a power-law fit. In these calculations we used $D = 40$, $L = 20$, $L_{\textrm{log}} / L = 0.2$, $W^* = W / 2$, $W = 8t_{\tt S}$ and $\Gamma = U = t_{\tt S}$.}
\label{fig:21}
\end{figure}

To illustrate the computational complexity of the algorithm, we have calculated the overall simulation walltime as a function of $\chi$. The results are shown in Fig.~\ref{fig:21} and exhibit the commonly-found polynomial complexity of time evolution in the class of tensor network algorithms. Furthermore, even though the bond dimension is homogeneous in the bulk of the system, we observe a lower scaling compared to the naively-expected $\chi^3$ power law for an algorithm dominated by singular value decomposition processes~\cite{schollwock:2011}. We associate this faster behaviour to the use of a divide-and-conquer decomposition algorithm~\cite{tnt_review1}, which rapidly converges deep within the time evolution. Thus, in spite of the polynomial growth of computational time as a function of the bond dimension, accurate approximations can be obtained within tractable computation times.

\section{CP symmetry}
\label{app:CP}

Here we prove that the energy current vanishes in the Heisenberg model described by Eq.~\eqref{eq:h_s_i} under conditions of combined charge conjugation-parity (CP) symmetry. The symmetry corresponds to a unitary transformation $\hat{\mathcal{C}}\hat{\mathcal{P}}$, with the particle-hole transformation $\hat{\mathcal{C}}$ and the parity transformation $\hat{\mathcal{P}}$.

In the bulk of the system, the parity and particle-hole transformations are respectively defined by
\begin{align}
    \label{P_bulk}
    \hat{\mathcal{P}}\hat{c}_j\hat{\mathcal{P}}^\dagger & = \hat{c}_{D-j+1}.\\
 \label{C_bulk}
    \hat{\mathcal{C}}\hat{c}_j\hat{\mathcal{C}}^\dagger & = (-1)^{j+1}\hat{c}_j^\dagger.
\end{align}
The phase factor in $\hat{\mathcal{C}}$ is defined so that particle excitations are mapped to hole excitations with the same kinetic energy. The bulk Hamiltonian in Eq.~\eqref{eq:h_s_i} is invariant under $\hat{\mathcal{P}}$, i.e.~$\hat{\mathcal{P}}\hat{H}_{\tt S}\hat{\mathcal{P}} = \hat{H}_{\tt S}$, and also invariant under $\hat{\mathcal{C}}$ so long as $\epsilon = -U$. 

The particle-hole transformation for the lead operators that is consistent with the action of $\hat{\mathcal{C}}$ in the bulk is of the form
\begin{align}
    \label{C_bath_L}
    \hat{\mathcal{C}}\hat{a}_{k,\tt L}\hat{\mathcal{C}}^\dagger & = -\hat{a}^\dagger_{L-k,\tt L},\\
    \label{C_bath_R}
    \hat{\mathcal{C}}\hat{a}_{k,\tt R}\hat{\mathcal{C}}^\dagger & = (-1)^D\hat{a}^\dagger_{L-k,\tt R},
\end{align}
while spatial reflection simply consists of the swap ${{\tt L}\leftrightarrow {\tt R}}$. With these conventions, the total Hamiltonian is invariant under $\hat{\mathcal{P}}$ if the left and right leads have identical spectra $\varepsilon_k$ and system-bath couplings $\kappa_{kp}$. The Hamiltonian is also invariant under $\hat{\mathcal{C}}$ if the lead spectra and couplings are symmetric around the centre of the band, i.e. $\varepsilon_k = -\varepsilon_{L-k}$ and $\kappa_{k,p} = \kappa_{L-k,p}$. Finally, the non-equilibrium forcing is CP-symmetric if the bath temperatures are equal, $T_{\tt L} = T_{\tt R}$, and the chemical potentials are opposite, $\mu_{\tt L}=-\mu_{\tt R}$, while the dissipation rates are invariant under spatial reflection and inversion about the centre of the band, i.e.~$\gamma_{k,\tt L} = \gamma_{L-k,\tt L} = \gamma_{k,\tt R}$. 

Under the above assumptions, the generator of the master equation is invariant under a combined CP transformation and therefore so is the steady state, i.e. $\hat{\mathcal{C}}\hat{\mathcal{P}}\hat{\rho}(\infty)(\hat{\mathcal{C}}\hat{\mathcal{P}})^\dagger = \hat{\rho}(\infty)$. At the particle-hole symmetric point of the Hamiltonian, with $\epsilon = -U$, the bulk energy current operator (defined in Sec.~\ref{ap:conteqs}) is odd under a CP transformation, in the sense that $\hat{\mathcal{C}}\hat{\mathcal{P}} \hat{J}^{\rm E}_{j-1\to j+1}(\hat{\mathcal{C}}\hat{\mathcal{P}})^\dagger = -\hat{J}^{\rm E}_{D-j\to D-j+2}$. It follows that
\begin{align}
\label{JE_is_zero}
\langle \hat{J}^{\rm E}_{j-1,j+1}\rangle = \langle  \hat{\mathcal{C}}\hat{\mathcal{P}}\hat{J}^{\rm E}_{j-1,j+1}(\hat{\mathcal{C}}\hat{\mathcal{P}})^\dagger\rangle = -\langle \hat{J}^{\rm E}_{D-j,D-j+2}\rangle,
\end{align}
and therefore $\langle \hat{J}^{\rm E}_{j-1,j+1}\rangle = -\langle \hat{J}^{\rm E}_{j-1,j+1}\rangle = 0$ because the mean current is homogeneous in the steady state. Note that the particle current operator is even and therefore is not constrained by CP symmetry. However, the particle density transforms as $\hat{\mathcal{C}}\hat{\mathcal{P}} \hat{n}_j(\hat{\mathcal{C}}\hat{\mathcal{P}})^\dagger = 1-\hat{n}_{D-j+1}$, so that in a CP-symmetric steady state we have $\langle \hat{n}_j\rangle  + \langle \hat{n}_{D-j+1}\rangle = 1$. In a ballistic regime with $\langle \hat{n}_j\rangle  = \rm const.$, we must therefore have $\langle \hat{n}_j\rangle = 0.5$, consistent with the trend in Fig.~\ref{fig:17} at high temperature.

\bibliography{bibliography.bib}

\end{document}